\documentclass{article}

\usepackage{float}
\usepackage{placeins}
\usepackage{amsmath}
\usepackage{amssymb}
\usepackage{bbm}
\usepackage{color}
\usepackage{geometry}
\usepackage[hidelinks]{hyperref}
\usepackage{physics}
\usepackage{slashed}

\usepackage{feynarts}

\def\SO10{\text{SO}(10)}

\def\SU{\,\text{SU}}

\newcommand{\ii}{\mathrm{i}}

\newcommand{\PL}{P_\text{L}}
\newcommand{\PR}{P_\text{R}}

\begin{document}

\vspace{0.25cm}

\begin{center}
{\Large One-Loop Matching Conditions in Neutrino Effective Theory}
\end{center}

\vspace{0.25cm}

\begin{center}
{\bf Tommy Ohlsson}$^{a,b}$~\footnote{E-mail: tohlsson@kth.se} {\bf and}
{\bf Marcus Pernow}$^{a,b}$~\footnote{E-mail: pernow@kth.se}
\\
\vspace{0.5cm}
{\small $^a$Department of Physics, School of Engineering Sciences, KTH Royal Institute of Technology, \\
AlbaNova University Center, Roslagstullsbacken 21, SE-106 91 Stockholm, Sweden \\
\smallskip
$^b$The Oskar Klein Centre for Cosmoparticle Physics, AlbaNova University Center, \\
Roslagstullsbacken 21, SE-106 91 Stockholm, Sweden}
\end{center}

\vspace{0.5cm}

\begin{abstract}
We investigate matching conditions and threshold corrections between full and effective theories based on the type I seesaw mechanism. In general, using an intuitive Feynman diagrammatical approach, we compute the amplitudes before and after integrating out heavy right-handed neutrinos at the matching scale. In particular, we derive the one-loop matching conditions between the full and the effective theories. The matching conditions of the parameters are influenced by one-loop corrections to the corresponding vertices as well as wave function corrections for the Higgs and the lepton fields. Our results are comparable to earlier results based on a functional approach.
\end{abstract}

\section{Introduction}

Neutrino masses can be generated in an effective field theory using the dimension-5 Weinberg operator~\cite{Weinberg:1979sa}. To relate the UV-completion of the effective operator to the values of the neutrino masses measured in experiments, one has to match the effective operator to the full theory as well as compute its renormalization group (RG) running. 

The RG running of the Weinberg operator is well known~\cite{Babu:1993qv,Chankowski:1993tx,Antusch:2001ck,Chankowski:2001hx}. In the case of the type~I seesaw mechanism with non-degenerate right-handed neutrino (RHN) masses, one must consider a sequence of effective field theories with a matching at each threshold and RG running between them~\cite{Antusch:2002rr}. The phenomenological consequences for some parameters of interest of this were further explored in Ref.~\cite{Antusch:2005gp}. This was extended to include the type II seesaw mechanism in Refs.~\cite{Chao:2006ye,Schmidt:2007nq}. For the RHN mass matrix, the two-loop RG equation was derived in Ref.~\cite{Ibarra:2020eia}. Two-loop RG equations for the RHN mass matrix have also found to be able to radiatively increase the rank of the light neutrino mass matrix from two to three (thereby generating a mass of the lightest neutrino in models with zero smallest mass)~\cite{Davidson:2006tg,Xing:2020ezi}. In the Standard Model (SM), RG equations at two-loop level were derived in the original papers~\cite{Machacek:1983tz,Machacek:1983fi,Machacek:1984zw} as well as in Refs.~\cite{Luo:2002ey,Luo:2002ti}. For more general dimension-5 and 7 operators, the renormalization was discussed in Ref.~\cite{Chala:2021juk} and the generation of neutrino masses from operators of dimension higher than five was considered in Ref.~\cite{Bonnet:2009ej}.

In general, the matching between two theories at some energy scale receives contributions at both tree- and loop-level~\cite{Weinberg:1980wa,Hall:1980kf}. Early studies on threshold effects of neutrino masses have focused on the effect due to RG running between thresholds \cite{Mohapatra:2005gs,Bergstrom:2010id,Gupta:2014lwa}. Loop-level threshold effects at $M_Z$ due to integrating out the Higgs and the weak gauge bosons were also studied in Refs.~\cite{Chankowski:2001hx,Chankowski:2001mx}.

In supersymmetric theories, in which the non-renormalization theorem makes it somewhat simpler, the matching conditions of the type I seesaw mechanism to the neutrino effective theory at one-loop level were computed in Ref.~\cite{Antusch:2015pda}. This has been used to analyze corrections to the smallest neutrino mass~\cite{Zhou:2021bqs}. In Ref.~\cite{Zhang:2021jdf}, a complete one-loop matching of the type I seesaw model onto the SM effective field theory has been performed using a functional approach and the results were presented in the so-called Green's and Warsaw bases. These results have been checked \cite{Carmona:2021xtq}.

In this work, following Refs.~\cite{Pich:1998xt,Manohar:2018aog,Cohen:2019wxr}, the approach is to match the amplitudes computed to one-loop level in the full type~I seesaw mechanism and the SM effective theories using Feynman diagrams and rules. As such, we draw and compute all one-light-particle-irreducible (1LPI) Feynman diagrams in both theories, including tree-level, one-loop, and one-loop counterterms. We assume the three RHNs to be degenerate.

This work is organized as follows. In Sec.~\ref{sec:basics}, we present the basics including the Lagrangians and the conventions of the full and the effective theories. Then, in Sec.~\ref{sec:tlfdm}, we discuss the matching at tree level. Next, in Sec.~\ref{sec:clfd}, we perform the computations of the loop diagrams at one-loop level for the full and the effective theories. The computations include the lepton and the Higgs propagators, the gauge couplings, the lepton Yukawa coupling, and the Higgs quartic coupling, as well as a discussion of the computations of the neutrino mass matrix in both the full and the effective theory. In Sec.~\ref{sec:mp}, we investigate in detail the matching procedure including wave function corrections, the Higgs mass, the U(1), SU(2), and SU(3) gauge couplings, the quark and the lepton Yukawa couplings, and the Higgs quartic coupling. Finally, in Sec.~\ref{sec:sc}, we summarize our results for the matching conditions and draw our conclusions.

\section{Basics}\label{sec:basics}

\subsection{Lagrangian}

\subsubsection{Standard Model Lagrangian}

To solidify formalism, notation and conventions, we give some of the terms of the Lagrangian. Notably, we work in the symmetric phase of the SM. The Lagrangian for the Higgs doublet field $\phi$ is given by
\begin{equation}
    \mathcal{L}_\text{Higgs} = \left(D_\mu\phi\right)^\dagger\left(D^\mu\phi\right) - m_\text{H}^2 \phi^\dagger\phi - \frac14\lambda\left(\phi^\dagger\phi\right)^2,
\end{equation}
where $m_\text{H}$ and $\lambda$ are parameters of the Higgs potential, $\lambda$ being the Higgs quartic coupling. We also have for the Yukawa Lagrangian
\begin{equation}
    \mathcal{L}_\text{Yukawa} = - \overline{\ell_R}Y_{\ell}\phi^\dagger L_L -  \overline{d_R} Y_d\phi^\dagger Q_L - \overline{u_R} Y_u \tilde{\phi}^\dagger Q_L + \text{h.c.},
\end{equation}
where $\tilde{\phi} = \ii\sigma_2\phi^*$, $\ell_R$ is the right-handed charged lepton, $L_L$ is the lepton doublet, $d_R$ is the right-handed down-type quark, $u_R$ is the right-handed up-type quark, and $Q_L$ is the quark doublet. The corresponding Yukawa couplings $Y_i$ ($i = \ell, d, u$) are $3\times3$ matrices in flavor space. These conventions are such that they agree with Refs.~\cite{Antusch:2001ck,Antusch:2002rr,Antusch:2005gp}.

\subsubsection{Right-Handed Neutrinos}

We add three generations of right-handed neutrinos $N$, which are all SM singlets. The resulting Yukawa and mass terms are 
\begin{equation}
    \mathcal{L}_N = - \overline{N_R} Y_\nu \tilde{\phi}^\dagger L_L - \overline{N_R} M N_R^C + \text{h.c.},
\end{equation}
where both $Y_\nu$ and $M$ are $3\times3$ matrices in flavor space. The Feynman rules with fermion number violating interactions are discussed in Ref.~\cite{Denner:1992vza}.

\subsubsection{Effective Theory}

After integrating out the RHNs, we are left with an effective theory which contains the Weinberg dimension-5 operator, namely
\begin{equation}
    \mathcal{L}_\kappa = \frac14 \kappa_{gf} \overline{L_L^C}_c^g \epsilon_{cd} \phi_d {L_L}_b^f \epsilon_{ba}\phi_a,
\end{equation}
where $\kappa$ is an effective coupling of the neutrino mass matrix that is symmetric under the interchange of $f$ and $g$ and $\epsilon$ is the two-index totally antisymmetric symbol.

\subsection{Conventions}

For computing the Feynman diagrams, we use the convention that the incoming lepton carries momentum $q_1$ and the incoming scalar carries momentum $q_2$ directed inward, whereas the outgoing lepton carries momentum $p_1$ outward and the outgoing scalar carries momentum $p_2$ outward. The labelling of the internal momentum will depend on the diagram.

For the Feynman diagrams involving gauge bosons, we use the generators $T^A_{ij}$, where $A\in\{0,1,2,3\}$ with a $0$ denoting hypercharge boson and the $1,2,3$ denoting the $\SU(2)$ bosons. We then use $\sum_A T^A_{ja}T^A_{ed} = Y_1 Y_2 \delta_{ja}\delta_{ed}$ for the $\text{U(1)}$ gauge boson exchange, with $Y_i$ denoting the hypercharge of the particle involved, and $\sum_A T^A_{ja}T^A_{ed} = \frac14(2\delta_{jd}\delta_{ae}-\delta_{ja}\delta_{ed})$ for the $\SU(2)$ gauge boson exchange. 

Computations are performed with the help of several packages. The amplitudes are generated using \texttt{FeynRules}~\cite{Alloul:2013bka} and \texttt{FeynArts}~\cite{Hahn:2000kx}. Computations of Feynman integrals are carried out in Mathematica with the help of \texttt{Package-X}~\cite{Patel:2015tea}. For dimensional regularization, we use $d=4-\epsilon$.

\FloatBarrier

\section{Tree-Level Feynman Diagrams and Matching}\label{sec:tlfdm}

At tree level, the only matching that occurs is of the effective neutrino mass matrix $\kappa$. All other quantities are the same in the full and effective theories at tree level.

\begin{figure}
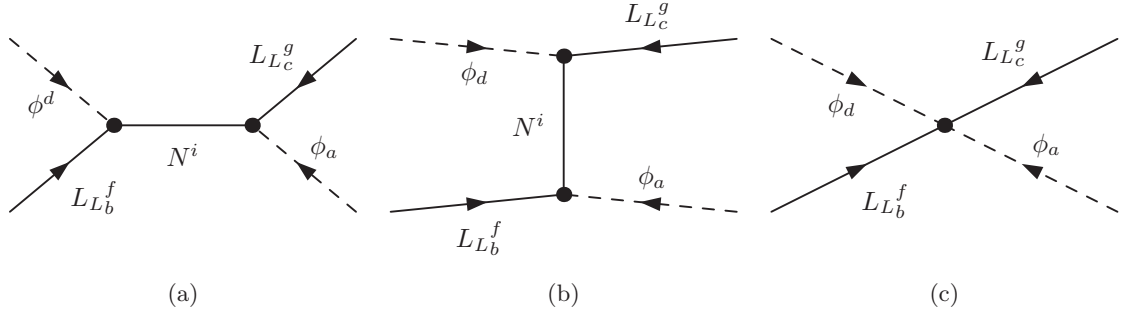

\vspace{-1cm}
\begin{center}
    \begin{feynartspicture}(432,168)(3,1.1)
        
        \FADiagram{(a)}
        \FAProp(0.,15.)(6.,10.)(0.,){ScalarDash}{1}
        \FALabel(2.64776,11.9813)[tr]{$\phi^d$}
        \FAProp(0.,5.)(6.,10.)(0.,){Straight}{1}
        \FALabel(3.51229,6.78926)[tl]{${L_L}_b^f$}
        \FAProp(20,15.)(14.,10.)(0.,){Straight}{1}
        \FALabel(16.4877,13.2107)[br]{${L_L}_c^g$}
        \FAProp(20,5.)(14.,10.)(0.,){ScalarDash}{1}
        \FALabel(17.3522,8.01869)[bl]{$\phi_a$}
        \FAProp(6.,10.)(14.,10.)(0.,){Straight}{0}
        \FALabel(10.,8.93)[t]{$N^i$}
        \FAVert(6.,10.){0}
        \FAVert(14.,10.){0}
        
        \FADiagram{(b)}
        \FAProp(0.,15.)(10.,14.)(0.,){ScalarDash}{1}
        \FALabel(4.87065,13.6865)[t]{$\phi_d$}
        \FAProp(0.,5.)(10.,6.)(0.,){Straight}{1}
        \FALabel(5.15423,4.43769)[t]{${L_L}_b^f$}
        \FAProp(20,15.)(10.,14.)(0.,){Straight}{1}
        \FALabel(14.8458,15.5623)[b]{${L_L}_c^g$}
        \FAProp(20,5.)(10.,6.)(0.,){ScalarDash}{1}
        \FALabel(15.1294,6.31355)[b]{$\phi_a$}
        \FAProp(10.,14.)(10.,6.)(0.,){Straight}{0}
        \FALabel(8.93,10.)[r]{$N^i$}
        \FAVert(10.,14.){0}
        \FAVert(10.,6.){0}
     
        \FADiagram{(c)}
        \FAProp(0.,15.)(10.,10.)(0.,){ScalarDash}{1}
        \FALabel(4.89862,11.8172)[tr]{$\phi_d$}
        \FAProp(0.,5.)(10.,10.)(0.,){Straight}{1}
        \FALabel(5.21318,6.59364)[tl]{${L_L}_b^f$}
        \FAProp(20,15.)(10.,10.)(0.,){Straight}{1}
        \FALabel(14.7868,13.4064)[br]{${L_L}_c^g$}
        \FAProp(20,5.)(10.,10.)(0.,){ScalarDash}{1}
        \FALabel(15.1014,8.18276)[bl]{$\phi_a$}
        \FAVert(10.,10.){0}
        
    \end{feynartspicture}
    \caption{Tree-level Feynman diagrams for matching full and effective theories. Diagrams (a) and (b) are contributions to the full theory, whereas diagram (c) is the contribution to the effective theory.\label{fig:tree}}
    \end{center}
\end{figure}

The relevant Feynman diagrams for the tree-level matching condition are shown in Fig.~\ref{fig:tree}, in which diagrams (a) and (b) are the two contributions in the full theory and diagram (c) is the corresponding contribution in the effective theory. 
To match them, we compute the amplitudes using the Feynman rules. For diagram (a), the amplitude is given by
\begin{equation}
    \ii \left(\Gamma_{(a)}\right)^{gf}_{abcd} = -\ii (Y_\nu^T)_{gi}\epsilon_{ca}\PL \frac{\slashed p + M}{p^2-M^2} (Y_\nu)_{if} (\epsilon^T)_{db} \PL,
    \label{eq:Gammaa}
\end{equation}
where $p$ is the internal momentum and $\PL$ is the projection operator onto left-handed chirality. Similarly, for diagram (b), it becomes
\begin{equation}
    \ii \left(\Gamma_{(b)}\right)^{gf}_{abcd} = -\ii (Y_\nu^T)_{gi}\epsilon_{cd}\PL \frac{\slashed p + M}{p^2-M^2} (Y_\nu)_{if} (\epsilon^T)_{ab} \PL.
    \label{eq:Gammab}
\end{equation}
In the limit $p\ll M$ (which is where the effective theory is valid), Eqs.~\eqref{eq:Gammaa} and~\eqref{eq:Gammab} become
\begin{equation}
    \ii \left(\Gamma_{(a)}+\Gamma_{(b)}\right)^{gf}_{abcd} \rightarrow \ii (Y_\nu^T)_{gi}M^{-1}(Y_\nu)_{if}(\epsilon_{ca}\epsilon_{bd}+\epsilon_{cd}\epsilon_{ba}) \PL.
    \label{eq:Gammaab}
\end{equation}

Diagram (c) is essentially only the Feynman rule for the four-point interaction and is therefore just given by
\begin{equation}
    \ii \left(\Gamma_{(c)}\right)^{gf}_{abcd} = \ii \kappa_{gf}\frac12 (\epsilon_{ca}\epsilon_{bd}+\epsilon_{cd}\epsilon_{ba}) \PL.
    \label{eq:Gammac}
\end{equation}
Matching Eqs.~\eqref{eq:Gammaab} and \eqref{eq:Gammac}, we obtain
\begin{equation}
    \kappa_{gf} = 2 (Y_\nu^T)_{gi}M^{-1}(Y_\nu)_{if}
\end{equation}
with no sum over $i$. Considering all three generations of heavy RHNs, we sum over $i$. This can be written as 
\begin{equation}\label{eq:tree_match}
    \kappa_{gf} = 2 (Y_\nu^T)_{gi}M_{ij}^{-1}(Y_\nu)_{jf}
\end{equation}
with $M_{ij}$ being a diagonal matrix. To generalize to any basis, we perform the transformations $M\rightarrow U^TMU$ and $Y\rightarrow U^TY$, where $U$ is a unitary matrix. This leaves the matching condition invariant, so Eq.~\eqref{eq:tree_match} is the tree-level matching condition for integrating out all three RHNs in a general basis.

\FloatBarrier

\section{Computations of Loop Feynman Diagrams}\label{sec:clfd}

\begin{figure}[h]
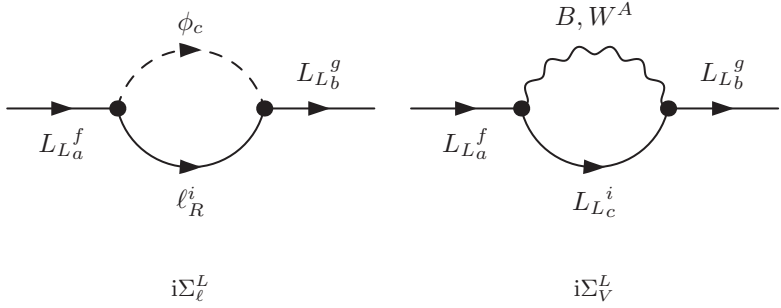

\vspace{-1cm}
\begin{center}
    \begin{feynartspicture}(432,168)(2,1.1)
        
        \FADiagram{$\ii\Sigma^L_\ell$}
        \FAProp(0.,10.)(6.,10.)(0.,){Straight}{1}
        \FALabel(3.,8.93)[t]{${L_L}_a^f$}
        \FAProp(20,10.)(14.,10.)(0.,){Straight}{-1}
        \FALabel(17.,11.07)[b]{${L_L}_b^g$}
        \FAProp(6.,10.)(14.,10.)(0.8,){Straight}{1}
        \FALabel(10.,5.73)[t]{$\ell_R^i$}
        \FAProp(6.,10.)(14.,10.)(-0.8,){ScalarDash}{1}
        \FALabel(10.,14.02)[b]{$\phi_c$}
        \FAVert(6.,10.){0}
        \FAVert(14.,10.){0}
        
        \FADiagram{$\ii\Sigma^L_V$}
        \FAProp(0.,10.)(6.,10.)(0.,){Straight}{1}
        \FALabel(3.,8.93)[t]{${L_L}_a^f$}
        \FAProp(20,10.)(14.,10.)(0.,){Straight}{-1}
        \FALabel(17.,11.07)[b]{${L_L}_b^g$}
        \FAProp(6.,10.)(14.,10.)(0.8,){Straight}{1}
        \FALabel(10.,5.73)[t]{${L_L}_c^i$}
        \FAProp(6.,10.)(14.,10.)(-0.8,){Sine}{0}
        \FALabel(10.,14.27)[b]{$B,W^A$}
        \FAVert(6.,10.){0}
        \FAVert(14.,10.){0}
        
    \end{feynartspicture}
    \caption{\label{fig:lepton_propagator_eft}Feynman diagrams for the lepton propagator at one-loop level in the effective theory.}
    \end{center}
\end{figure}

\begin{figure}[h]
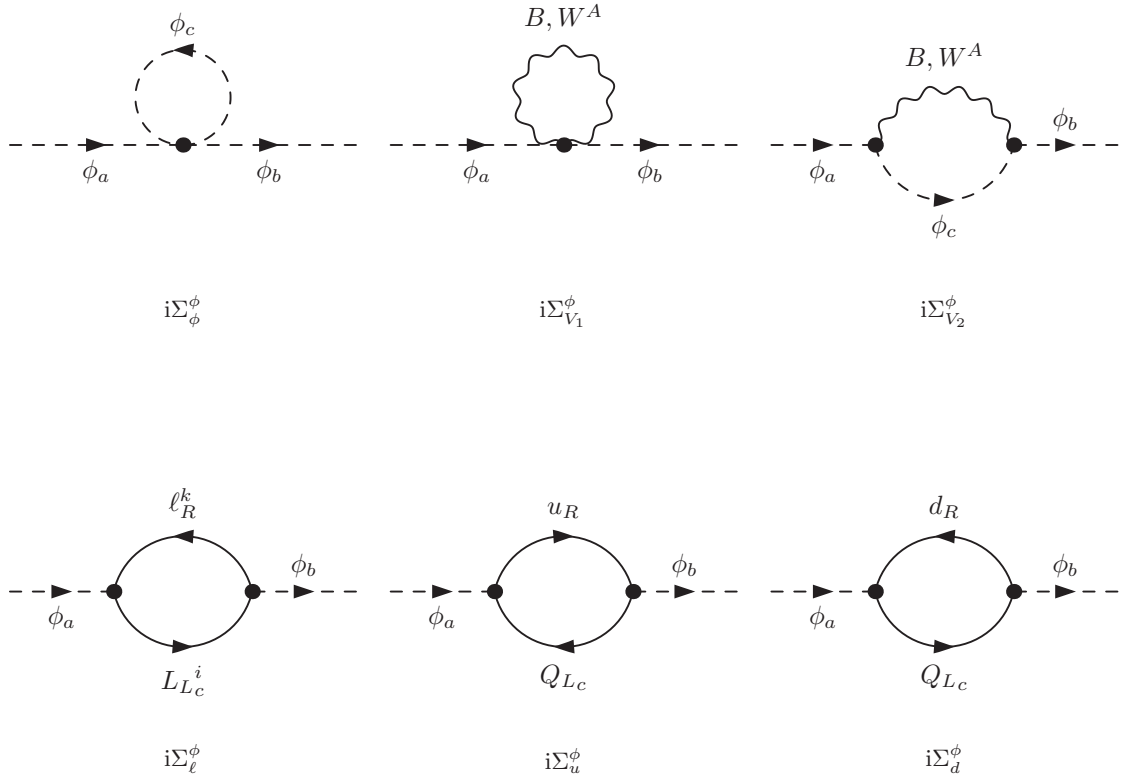

\vspace{-1cm}
\begin{center}
    \begin{feynartspicture}(432,168)(3,1.1)
        
        \FADiagram{$\ii\Sigma^\phi_\phi$}
        \FAProp(0.,10.)(10.,10.)(0.,){ScalarDash}{1}
        \FALabel(5.,9.18)[t]{$\phi_a$}
        \FAProp(20,10.)(10.,10.)(0.,){ScalarDash}{-1}
        \FALabel(15.,9.18)[t]{$\phi_b$}
        \FAProp(10.,10.)(10.,10.)(10.,15.5){ScalarDash}{1}
        \FALabel(10.,16.32)[b]{$\phi_c$}
        \FAVert(10.,10.){0}
        
        \FADiagram{$\ii\Sigma^\phi_{V_1}$}
        \FAProp(0.,10.)(10.,10.)(0.,){ScalarDash}{1}
        \FALabel(5.,9.18)[t]{$\phi_a$}
        \FAProp(20,10.)(10.,10.)(0.,){ScalarDash}{-1}
        \FALabel(15.,9.18)[t]{$\phi_b$}
        \FAProp(10.,10.)(10.,10.)(10.,15.5){Sine}{0}
        \FALabel(10.,16.57)[b]{$B,W^A$}
        \FAVert(10.,10.){0}
        
        \FADiagram{$\ii\Sigma^\phi_{V_2}$}
        \FAProp(0.,10.)(6.,10.)(0.,){ScalarDash}{1}
        \FALabel(3.,9.18)[t]{$\phi_a$}
        \FAProp(20,10.)(14.,10.)(0.,){ScalarDash}{-1}
        \FALabel(17.,10.82)[b]{$\phi_b$}
        \FAProp(6.,10.)(14.,10.)(0.8,){ScalarDash}{1}
        \FALabel(10.,5.98)[t]{$\phi_c$}
        \FAProp(6.,10.)(14.,10.)(-0.8,){Sine}{0}
        \FALabel(10.,14.27)[b]{$B,W^A$}
        \FAVert(6.,10.){0}
        \FAVert(14.,10.){0}

	\end{feynartspicture}
        \begin{feynartspicture}(432,168)(3,1.1)

        \FADiagram{$\ii\Sigma^\phi_\ell$}
        \FAProp(0.,10.)(6.,10.)(0.,){ScalarDash}{1}
        \FALabel(3.,9.18)[t]{$\phi_a$}
        \FAProp(20,10.)(14.,10.)(0.,){ScalarDash}{-1}
        \FALabel(17.,10.82)[b]{$\phi_b$}
        \FAProp(6.,10.)(14.,10.)(0.8,){Straight}{1}
        \FALabel(10.,5.73)[t]{${L_L}_c^i$}
        \FAProp(6.,10.)(14.,10.)(-0.8,){Straight}{-1}
        \FALabel(10.,14.27)[b]{$\ell_R^k$}
        \FAVert(6.,10.){0}
        \FAVert(14.,10.){0}
        
        \FADiagram{$\ii\Sigma^\phi_u$}
        \FAProp(0.,10.)(6.,10.)(0.,){ScalarDash}{1}
        \FALabel(3.,9.18)[t]{$\phi_a$}
        \FAProp(20,10.)(14.,10.)(0.,){ScalarDash}{-1}
        \FALabel(17.,10.82)[b]{$\phi_b$}
        \FAProp(6.,10.)(14.,10.)(0.8,){Straight}{-1}
        \FALabel(10.,5.73)[t]{${Q_L}_c$}
        \FAProp(6.,10.)(14.,10.)(-0.8,){Straight}{1}
        \FALabel(10.,14.27)[b]{$u_R$}
        \FAVert(6.,10.){0}
        \FAVert(14.,10.){0}
        
        \FADiagram{$\ii\Sigma^\phi_d$}
        \FAProp(0.,10.)(6.,10.)(0.,){ScalarDash}{1}
        \FALabel(3.,9.18)[t]{$\phi_a$}
        \FAProp(20,10.)(14.,10.)(0.,){ScalarDash}{-1}
        \FALabel(17.,10.82)[b]{$\phi_b$}
        \FAProp(6.,10.)(14.,10.)(0.8,){Straight}{1}
        \FALabel(10.,5.73)[t]{${Q_L}_c$}
        \FAProp(6.,10.)(14.,10.)(-0.8,){Straight}{-1}
        \FALabel(10.,14.27)[b]{$d_R$}
        \FAVert(6.,10.){0}
        \FAVert(14.,10.){0}
        
        \end{feynartspicture}
    \caption{\label{fig:higgs_propagator_eft}Feynman diagrams for the Higgs propagator at one-loop level in the effective theory.}
    \end{center}
\end{figure}

For the matching at loop level, we need to compute all relevant loop Feynman diagrams. At loop level, the propagators of the lepton doublet and the Higgs doublet both have non-trivial matching conditions between the full and effective theories due to loops involving the RHN. For the lepton doublet, the one-loop contributions are shown in Figs.~\ref{fig:lepton_propagator_full} and \ref{fig:lepton_propagator_eft}, whereas for the Higgs doublet, they are shown in Figs.~\ref{fig:higgs_propagator_full} and \ref{fig:higgs_propagator_eft}. These will be used on the external legs of the physical processes, as well as Higgs mass corrections.

\subsection{Propagators}

\begin{figure}[h]
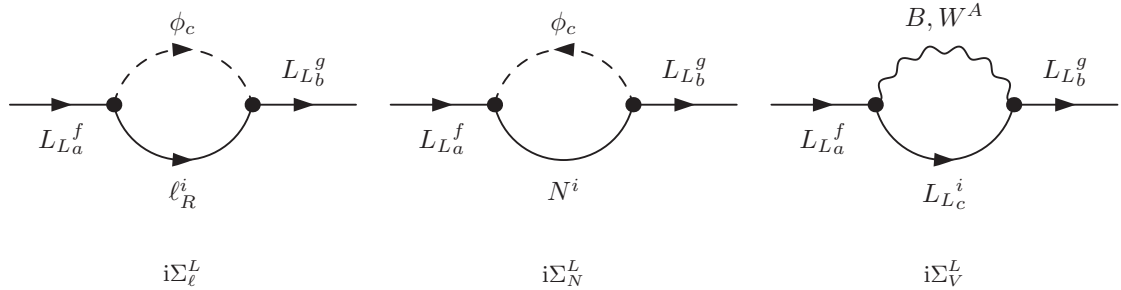

\vspace{-1cm}
\begin{center}
    \begin{feynartspicture}(432,168)(3,1.1)
        
        \FADiagram{$\ii\Sigma^L_\ell$}
        \FAProp(0.,10.)(6.,10.)(0.,){Straight}{1}
        \FALabel(3.,8.93)[t]{${L_L}_a^f$}
        \FAProp(20,10.)(14.,10.)(0.,){Straight}{-1}
        \FALabel(17.,11.07)[b]{${L_L}_b^g$}
        \FAProp(6.,10.)(14.,10.)(0.8,){Straight}{1}
        \FALabel(10.,5.73)[t]{$\ell_R^i$}
        \FAProp(6.,10.)(14.,10.)(-0.8,){ScalarDash}{1}
        \FALabel(10.,14.02)[b]{$\phi_c$}
        \FAVert(6.,10.){0}
        \FAVert(14.,10.){0}
        
        \FADiagram{$\ii\Sigma^L_N$}
        \FAProp(0.,10.)(6.,10.)(0.,){Straight}{1}
        \FALabel(3.,8.93)[t]{${L_L}_a^f$}
        \FAProp(20,10.)(14.,10.)(0.,){Straight}{-1}
        \FALabel(17.,11.07)[b]{${L_L}_b^g$}
        \FAProp(6.,10.)(14.,10.)(0.8,){Straight}{0}
        \FALabel(10.,5.73)[t]{$N^i$}
        \FAProp(6.,10.)(14.,10.)(-0.8,){ScalarDash}{-1}
        \FALabel(10.,14.02)[b]{$\phi_c$}
        \FAVert(6.,10.){0}
        \FAVert(14.,10.){0}
        
        \FADiagram{$\ii\Sigma^L_V$}
        \FAProp(0.,10.)(6.,10.)(0.,){Straight}{1}
        \FALabel(3.,8.93)[t]{${L_L}_a^f$}
        \FAProp(20,10.)(14.,10.)(0.,){Straight}{-1}
        \FALabel(17.,11.07)[b]{${L_L}_b^g$}
        \FAProp(6.,10.)(14.,10.)(0.8,){Straight}{1}
        \FALabel(10.,5.73)[t]{${L_L}_c^i$}
        \FAProp(6.,10.)(14.,10.)(-0.8,){Sine}{0}
        \FALabel(10.,14.27)[b]{$B,W^A$}
        \FAVert(6.,10.){0}
        \FAVert(14.,10.){0}

    \end{feynartspicture}
    \caption{\label{fig:lepton_propagator_full}Feynman diagrams for the lepton propagator at one-loop level in the full theory.}
    \end{center}
\end{figure}

\begin{figure}[h]
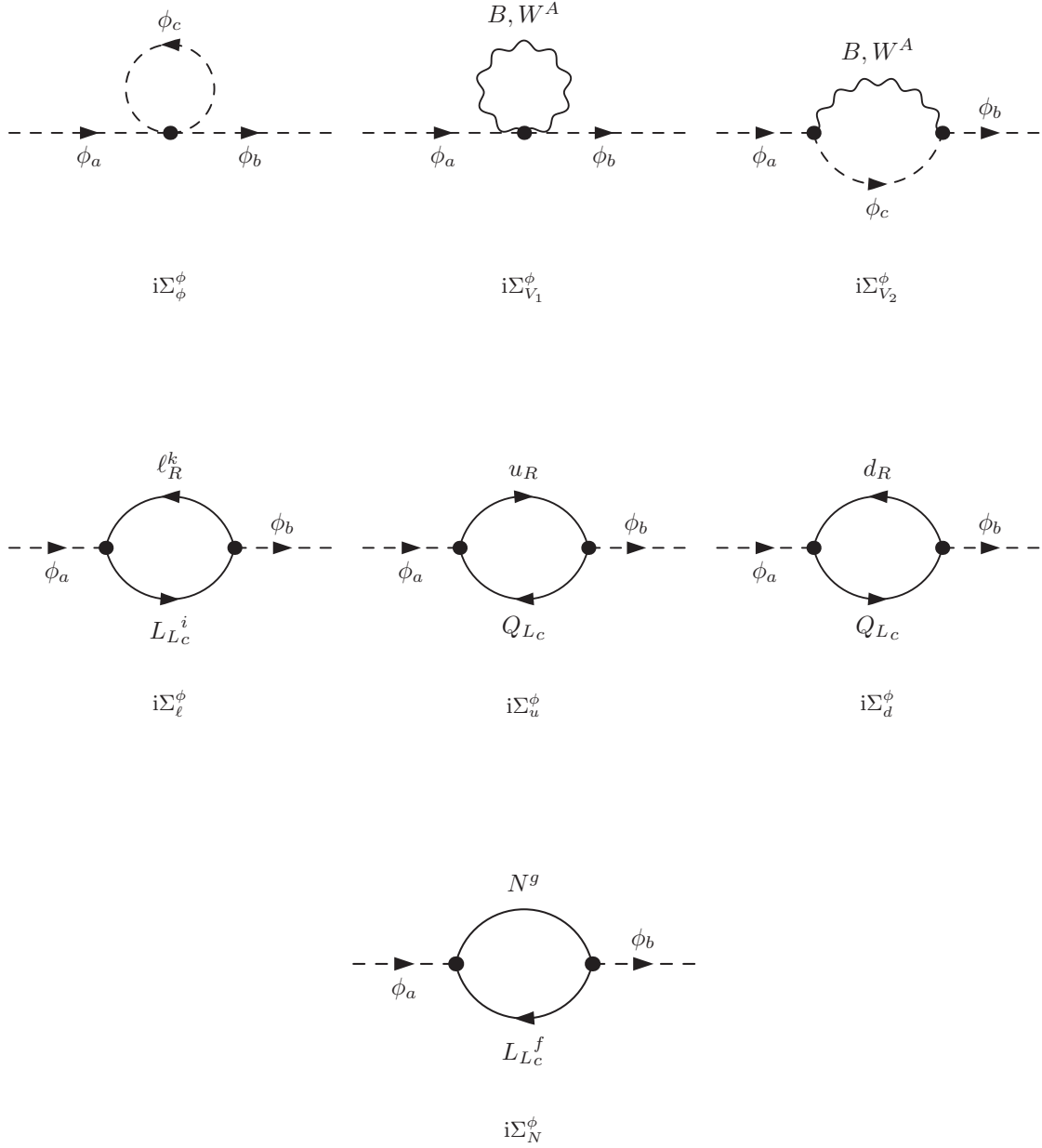

\vspace{-1cm}
\begin{center}
    \begin{feynartspicture}(432,168)(3,1.1)
        
        \FADiagram{$\ii\Sigma^\phi_\phi$}
        \FAProp(0.,10.)(10.,10.)(0.,){ScalarDash}{1}
        \FALabel(5.,9.18)[t]{$\phi_a$}
        \FAProp(20,10.)(10.,10.)(0.,){ScalarDash}{-1}
        \FALabel(15.,9.18)[t]{$\phi_b$}
        \FAProp(10.,10.)(10.,10.)(10.,15.5){ScalarDash}{1}
        \FALabel(10.,16.32)[b]{$\phi_c$}
        \FAVert(10.,10.){0}
        
        \FADiagram{$\ii\Sigma^\phi_{V_1}$}
        \FAProp(0.,10.)(10.,10.)(0.,){ScalarDash}{1}
        \FALabel(5.,9.18)[t]{$\phi_a$}
        \FAProp(20,10.)(10.,10.)(0.,){ScalarDash}{-1}
        \FALabel(15.,9.18)[t]{$\phi_b$}
        \FAProp(10.,10.)(10.,10.)(10.,15.5){Sine}{0}
        \FALabel(10.,16.57)[b]{$B,W^A$}
        \FAVert(10.,10.){0}
        
        \FADiagram{$\ii\Sigma^\phi_{V_2}$}
        \FAProp(0.,10.)(6.,10.)(0.,){ScalarDash}{1}
        \FALabel(3.,9.18)[t]{$\phi_a$}
        \FAProp(20,10.)(14.,10.)(0.,){ScalarDash}{-1}
        \FALabel(17.,10.82)[b]{$\phi_b$}
        \FAProp(6.,10.)(14.,10.)(0.8,){ScalarDash}{1}
        \FALabel(10.,5.98)[t]{$\phi_c$}
        \FAProp(6.,10.)(14.,10.)(-0.8,){Sine}{0}
        \FALabel(10.,14.27)[b]{$B,W^A$}
        \FAVert(6.,10.){0}
        \FAVert(14.,10.){0}

        \end{feynartspicture}
        \begin{feynartspicture}(432,168)(3,1.1)

        \FADiagram{$\ii\Sigma^\phi_\ell$}
        \FAProp(0.,10.)(6.,10.)(0.,){ScalarDash}{1}
        \FALabel(3.,9.18)[t]{$\phi_a$}
        \FAProp(20,10.)(14.,10.)(0.,){ScalarDash}{-1}
        \FALabel(17.,10.82)[b]{$\phi_b$}
        \FAProp(6.,10.)(14.,10.)(0.8,){Straight}{1}
        \FALabel(10.,5.73)[t]{${L_L}_c^i$}
        \FAProp(6.,10.)(14.,10.)(-0.8,){Straight}{-1}
        \FALabel(10.,14.27)[b]{$\ell_R^k$}
        \FAVert(6.,10.){0}
        \FAVert(14.,10.){0}
        
        \FADiagram{$\ii\Sigma^\phi_u$}
        \FAProp(0.,10.)(6.,10.)(0.,){ScalarDash}{1}
        \FALabel(3.,9.18)[t]{$\phi_a$}
        \FAProp(20,10.)(14.,10.)(0.,){ScalarDash}{-1}
        \FALabel(17.,10.82)[b]{$\phi_b$}
        \FAProp(6.,10.)(14.,10.)(0.8,){Straight}{-1}
        \FALabel(10.,5.73)[t]{${Q_L}_c$}
        \FAProp(6.,10.)(14.,10.)(-0.8,){Straight}{1}
        \FALabel(10.,14.27)[b]{$u_R$}
        \FAVert(6.,10.){0}
        \FAVert(14.,10.){0}
        
        \FADiagram{$\ii\Sigma^\phi_d$}
        \FAProp(0.,10.)(6.,10.)(0.,){ScalarDash}{1}
        \FALabel(3.,9.18)[t]{$\phi_a$}
        \FAProp(20,10.)(14.,10.)(0.,){ScalarDash}{-1}
        \FALabel(17.,10.82)[b]{$\phi_b$}
        \FAProp(6.,10.)(14.,10.)(0.8,){Straight}{1}
        \FALabel(10.,5.73)[t]{${Q_L}_c$}
        \FAProp(6.,10.)(14.,10.)(-0.8,){Straight}{-1}
        \FALabel(10.,14.27)[b]{$d_R$}
        \FAVert(6.,10.){0}
        \FAVert(14.,10.){0}
        
       \end{feynartspicture}
       \begin{feynartspicture}(432,168)(1,1.1)
       
        \FADiagram{$\ii\Sigma^\phi_N$}
        \FAProp(0.,10.)(6.,10.)(0.,){ScalarDash}{1}
        \FALabel(3.,9.18)[t]{$\phi_a$}
        \FAProp(20,10.)(14.,10.)(0.,){ScalarDash}{-1}
        \FALabel(17.,10.82)[b]{$\phi_b$}
        \FAProp(6.,10.)(14.,10.)(0.8,){Straight}{-1}
        \FALabel(10.,5.73)[t]{${L_L}_c^f$}
        \FAProp(6.,10.)(14.,10.)(-0.8,){Straight}{0}
        \FALabel(10.,14.27)[b]{$N^g$}
        \FAVert(6.,10.){0}
        \FAVert(14.,10.){0}
        
    \end{feynartspicture}
    \caption{\label{fig:higgs_propagator_full}Feynman diagrams for the Higgs propagator at one-loop level in the full theory.}
    \end{center}
\end{figure}

We only need to compute the propagator correction that differs between the two theories. For the lepton propagator, this is the one with $N$ in the loop in Fig.~\ref{fig:lepton_propagator_full}, denoted $\ii \Sigma^L_N$. Using the Feynman rules, it evaluates to
\begin{align}
    -\ii \left(\Sigma^L_N\right)^{gf}_{ba} &= \left(-\ii{\bar\mu}^{\epsilon/2} (Y_\nu^\dagger)_{gi}\epsilon_{bc}P_\text{R}\right)\int\frac{\dd^dk}{(2\pi)^d}\frac{\ii{\slashed k}}{k^2-M^2} \frac{\ii}{(k-p)^2-m^2}\left(-\ii{\bar\mu}^{\epsilon/2} (Y_\nu)_{if}(\epsilon^T)_{ca}\PL\right) \nonumber\\
    &= - \frac{\ii}{16\pi^2}\delta_{ab}{\slashed p}\PL (Y_\nu^\dagger Y_\nu)_{gf} B_1(0;M,m),
\end{align}
where $B_1$ is a Passarino--Veltman function, $m$ is the mass of the Higgs field,  and we used the fact that $p^2=0$ for the external lepton. We are interested in the finite part (since the divergent part is subtracted by counterterms) in the limit $M^2 \gg m^2$ and at the energy scale ${\bar\mu}=M$, where the matching occurs, which gives 
\begin{equation}
    B_1(0;M,m)|_\text{finite} \rightarrow -\frac34 - \frac12 \frac{m^2}{M^2}\left(1+4\ln\frac{m}{M}\right),
\end{equation}
using the $\overline{\text{MS}}$ scheme. This leads to the correction to the lepton propagator
\begin{equation}
    -\ii \left(\Sigma^L_N|_\text{finite}\right)^{gf}_{ba} = \frac{\ii}{32\pi^2}\delta_{ab} (Y_\nu^\dagger Y_\nu)_{gf}{\slashed p}\PL\left[\frac32 + \frac{m^2}{M^2}\left(1+4\ln\frac{m}{M}\right)\right]\simeq\frac{3\ii}{64\pi^2}\delta_{ab} (Y_\nu^\dagger Y_\nu)_{gf}{\slashed p}\PL.
    \label{eq:SLNfinite}
\end{equation}

Similarly, for the Higgs propagator, the relevant Feynman diagram is the one denoted $\ii \Sigma^\phi_N$ in Fig.~\ref{fig:higgs_propagator_full}. Again, using the Feynman rules, it evaluates to
\begin{align}
    -\ii \left(\Sigma^\phi_N\right)_{ba} &= (-1) \Tr\left[\left(-\ii{\bar\mu}^{\epsilon/2}(Y_\nu^\dagger)_{fg}\epsilon_{cb}\PR\right) \int \frac{\dd^dk}{(2\pi)^d} \frac{\ii\slashed{k}}{k^2-M^2} \left(-\ii{\bar\mu}^{\epsilon/2}(Y_\nu)_{gf}(\epsilon^T)_{ac}\PL\right)\frac{\ii({\slashed k}-\slashed{p})}{(k-p)^2}\right] \nonumber\\
    &= -\frac{\ii}{16\pi^2} \Tr\left(Y_\nu^\dagger Y_\nu\right)\delta_{ab}\left[(M^2-p^2)B_0(p^2;M,0)+A_0(M)+A_0(0)\right],
\end{align}
where $A_0$ and $B_0$ are Passarino--Veltman functions. In the limit $M^2 \gg m^2$ and ${\bar\mu}=M$, we have the finite part
\begin{equation}\label{eq:sigma_phi}
    -\ii \left(\Sigma^\phi_N|_\text{finite}\right)_{ba} = \frac{\ii}{96\pi^2} \Tr(Y_\nu^\dagger Y_\nu)\delta_{ab} \left[3p^2 - 12M^2 + 2 \frac{(p^2)^2}{M^2}\right]\simeq \frac{\ii}{32\pi^2} \Tr(Y_\nu^\dagger Y_\nu)\delta_{ab} \left(p^2 - 4M^2\right).
\end{equation}
Furthermore, assuming $M^2 \gg p^2$, we obtain the correction to the Higgs propagator as
\begin{equation}
    -\ii \left(\Sigma^\phi_N|_\text{finite}\right)_{ba} \simeq - \frac{\ii}{8\pi^2} M^2 \Tr(Y_\nu^\dagger Y_\nu)\delta_{ab}.
    \label{eq:sigma_phi2}
\end{equation}

\FloatBarrier

\subsection{Gauge Couplings}

\begin{figure}[h]
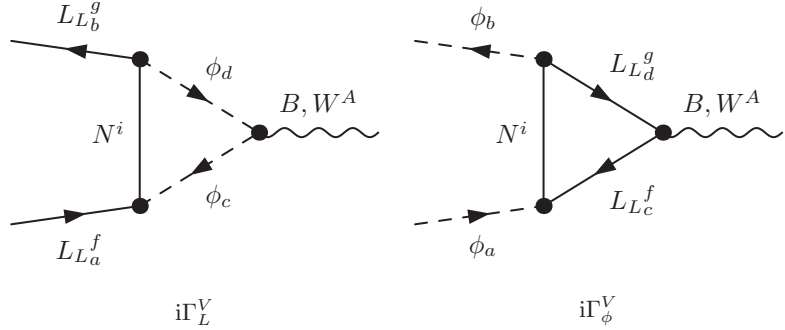

\vspace{-1cm}
\begin{center}
    \begin{feynartspicture}(432,168)(2,1.1)
        
        \FADiagram{$\ii\Gamma^V_L$}
        \FAProp(0.,15.)(7.,14.)(0.,){Straight}{-1}
        \FALabel(3.7192,15.5544)[b]{${L_L}_b^g$}
        \FAProp(0.,5.)(7.,6.)(0.,){Straight}{1}
        \FALabel(3.7192,4.44558)[t]{${L_L}_a^f$}
        \FAProp(20,10.)(13.5,10.)(0.,){Sine}{0}
        \FALabel(16.75,10.82)[b]{$B,W^A$}
        \FAProp(7.,14.)(7.,6.)(0.,){Straight}{0}
        \FALabel(6.18,10.)[r]{$N^i$}
        \FAProp(7.,14.)(13.5,10.)(0.,){ScalarDash}{1}
        \FALabel(10.5824,12.8401)[bl]{$\phi_d$}
        \FAProp(7.,6.)(13.5,10.)(0.,){ScalarDash}{-1}
        \FALabel(10.5824,7.15993)[tl]{$\phi_c$}
        \FAVert(7.,14.){0}
        \FAVert(7.,6.){0}
        \FAVert(13.5,10.){0}

        \FADiagram{$\ii\Gamma^V_\phi$}
        \FAProp(0.,15.)(7.,14.)(0.,){ScalarDash}{-1}
        \FALabel(3.7192,15.5544)[b]{$\phi_b$}
        \FAProp(0.,5.)(7.,6.)(0.,){ScalarDash}{1}
        \FALabel(3.7192,4.44558)[t]{$\phi_a$}
        \FAProp(20,10.)(13.5,10.)(0.,){Sine}{0}
        \FALabel(16.75,10.82)[b]{$B,W^A$}
        \FAProp(7.,14.)(7.,6.)(0.,){Straight}{0}
        \FALabel(6.18,10.)[r]{$N^i$}
        \FAProp(7.,14.)(13.5,10.)(0.,){Straight}{1}
        \FALabel(10.5824,12.8401)[bl]{${L_L}_d^g$}
        \FAProp(7.,6.)(13.5,10.)(0.,){Straight}{-1}
        \FALabel(10.5824,7.15993)[tl]{${L_L}_c^f$}
        \FAVert(7.,14.){0}
        \FAVert(7.,6.){0}
        \FAVert(13.5,10.){0}

\end{feynartspicture}
\caption{\label{fig:gauge}Feynman diagrams for the corrections to the gauge couplings $g_1$ and $g_2$ defined through lepton or Higgs interactions.}
\end{center}
\end{figure}

The gauge couplings $g_1$ and $g_2$ have loop corrections that are different in the two theories, while the loop corrections to $g_3$ are the same in the two theories. One can define the couplings $g_1$ and $g_2$ either through interactions with the lepton doublet or the Higgs doublet. We display both diagrams in Fig.~\ref{fig:gauge}.

Inserting the Feynman rules for the interactions with the lepton doublet, we obtain
\begin{align}
    {\bar\mu}^{\epsilon/2} \ii \left(\Gamma_{L,\mu}^V\right)^{gf}_{ab} &= \int\frac{\dd^dk}{(2\pi)^d} \left(-\ii{\bar\mu}^{\epsilon/2} (Y_\nu^\dagger)_{gi}\epsilon_{bd}\PR\right) \frac{\ii(\slashed{k}+M)}{k^2-M^2} \left(-\ii{\bar\mu}^{\epsilon/2} (Y_\nu)_{if}(\epsilon^T)_{ca}\PL\right)\frac{\ii}{(k+q_2)^2-m^2} \nonumber\\ 
    &\quad\times\frac{\ii}{(k-q_1)^2-m^2} \left(-\ii{\bar\mu}^{\epsilon/2} g_A T^A_{cd} (2k-q_1+q_2)_\mu\right) \nonumber\\
    &= {\bar\mu}^{3\epsilon/2} (Y_\nu^\dagger Y_\nu)_{gf} g_A \epsilon_{ac}\epsilon_{bd} T^A_{cd}\PR \int\frac{\dd^dk}{(2\pi)^d} \frac{\slashed{k}(2k-q_1+q_2)_\mu}{(k^2-M^2)[(k+q_2)^2-m^2][(k-q_1)^2-m^2]},
\end{align}
where $g_A$ is the gauge coupling and $\slashed{k}=k^\mu \gamma_\mu$ for a momentum $k$ and Dirac matrix $\gamma_\mu$. After evaluating the integral and taking the limit in which $M\gg m, q_1, q_2$, we find that
\begin{equation}
    {\bar\mu}^{\epsilon/2} \ii \left(\Gamma_{L,\mu}^V\right)^{gf}_{ab} \simeq \frac{3\ii{\bar\mu}^{\epsilon/2}}{64\pi^2}g_A(Y_\nu^\dagger Y_\nu)_{gf} \epsilon_{ac}\epsilon_{bd}T^A_{cd} \gamma_\mu \PL.
    \label{eq:GVL}
\end{equation}
Similarly, for the interactions with the Higgs doublet, we obtain
\begin{align}
    {\bar\mu}^{\epsilon/2} \ii \left(\Gamma_{\phi,\mu}^V\right)_{ab} &= (-1)\Tr\left[\int\frac{\dd^dk}{(2\pi)^d}\left(-\ii{\bar\mu}^{\epsilon/2} (Y_\nu^\dagger)_{gi}\epsilon_{db}\PR\right) \frac{\ii(\slashed{k}+M)}{k^2-M^2} \left(-\ii{\bar\mu}^{\epsilon/2} (Y_\nu)_{if}(\epsilon^T)_{ac}\PL\right)\right.\nonumber\\
    &\quad\times\left. \frac{\ii(\slashed{k}-\slashed{q_1})}{(k-q_1)^2} \left(-\ii{\bar\mu}^{\epsilon/2} g_A \delta_{gf}T^A_{cd}\gamma_\mu\PL\right)\frac{\ii(\slashed{k}+\slashed{q_2})}{(k+q_2)^2}\right] \nonumber\\
    &= -{\bar\mu}^{3\epsilon/2} \Tr(Y_\nu^\dagger Y_\nu) g_A \epsilon_{ac}\epsilon_{bd}T^A_{cd} \int\frac{\dd^dk}{(2\pi)^d}\frac{\Tr\left[\slashed{k}(\slashed{k}-\slashed{q_1})\gamma_\mu\PL(\slashed{k}+\slashed{q_2})\right]}{(k^2-M^2)(k-q_1)^2(k+q_2)^2}.
\end{align}
After evaluating the integral and taking the limit in which $M\gg m$, we find that
\begin{equation}
    {\bar\mu}^{\epsilon/2} \ii \left(\Gamma_{\phi,\mu}^V\right)_{ab} \simeq \frac{\ii{\bar\mu}^{\epsilon/2}}{32\pi^2}g_A\Tr(Y_\nu^\dagger Y_\nu) \epsilon_{ac}\epsilon_{bd}T^A_{cd} (q_1-q_2)_\mu.
    \label{eq:GVP}
\end{equation}

\FloatBarrier

\subsection{Lepton Yukawa Coupling}

\begin{figure}[h]
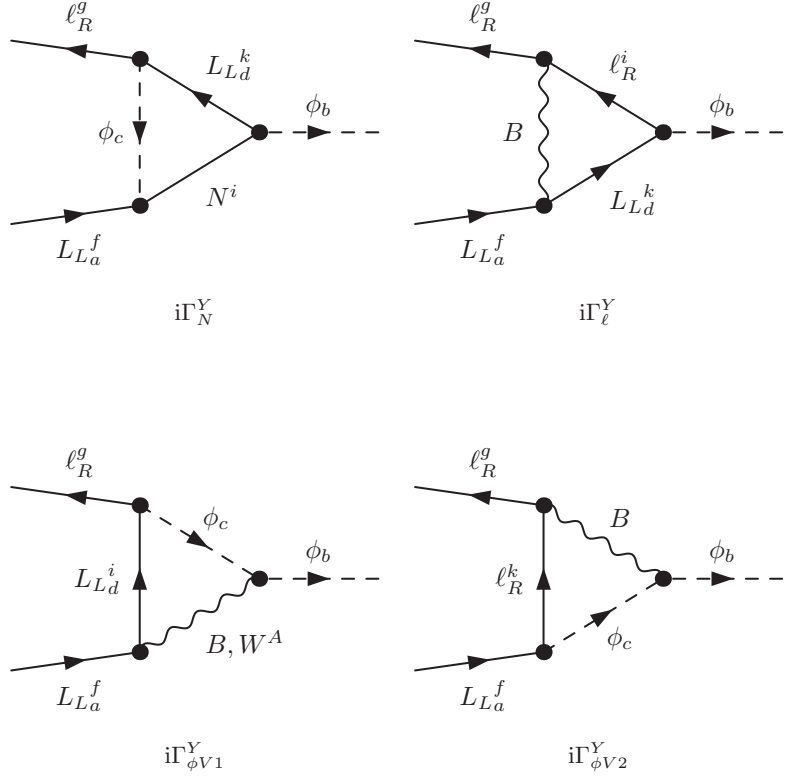

\vspace{-1cm}
\begin{center}
\begin{feynartspicture}(432,168)(2,1.1)
    
    \FADiagram{$\ii\Gamma^Y_N$}
    \FAProp(0.,15.)(7.,14.)(0.,){Straight}{-1}
    \FALabel(3.7192,15.5544)[b]{$\ell_R^g$}
    \FAProp(0.,5.)(7.,6.)(0.,){Straight}{1}
    \FALabel(3.7192,4.44558)[t]{${L_L}_a^f$}
    \FAProp(20,10.)(13.5,10.)(0.,){ScalarDash}{-1}
    \FALabel(16.75,10.82)[b]{$\phi_b$}
    \FAProp(7.,14.)(7.,6.)(0.,){ScalarDash}{1}
    \FALabel(6.18,10.)[r]{$\phi_c$}
    \FAProp(7.,14.)(13.5,10.)(0.,){Straight}{-1}
    \FALabel(10.5824,12.8401)[bl]{${L_L}_d^k$}
    \FAProp(7.,6.)(13.5,10.)(0.,){Straight}{0}
    \FALabel(10.5824,7.15993)[tl]{$N^i$}
    \FAVert(7.,14.){0}
    \FAVert(7.,6.){0}
    \FAVert(13.5,10.){0}
    
    \FADiagram{$\ii\Gamma^Y_{\ell}$}
    \FAProp(0.,15.)(7.,14.)(0.,){Straight}{-1}
    \FALabel(3.7192,15.5544)[b]{$\ell_R^g$}
    \FAProp(0.,5.)(7.,6.)(0.,){Straight}{1}
    \FALabel(3.7192,4.44558)[t]{${L_L}_a^f$}
    \FAProp(20,10.)(13.5,10.)(0.,){ScalarDash}{-1}
    \FALabel(16.75,10.82)[b]{$\phi_b$}
    \FAProp(7.,14.)(7.,6.)(0.,){Sine}{0}
    \FALabel(5.93,10.)[r]{$B$}
    \FAProp(7.,14.)(13.5,10.)(0.,){Straight}{-1}
    \FALabel(10.5824,12.8401)[bl]{$\ell_R^i$}
    \FAProp(7.,6.)(13.5,10.)(0.,){Straight}{1}
    \FALabel(10.5824,7.15993)[tl]{${L_L}_d^k$}
    \FAVert(7.,14.){0}
    \FAVert(7.,6.){0}
    \FAVert(13.5,10.){0}
    
    \end{feynartspicture}
    \begin{feynartspicture}(432,168)(2,1.1)
    
    \FADiagram{$\ii\Gamma^Y_{\phi V1}$}
    \FAProp(0.,15.)(7.,14.)(0.,){Straight}{-1}
    \FALabel(3.7192,15.5544)[b]{$\ell_R^g$}
    \FAProp(0.,5.)(7.,6.)(0.,){Straight}{1}
    \FALabel(3.7192,4.44558)[t]{${L_L}_a^f$}
    \FAProp(20,10.)(13.5,10.)(0.,){ScalarDash}{-1}
    \FALabel(16.75,10.82)[b]{$\phi_b$}
    \FAProp(7.,14.)(7.,6.)(0.,){Straight}{-1}
    \FALabel(5.93,10.)[r]{${L_L}_d^i$}
    \FAProp(7.,14.)(13.5,10.)(0.,){ScalarDash}{1}
    \FALabel(10.4513,12.6272)[bl]{$\phi_c$}
    \FAProp(7.,6.)(13.5,10.)(0.,){Sine}{0}
    \FALabel(10.5824,7.15993)[tl]{$B,W^A$}
    \FAVert(7.,14.){0}
    \FAVert(7.,6.){0}
    \FAVert(13.5,10.){0}
    
    \FADiagram{$\ii\Gamma^Y_{\phi V2}$}
    \FAProp(0.,15.)(7.,14.)(0.,){Straight}{-1}
    \FALabel(3.7192,15.5544)[b]{$\ell_R^g$}
    \FAProp(0.,5.)(7.,6.)(0.,){Straight}{1}
    \FALabel(3.7192,4.44558)[t]{${L_L}_a^f$}
    \FAProp(20,10.)(13.5,10.)(0.,){ScalarDash}{-1}
    \FALabel(16.75,10.82)[b]{$\phi_b$}
    \FAProp(7.,14.)(7.,6.)(0.,){Straight}{-1}
    \FALabel(5.93,10.)[r]{$\ell_R^k$}
    \FAProp(7.,14.)(13.5,10.)(0.,){Sine}{0}
    \FALabel(10.5824,12.8401)[bl]{$B$}
    \FAProp(7.,6.)(13.5,10.)(0.,){ScalarDash}{1}
    \FALabel(10.4513,7.37284)[tl]{$\phi_c$}
    \FAVert(7.,14.){0}
    \FAVert(7.,6.){0}
    \FAVert(13.5,10.){0}

\end{feynartspicture}
\caption{\label{fig:yukawa_full}Feynman diagrams for the lepton Yukawa coupling at one-loop level in the full theory.}
\end{center}
\end{figure}

\begin{figure}[h]
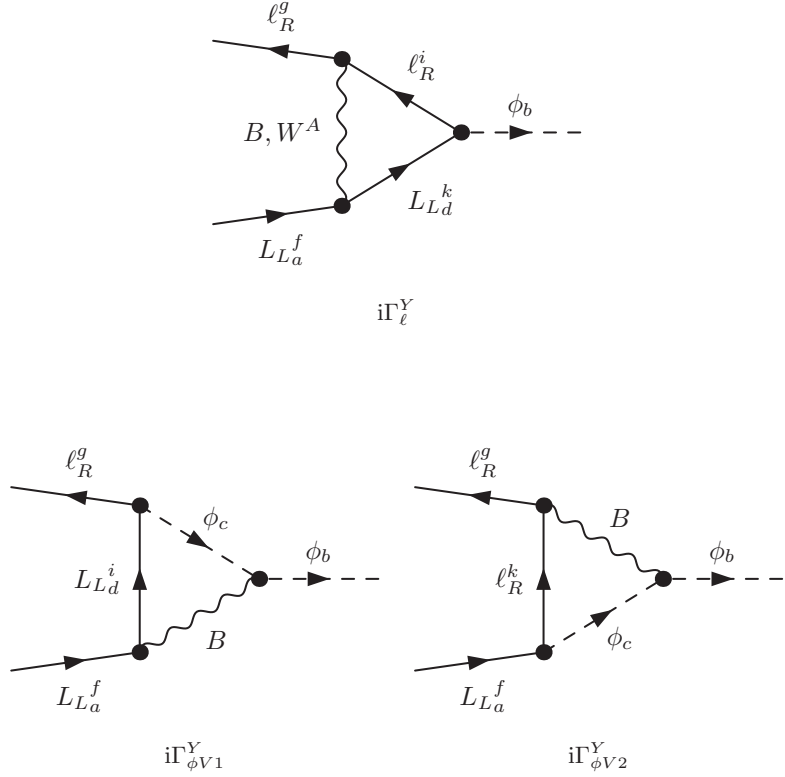

\vspace{-1cm}
\begin{center}
\begin{feynartspicture}(432,168)(1,1.1)
        
    \FADiagram{$\ii\Gamma^Y_{\ell}$}
    \FAProp(0.,15.)(7.,14.)(0.,){Straight}{-1}
    \FALabel(3.7192,15.5544)[b]{$\ell_R^g$}
    \FAProp(0.,5.)(7.,6.)(0.,){Straight}{1}
    \FALabel(3.7192,4.44558)[t]{${L_L}_a^f$}
    \FAProp(20,10.)(13.5,10.)(0.,){ScalarDash}{-1}
    \FALabel(16.75,10.82)[b]{$\phi_b$}
    \FAProp(7.,14.)(7.,6.)(0.,){Sine}{0}
    \FALabel(5.93,10.)[r]{$B,W^A$}
    \FAProp(7.,14.)(13.5,10.)(0.,){Straight}{-1}
    \FALabel(10.5824,12.8401)[bl]{$\ell_R^i$}
    \FAProp(7.,6.)(13.5,10.)(0.,){Straight}{1}
    \FALabel(10.5824,7.15993)[tl]{${L_L}_d^k$}
    \FAVert(7.,14.){0}
    \FAVert(7.,6.){0}
    \FAVert(13.5,10.){0}
    
    \end{feynartspicture}
    \begin{feynartspicture}(432,168)(2,1.1)
    
    \FADiagram{$\ii\Gamma^Y_{\phi V1}$}
    \FAProp(0.,15.)(7.,14.)(0.,){Straight}{-1}
    \FALabel(3.7192,15.5544)[b]{$\ell_R^g$}
    \FAProp(0.,5.)(7.,6.)(0.,){Straight}{1}
    \FALabel(3.7192,4.44558)[t]{${L_L}_a^f$}
    \FAProp(20,10.)(13.5,10.)(0.,){ScalarDash}{-1}
    \FALabel(16.75,10.82)[b]{$\phi_b$}
    \FAProp(7.,14.)(7.,6.)(0.,){Straight}{-1}
    \FALabel(5.93,10.)[r]{${L_L}_d^i$}
    \FAProp(7.,14.)(13.5,10.)(0.,){ScalarDash}{1}
    \FALabel(10.4513,12.6272)[bl]{$\phi_c$}
    \FAProp(7.,6.)(13.5,10.)(0.,){Sine}{0}
    \FALabel(10.5824,7.15993)[tl]{$B$}
    \FAVert(7.,14.){0}
    \FAVert(7.,6.){0}
    \FAVert(13.5,10.){0}
    
    \FADiagram{$\ii\Gamma^Y_{\phi V2}$}
    \FAProp(0.,15.)(7.,14.)(0.,){Straight}{-1}
    \FALabel(3.7192,15.5544)[b]{$\ell_R^g$}
    \FAProp(0.,5.)(7.,6.)(0.,){Straight}{1}
    \FALabel(3.7192,4.44558)[t]{${L_L}_a^f$}
    \FAProp(20,10.)(13.5,10.)(0.,){ScalarDash}{-1}
    \FALabel(16.75,10.82)[b]{$\phi_b$}
    \FAProp(7.,14.)(7.,6.)(0.,){Straight}{-1}
    \FALabel(5.93,10.)[r]{$\ell_R^k$}
    \FAProp(7.,14.)(13.5,10.)(0.,){Sine}{0}
    \FALabel(10.5824,12.8401)[bl]{$B$}
    \FAProp(7.,6.)(13.5,10.)(0.,){ScalarDash}{1}
    \FALabel(10.4513,7.37284)[tl]{$\phi_c$}
    \FAVert(7.,14.){0}
    \FAVert(7.,6.){0}
    \FAVert(13.5,10.){0}
        
\end{feynartspicture}     
\caption{\label{fig:yukawa_eft}Feynman diagrams for the lepton Yukawa coupling at one-loop level in the effective theory.}
\end{center}
\end{figure}

The lepton Yukawa coupling also needs to be matched. At one-loop level, the lepton Yukawa coupling receives an additional contribution in the full theory as compared to the effective theory. In the full theory, the contributions are shown in Fig.~\ref{fig:yukawa_full}, whereas the ones in the effective theory are shown in Fig.~\ref{fig:yukawa_eft}.

The only Feynman diagram that needs to be computed is $\ii\Gamma^Y_N$ in Fig.~\ref{fig:yukawa_full}. Using the Feynman rules, it evaluates to
\begin{align}
    {\bar\mu}^{\epsilon/2} \ii \left(\Gamma^Y_N\right)^{gf}_{ba} &= \int\frac{\dd^dk}{(2\pi^d)}\left(-\ii{\bar\mu}^{\epsilon/2} (Y_{\ell})_{gk}\delta_{cd} \PL\right) \frac{\ii(\slashed{k}-\slashed{q_1}-\slashed{q_2})}{(k-q_1-q_2)^2} \left(-\ii{\bar\mu}^{\epsilon/2} (Y_\nu^\dagger)_{ki}\epsilon_{db}\PR\right)\nonumber\\
    &\quad\times\frac{\ii(\slashed{k}+M)}{k^2-M^2} \left(-\ii{\bar\mu}^{\epsilon/2}(Y_\nu)_{if}(\epsilon^T)_{ca}\PL\right)\frac{\ii}{(k-q_1)^2-m^2} \nonumber\\
    &= -\frac{\ii{\bar\mu}^{\epsilon/2}}{16\pi^2} (Y_{\ell}Y_\nu^\dagger Y_\nu)_{gf} \delta_{ab} \PL \left[B_0(0;m,0)+m^2 C_2\left(0,0,m^2;M,m,0\right)\right.\nonumber\\
    &\quad\left. +m^2C_1\left(0,0,m^2;M,m,0\right)+M^2 C_0\left(0,0,m^2;M,m,0\right) \right. \nonumber\\
    &\quad\left. -\left(\slashed{q_1}\slashed{q_2}\right) C_1\left(0,0,m^2;M,m,0\right)\right],
\end{align}
where $B_0$, $C_0$, $C_1$, and $C_2$ are Passarino--Veltman functions. The finite part of this diagram in the limit of heavy RHNs is
\begin{equation}
    {\bar\mu}^{\epsilon/2} \ii \left(\Gamma^Y_N|_\text{finite}\right)^{gf}_{ba} = -\frac{\ii{\bar\mu}^{\epsilon/2}}{32\pi^2} (Y_{\ell}Y_\nu^\dagger Y_\nu)_{gf} \delta_{ab} \PL\left[2 + \frac{\slashed{q_1}\slashed{q_2}}{M^2}\left(1+\ln\frac{m^2}{M^2}\right)\right].
\end{equation}
Thus, since $q_1q_2 \ll M^2$, we obtain the correction to the lepton Yukawa coupling as
\begin{equation}
    {\bar\mu}^{\epsilon/2} \ii \left(\Gamma^Y_N|_\text{finite}\right)^{gf}_{ba} \simeq -\frac{\ii{\bar\mu}^{\epsilon/2}}{16\pi^2} (Y_{\ell}Y_\nu^\dagger Y_\nu)_{gf} \delta_{ab} \PL .
    \label{eq:GYN}
\end{equation}

\FloatBarrier

\subsection{Higgs Quartic Coupling}

The Higgs quartic coupling receives several corrections at one-loop level. In the full and the effective theories, there exist some Feynman diagrams that appear in both theories (but with the matched effective couplings in the effective theory), as well as some diagrams that contain $\kappa$ in the effective theory and some diagrams with an internal $N$ propagator in the full theory.

In the full theory, these diagrams are displayed in Fig.~\ref{fig:higgs2}. There also exist  diagrams involving gauge bosons, but these are not 1LPI and therefore do not generate unique matching. In the effective theory, there is only one diagram, displayed in Fig.~\ref{fig:higgs_eff}. These, as well as the computations of the diagrams, are presented in App.~\ref{app:lambda_loops}.

After all individual amplitudes have been computed, they need to be put together and evaluated in the limit of heavy RHNs. This procedure proves to be technically involved and the following steps are followed:
\begin{enumerate}
    \item Express all integrals in terms of Passarino--Veltman functions.
    \item Add all amplitudes together to find the full amplitude.
    \item Divide the total amplitude into several terms, each with a different Passarino--Veltman function.
    \item Expand each term separately to second order in the small parameters $m/M$, $p\cdot q_1/M^2$, and $p \cdot q_2/M^2$.
    \item Add the expanded terms together and retain only the lowest non-trivial order in the expansion parameter.
\end{enumerate}

\FloatBarrier

\subsection{Neutrino Mass Matrix}

\subsubsection{Full Theory}

In the full theory, the one-loop Feynman diagrams can be split into several classes. We organize the diagrams according to which counterterm diagram they correspond to. They are all reproduced in App.~\ref{app:kappa_loops_full}.

The divergent diagrams all have corresponding counterterm insertions. For the diagrams with $s$-channel RHN exchange, the class of diagrams with a corresponding counterterm insertion on the right vertex are shown in Fig.~\ref{fig:RHN_s_right_ct}. Those with an insertion on the left vertex are shown in Fig.~\ref{fig:RHN_s_left_ct} and those with a counterterm insertion on the RHN propagator are in Fig.~\ref{fig:RHN_s_mid_ct}. Similarly for the $t$-channel exchange of RHNs, we have a class of diagrams corresponding to a counterterm on the top vertex in Fig.~\ref{fig:RHN_t_top_ct}, those with a counterterm on the bottom vertex in Fig.~\ref{fig:RHN_t_bottom_ct} and on the propagator in Fig.~\ref{fig:RHN_t_mid_ct}. The finite diagrams that involve gauge bosons are shown in Fig.~\ref{fig:RHN_finite_gauge} and those without gauge bosons in Fig.~\ref{fig:RHN_finite_other}. Due to the large number of diagrams in Figs.~\ref{fig:RHN_s_right_ct}--\ref{fig:RHN_finite_other}, we do not reproduce their expressions here.

\subsubsection{Effective Theory}\label{sec:kappa_loops}

In the effective theory, the divergent diagrams, which correspond to the counterterm, are given in Fig.~\ref{fig:EFT_divergent}. There are no finite diagrams appearing in the effective theory. All diagrams are given and computed in App.~\ref{app:kappa_loops_EFT}. To verify that the calculations are correct, we also derive the counterterm for $\kappa$ in App.~\ref{sec:counterterm}.

\FloatBarrier

\section{Matching Procedure}\label{sec:mp}

The matching procedure is outlined in Ref.~\cite{Antusch:2015pda}. It amounts to equating the amplitudes in the two theories at the matching scale to the order in perturbation theory that we are interested in. 

\subsection{General Matching of Couplings}

Let $Q^\text{EFT}$ and $Q^\text{f.t.}$ denote some amplitude in the effective and the full theories, respectively, to the order of interest. We also need to include the external fields for the amplitudes. Therefore, we have the matching condition
\begin{equation}
    \Phi^\text{EFT}_{i_1}\cdots \Phi^\text{EFT}_{i_m} Q^\text{EFT} \Phi^\text{EFT}_{i_{m+1}}\cdots \Phi^\text{EFT}_{i_n} = \Phi^\text{f.t.}_{i_1}\cdots \Phi^\text{f.t.}_{i_m} Q^\text{f.t.}\Phi^\text{f.t.}_{i_{m+1}}\cdots \Phi^\text{f.t.}_{i_n},
\end{equation}
where the possible matrix structure of $Q$ is preserved by writing the fields and couplings in the same order as they appear in the Lagrangian. Then, using
\begin{equation}
\Phi^\text{EFT}_{i} = \left(1+\frac12 \Delta\Phi_{i}\right) \Phi^\text{f.t.}_{i}
\quad \Leftrightarrow \quad
\Phi^\text{f.t.}_{i} \simeq \left(1-\frac12 \Delta\Phi_{i}\right) \Phi^\text{EFT}_{i},
\end{equation}
we find that
\begin{equation}
    Q^\text{EFT} = \left(1-\frac12 \Delta\Phi_{i_1}\right) \cdots \left(1-\frac12 \Delta\Phi_{i_m}\right)Q^\text{f.t.} \left(1-\frac12 \Delta\Phi_{i_{m+1}}\right) \cdots \left(1-\frac12 \Delta\Phi_{i_n}\right).
\end{equation}
Next, expanding to first order in $\Delta\Phi$, we have
\begin{equation}
    Q^\text{EFT} = \left(1-\frac12 \sum_{i\in\{i_1,\dots,i_m\}}\Delta\Phi_i\right) Q^\text{f.t.}\left(1-\frac12 \sum_{i\in\{i_{m+1},\dots,i_n\}}\Delta\Phi_i\right).
\end{equation}
Finally, note that $Q^\text{f.t.} = Q^\text{f.t.}_\text{tree} + Q^\text{f.t.}_\text{1 loop}$ and that to the appropriate order, we obtain
\begin{equation}\label{eq:matching}
    Q^\text{EFT} = Q^\text{f.t.}_\text{tree} - \frac12 \left(\sum_{i\in\{i_1,\dots,i_m\}}\Delta\Phi_i\right)Q^\text{f.t.}_\text{tree} - \frac12 Q^\text{f.t.}_\text{tree}\left(\sum_{i\in\{i_1,\dots,i_m\}}\Delta\Phi_i\right) + Q^\text{f.t.}_\text{1 loop}.
\end{equation}

\subsection{Matching Wave Function Corrections}

In order to compute the matching of some amplitude according to Eq.~\eqref{eq:matching}, we first need to relate the fields in the two theories. This can be performed by looking at two-point functions, i.e.~the propagator, at loop level. In general, the propagator receives self-energy correction $-\ii\Sigma(p)$. Thus, following Ref.~\cite{Grozin:2012ec}, we find a series for the propagator $\ii\Delta(p)$ as
\begin{align}
    \ii\Delta(p) &= \ii\Delta_0(p) + \ii\Delta_0(p)\left[-\ii\Sigma(p)\right]\ii\Delta_0(p) + \dots \nonumber\\
    &= \ii\Delta_0(p) \left[1 + \Sigma(p)\Delta_0(p) + \Sigma(p)\Delta_0(p)\Sigma(p)\Delta_0(p) + \dots\right] \nonumber\\
    &= \frac{\ii\Delta_0(p)}{1 - \Sigma(p)\Delta_0(p)} = \frac{\ii}{\Delta_0^{-1}(p) - \Sigma(p)},
\end{align}
where $\ii\Delta_0(p)$ is the free propagator.

Now, we specialize to the fermion case. We have massless fermions with the free propagator $\Delta_0(p)=\slashed{p}/p^2$. Thus, we obtain
\begin{equation}
    \Delta(p) = \frac{1}{\left(\frac{\slashed{p}}{p^2}\right)^{-1} - \Sigma(p)} = \frac{\slashed{p}/p^2}{1-\frac{\slashed{p}}{p^2}\Sigma(p)} = \frac{\slashed{p}}{\slashed{p}\left(\slashed{p}-\Sigma(p)\right)} = \frac{1}{\slashed{p}-\Sigma(p)}.
\end{equation}
Then, we use the fact that the fermion self-energy is proportional to $\slashed{p}$, so we can write $\Sigma(p) = \slashed{p}\Sigma'(p)$. Inserting this allows us to write
\begin{equation}
    \Delta(p) = \frac{1}{\slashed{p}\left(1-\Sigma'(p)\right)} = \frac{1 + \Sigma'(p)}{\slashed{p}} \equiv \Delta_R(p) Z,
\end{equation}
which defines the renormalized propagator $\Delta_R(p) \equiv \slashed{p}/p^2$, which is the same as the free propagator, and the renormalization constant $Z \equiv 1+\Sigma'(p)$.

We can perform a similar procedure for scalars, but we need also to take into account the mass renormalization. In this case, $\Delta_0(p^2) = 1/(p^2-m^2)$ and we find
\begin{equation}
    \Delta(p^2) = \frac{1}{p^2-m^2-\Sigma(p^2)}.
    \label{eq:DRp2}
\end{equation}
Now, expanding $\Sigma(p^2)$ around the physical mass $m_p^2$ as
\begin{equation}
    \Sigma(p^2) = \Sigma(m_p^2) + (p^2-m_p^2)\Sigma'(m_p^2) + \tilde{\Sigma}(p^2),
\end{equation}
where the last term contains the rest of the terms in the expansion, we can write Eq.~\eqref{eq:DRp2} as
\begin{equation}
    \Delta(p^2) = \frac{1}{p^2 - m^2 -\Sigma(m_p^2) - (p^2-m_p^2)\Sigma'(m_p^2) - \tilde{\Sigma}(p^2)}.
    \label{eq:DRp22}
\end{equation}
Then, defining the physical mass through $m_p^2 = m^2 + \Sigma(m_p^2)$ and writing $\tilde{\Sigma}(p^2)\simeq \left[1 - \Sigma'(m_p^2)\right]\tilde{\Sigma}(p^2)$ (which holds to the desired order in perturbation theory), we can write Eq.~\eqref{eq:DRp22} as
\begin{align}
    \Delta(p^2) &= \frac{1}{(p^2-m_p^2)\left[1 -\Sigma'(m_p^2)\right] - \tilde{\Sigma}(p^2)} \simeq \frac{1}{\left[p^2-m_p^2 - \tilde{\Sigma}(p^2)\right]\left[1 -\Sigma'(m_p^2)\right]} \nonumber\\
    &\simeq \frac{1 + \Sigma'(m_p^2)}{p^2-m_p^2 - \tilde{\Sigma}(p^2)} \equiv \frac{Z}{p^2 - (m^2 +\delta m^2)} \equiv \Delta_R(p^2) Z,
\end{align}
which defines the renormalization constant $Z \equiv 1 + \Sigma'(m_p^2)$, the mass shift $\delta m^2 \equiv \Sigma(m_p^2)$, as well as the renormalized propagator $\Delta_R(p^2) \equiv [p^2 - (m^2 +\delta m^2)]^{-1}$.

The quantity $Z$ is different in the two theories, since they have different corrections in $\Sigma(p)$. This leads to the fields having different normalizations in the two theories, which can be seen by simply matching the two-point function. Matching the two-point function for a field $\Phi$ in the two theories yields the relation
\begin{equation}
    {\Phi^\text{EFT}}^\dagger \Delta_R^\text{EFT}\Phi^\text{EFT} = {\Phi^\text{f.t.}}^\dagger \Delta_R^\text{f.t.}\Phi^\text{f.t.}.
\end{equation}
Rescaling the propagator by a factor of $Z$ in both sides gives
\begin{equation}
    {\Phi^\text{EFT}}^\dagger {(Z_\Phi^\text{EFT})^{-1/2}}^\dagger\Delta(Z_\Phi^\text{EFT})^{-1/2}\Phi^\text{EFT} = {\Phi^\text{f.t.}}^\dagger {(Z_\Phi^\text{f.t.})^{-1/2}}^\dagger\Delta(Z_\Phi^\text{f.t.})^{-1/2}\Phi^\text{f.t.}.
\end{equation}
Since the free propagator $\Delta$ is the same on both sides, this leads to a matching
\begin{equation}
    (Z_\Phi^\text{EFT})^{-1/2}\Phi^\text{EFT} = (Z_\Phi^\text{f.t.})^{-1/2}\Phi^\text{f.t.}
\end{equation}
such that 
\begin{equation}\label{eq:wf_matching0}
    \Phi^\text{EFT} = (Z_\Phi^\text{EFT})^{1/2}(Z_\Phi^\text{f.t.})^{-1/2}\Phi^\text{f.t.}.
\end{equation}
Expanding the right-hand side to first order in $\Sigma(p)$, we find that
\begin{equation}\label{eq:wf_matching}
    \Phi^\text{EFT} \simeq \left(1+\frac12 \Delta\Phi\right)\Phi^\text{f.t.},
\end{equation}
which defines $\Delta\Phi$ in terms of the renormalization constants.

The two fields that have such corrections are the lepton and the Higgs doublets. Starting with the lepton doublet and expanding Eq.~\eqref{eq:wf_matching0} with $\Sigma$ as a perturbation and comparing to Eq.~\eqref{eq:wf_matching}, we obtain the matching as
\begin{equation}
    1+\frac12 \Delta L \simeq 1+\frac12 {\Sigma_L^\text{EFT}}'(p) - \frac12 {\Sigma_L^\text{f.t.}}'(p).
\end{equation}
Since the two $\Sigma_L$ quantities differ in only $\Sigma^L_N$, which appears in $\Sigma_L^\text{f.t.}$ and not in $\Sigma_L^\text{EFT}$, we have, using Eq.~\eqref{eq:SLNfinite}, the lepton doublet wave function correction
\begin{equation}
    \Delta L^{gf}_{ba} = -\left({\Sigma^L_N}'(p)\right)^{gf}_{ba} \simeq \frac{1}{32\pi^2}\delta_{ab} (Y_\nu^\dagger Y_\nu)_{gf} \left[\frac32 + \frac{m^2}{M^2}\left(1+4\ln\frac{m}{M}\right)\right] \PL \sim \frac{3}{64\pi^2}\delta_{ab} (Y_\nu^\dagger Y_\nu)_{gf} \PL.
    \label{eq:DL}
\end{equation}
Similarly, the conjugated lepton doublet has 
\begin{equation}
    \Delta\overline{L}^{gf}_{ba} \simeq \frac{1}{32\pi^2}\delta_{ab} (Y_\nu^\dagger Y_\nu)_{gf}\left[\frac32 + \frac{m^2}{M^2}\left(1+4\ln\frac{m}{M}\right)\right]\PR \sim \frac{3}{64\pi^2}\delta_{ab} (Y_\nu^\dagger Y_\nu)_{gf} \PR.
    \label{eq:DLbar}
\end{equation}

By the same argument for the Higgs doublet as for the lepton and the conjugated lepton doublets, we obtain the matching as
\begin{equation}
    1+\frac12 \Delta \phi \simeq 1+\frac12 {\Sigma_\phi^\text{EFT}}'(m_p^2) - \frac12 {\Sigma_\phi^\text{f.t.}}'(m_p^2),
\end{equation}
which, using Eq.~\eqref{eq:sigma_phi}, gives the Higgs doublet wave function correction
\begin{equation}
    \Delta \phi_{ba} = -\left({\Sigma^\phi_N}'(m_p^2)\right)_{ba} \simeq \frac{1}{32\pi^2} \Tr(Y_\nu^\dagger Y_\nu)\delta_{ab}.
    \label{eq:DP}
\end{equation}

\subsection{Higgs Mass}

With the renormalization of the propagator as above, we shift the position of the pole and hence the physical mass of the Higgs boson. Thus, we have the physical mass 
\begin{equation}
    m_p^2 = m^2 + \delta m^2 = m^2 + \Sigma(m_p^2).
\end{equation}
The last term on the right-hand side will contain contributions from all self-energy diagrams. Comparing the full theory to the effective theory, we observe that they only differ in the Feynman diagram with the RHN. Thus, assuming $m_p^2 \ll M^2$ and using Eq.~\eqref{eq:sigma_phi2}, we find that
\begin{equation}
    (m_p^\text{EFT})^2 \simeq (m_p^\text{f.t.})^2 - \frac{1}{8\pi^2} M^2 \Tr(Y_\nu^\dagger Y_\nu).
\end{equation}

\subsection{U(1) and SU(2) Gauge Couplings}

These two couplings receive contributions from a RHN running in the loop. At one-loop level, they differ by an additional Feynman diagram in the full theory compared to the effective theory. 

\subsubsection{Lepton-Lepton-Gauge Couplings} \label{sec:llg}

We first use the vertex with two lepton doublets and one vector boson to define the gauge couplings. In the full theory, we split the loop-level amplitudes into $\Gamma_N$, which contains the RHNs, and $\Gamma_\text{loop}$, which contains the rest. We then have for $g_1$ the following matching relation
\begin{align}
    \Gamma_\text{tree}(g_1^{\text{EFT}}) + \Gamma_\text{loop}^\text{EFT}(g_1^{\text{EFT}}) &\simeq \Gamma_\text{tree}(g_1^{\text{f.t.}}) -\frac12 \Delta \overline{L} \Gamma_\text{tree}(g_1^{\text{f.t.}}) - \frac12 \Gamma_\text{tree}(g_1^{\text{f.t.}}) \Delta L \nonumber \\
    &\quad+ \Gamma_\text{loop}^\text{f.t.}(g_1^{\text{f.t.}}) + \Gamma_N(g_1^{\text{f.t.}}),
\end{align}
where $\Gamma_\text{loop}^\text{EFT}$ contains couplings in the effective theory, but is otherwise the same as $\Gamma_\text{loop}^\text{f.t.}$. To one-loop level, the loop-level terms are the same on the two sides, giving rise to
\begin{equation}
    \Gamma_\text{tree}(g_1^{\text{EFT}}) \simeq \Gamma_\text{tree}(g_1^{\text{f.t.}})-\frac12 \Delta \overline{L} \Gamma_\text{tree}(g_1^{\text{f.t.}}) - \frac12 \Gamma_\text{tree}(g_1^{\text{f.t.}}) \Delta L + \Gamma_N(g_1^{\text{f.t.}}).
\end{equation}
Using the same argument for $g_2$ leads to
\begin{equation}
     \Gamma_\text{tree}(g_2^{\text{EFT}}) \simeq \Gamma_\text{tree}(g_2^{\text{f.t.}})-\frac12 \Delta \overline{L} \Gamma_\text{tree}(g_2^{\text{f.t.}}) - \frac12 \Gamma_\text{tree}(g_2^{\text{f.t.}}) \Delta L + \Gamma_N(g_2^{\text{f.t.}}).
\end{equation}
Now, we have the Feynman rule at tree level
\begin{equation}
    \ii \left(\Gamma_\text{tree}(g_A)_\mu\right)^{gf}_{ba} = -\ii g_A \delta_{gf} T^A_{ba} \gamma_\mu \PL
\end{equation}
in both the effective and the full theory with the only difference being that $g_A$ equals to $g_A^\text{EFT}$ or $g_A^\text{f.t.}$.

Thus, using these intermediate results [together with Eqs.~\eqref{eq:GVL}, \eqref{eq:DL}, and~\eqref{eq:DLbar}] gives
\begin{align}
&-g_A^{\text{EFT}} \delta_{gf} \gamma_\mu \PL T^{A(L)}_{ba} \simeq -g_A^{\text{f.t.}}\delta_{gf} \gamma_\mu \PL T^{A(L)}_{ba} - \frac12 \left(\frac{3}{64\pi^2}\delta_{bc} (Y_\nu^\dagger Y_\nu)_{gi} \PR \right) \left(-g_A^{\text{f.t.}}\delta_{if}T^{A(L)}_{ca}\gamma_\mu \PL\right) \nonumber\\
&- \frac12 \left(-g_A^{\text{f.t.}}\delta_{gi}T^{A(L)}_{bc}\gamma_\mu \PL\right)\left(\frac{3}{64\pi^2}\delta_{ca} (Y_\nu^\dagger Y_\nu)_{if} \PL \right) + \frac{3}{64\pi^2} g_A^{\text{f.t.}}(Y_\nu^\dagger Y_\nu)_{gf}\epsilon_{ac}\epsilon_{bd}T^{A(\phi)}_{cd}\gamma_\mu \PL.
\end{align}
Multiplying both sides by $\delta_{gf}$ and summing over flavor indices, we obtain
\begin{equation}
    g_A^{\text{EFT}} T^{A(L)}_{ba} \simeq g_A^{\text{f.t.}}\left[T^{A(L)}_{ba} - \frac{1}{64\pi^2} \Tr(Y_\nu^\dagger Y_\nu) \left(T^{A(L)}_{ba} + \epsilon_{ac}\epsilon_{bd}T^{A(\phi)}_{cd}\right) \right].
    \label{eq:gAEFTL}
\end{equation}

We now define the quantity $X^A_{bc} \equiv T^{A(L)}_{ba} + \epsilon_{ac}\epsilon_{bd}T^{A(\phi)}_{cd}$ that will give the gauge structure for the matching. Starting with the $\text{U}(1)$ case, we have $T^{0(L)}_{ba} = Y_L\delta_{ba}=-\frac12\delta_{ba}$ and $T^{0(\phi)}_{cd} =Y_\phi\delta_{cd} = +\frac12\delta_{cd}$, since, in the first case, we have lepton doublets coupling to the gauge boson and, in the second case, we have Higgs coupling to the gauge boson. Thus, we find that
\begin{equation}
    X^0_{ba} = -\frac12\delta_{ba} + \epsilon_{ac}\epsilon_{bd} \frac12\delta_{cd} = \frac12\left(-\delta_{ba} + \epsilon_{bc} (\epsilon^T)_{ca}\right)= \frac12(-\delta_{ba} + \delta_{ba}) = 0.
\end{equation}
Meanwhile, for the $\SU(2)$ case, we have $T^A_{ba} = T^{A(L)}_{ba} = T^{A(\phi)}_{ba} = \frac12\tau^A_{ba}$, where $\tau^A$ are the Pauli matrices. We can then write
\begin{equation}
    X^A_{ba} = \frac12\tau^A_{ba} + \epsilon_{ac}\epsilon_{bd}\frac12 \tau^A_{cd} = \frac12\tau^A_{ba} + \frac12 \left(\epsilon(\tau^A)^T\epsilon^T\right)_{ba} = \frac12\tau^A_{ba} + \frac12 (-\tau^A)_{ba} = 0.
    \label{eq:XAba}
\end{equation}
Finally, using Eqs.~\eqref{eq:gAEFTL}--\eqref{eq:XAba}, we obtain the matching conditions for the two couplings $g_1$ and $g_2$ as
\begin{equation}
    g_1^{\text{EFT}} \simeq g_1^{\text{f.t.}}, \quad
    g_2^{\text{EFT}} \simeq g_2^{\text{f.t.}},
    \label{eq:g1g2LLG}
\end{equation}
respectively, which means that the matching conditions are trivial.

\subsubsection{Higgs-Higgs-Gauge Couplings}

In Sec.~\ref{sec:llg}, we defined the gauge couplings using the vertex with two lepton doublets and one vector boson. Instead, defining the gauge couplings through the interaction of the Higgs doublet with the gauge boson, the matching will be 
\begin{equation}
    \Gamma_\text{tree}(g_A^{\text{EFT}}) \simeq \Gamma_\text{tree}(g_A^{\text{f.t.}}) -\frac12\Delta\phi\Gamma_\text{tree}(g_A^{\text{f.t.}}) - \frac12 \Gamma_\text{tree}(g_A^{\text{f.t.}})\Delta\phi + \Gamma_N(g_A^{\text{f.t.}}),
    \label{eq:GgAEFT}
\end{equation}
where the different quantities [see Eqs.~\eqref{eq:GVP} and~\eqref{eq:DP}] are given by
\begin{align}
    \ii \left(\Gamma_\text{tree}(g_A)_\mu\right)_{ba} &= -\ii g_A (q_1-q_2)_\mu T^{A}_{ba}, \label{eq:Gtreeba}\\
    \Delta\phi_{ba} &\simeq \frac{1}{32\pi^2}\Tr(Y_\nu^\dagger Y_\nu)\delta_{ba},\\
    \ii \left(\Gamma_N(g_A)_\mu\right)_{ba} &\simeq \frac{\ii}{32\pi^2} g_A \Tr(Y_\nu^\dagger Y_\nu) (q_1-q_2)_\mu \epsilon_{ac}\epsilon_{bd} T^{A}_{cd}. \label{eq:GNba}
\end{align}
Inserting Eqs.~\eqref{eq:Gtreeba}--\eqref{eq:GNba} into Eq.~\eqref{eq:GgAEFT} yields
\begin{align}
&-g_A^{\text{EFT}} (q_1-q_2)_\mu T^{A(\phi)}_{ba} \simeq -g_A^{\text{f.t.}}(q_1-q_2)_\mu T^{A(\phi)}_{ba} - \frac12 \left[\frac{1}{32\pi^2}\Tr(Y_\nu^\dagger Y_\nu) \delta_{bc}\right] \left[-g_A^{\text{f.t.}}(q_1-q_2)_\mu T^{A(\phi)}_{ca}\right] \nonumber\\
&- \frac12\left[-g_A^{\text{f.t.}}(q_1-q_2)_\mu T^{A(\phi)}_{bc}\right]\left[\frac{1}{32\pi^2}\Tr(Y_\nu^\dagger Y_\nu) \delta_{ca}\right] + \frac{1}{32\pi^2}g_A^{\text{f.t.}}\Tr(Y_\nu^\dagger Y_\nu)(q_1-q_2)_\mu \epsilon_{ac}\epsilon_{bd}T^{A(L)}_{cd}.
\end{align}

The situation is similar to that for the lepton-lepton-gauge coupling, but with the difference in the $\text{U}(1)$ case that $T^{0(\phi)}_{ba} = Y_\phi\delta_{ba} = +\frac12 \delta_{ba}$ and $T^{0(L)}_{cd} = Y_L\delta_{cd}=- \frac12 \delta_{cd}$. The resulting matching is then
\begin{equation}
    g_1^{\text{EFT}} \simeq g_1^{\text{f.t.}}, \quad g_2^{\text{EFT}} \simeq g_2^{\text{f.t.}}.
    \label{eq:g1g2HHG}
\end{equation}
Finally, note that the result in Eq.~\eqref{eq:g1g2HHG} is exactly the same as that derived from the interaction with leptons presented in Sec.~\ref{sec:llg} [see Eq.~\eqref{eq:g1g2LLG}].

\subsection{SU(3) Gauge Coupling}

The QCD coupling constant receives no matching, since the Feynman diagrams at one-loop level are the same in both the full and the effective theories, and the external legs (quark doublet $Q_L$ and right-handed $u_R$ and $d_R$) are also the same in both theories. Hence, we have
\begin{equation}
    g_3^\text{EFT} = g_3^\text{f.t.}.
\end{equation}

\subsection{Quark Yukawa Couplings}

The quark Yukawa couplings $Y_u$ and $Y_d$ have the same one-loop Feynman diagrams in the two theories. However, due to the corrections to the external $\phi$ leg of the diagrams, it still receives a matching at one-loop level.

First, consider $Y_u$. Since all Feynman diagrams are proportional to $Y_u$, we can factor it out and also split the amplitudes up into tree-level and loop-level. Equation~\eqref{eq:matching} then gives the matching
\begin{equation}
    \Gamma_\text{tree}(Y_u^\text{EFT}) + \Gamma_\text{loop}^\text{EFT}(Y_u^\text{EFT}) \simeq \left(1-\frac12 \Delta\phi\right) \Gamma_\text{tree}(Y_u^\text{f.t.}) + \Gamma_\text{loop}^\text{f.t.}(Y_u^\text{f.t.}),
\end{equation}
where $\Gamma_\text{tree}$ is the same on both sides, while $\Gamma_\text{loop}$ differs in the couplings that appear in either theory. However, they are proportional to the same kinematic quantity in both sides with the only difference being the coupling. Considering the EFT couplings to be of the form $\eta^\text{EFT} = \eta^\text{f.t.}(1+\Delta \eta)$ with $\Delta \eta$ being a loop-level correction (and similar for any other couplings that may appear inside $\Gamma_\text{loop}^\text{EFT}$), keeping to one-loop order in perturbation theory means that the loop-level terms are the same in the two theories. This leaves us with
\begin{equation}
    Y_u^\text{EFT} \simeq \left(1-\frac12\Delta \phi\right) Y_u^\text{f.t.}.
    \label{eq:YuEFT}
\end{equation}
Second, consider $Y_d$. By the same argument as for $Y_u$, we also have
\begin{equation}
    Y_d^\text{EFT} \simeq \left(1-\frac12\Delta \phi\right)Y_d^\text{f.t.}.
    \label{eq:YdEFT}
\end{equation}
Now, we can simply insert $\Delta\phi \simeq \frac{1}{32\pi^2}\Tr(Y_\nu^\dagger Y_\nu)$ [cf.~Eq.~\eqref{eq:DP}] into Eqs.~\eqref{eq:YuEFT} and~\eqref{eq:YdEFT} to obtain the matching conditions
\begin{align}
    Y_u^\text{EFT} &\simeq Y_u^\text{f.t.}\left[1-\frac{1}{64\pi^2}\Tr(Y_\nu^\dagger Y_\nu)\right], \\
    Y_d^\text{EFT} &\simeq Y_d^\text{f.t.}\left[1-\frac{1}{64\pi^2}\Tr(Y_\nu^\dagger Y_\nu)\right].
\end{align}

\subsection{Lepton Yukawa Coupling}

Just as for the gauge couplings $g_1$ and $g_2$, the difference in the two theories is one additional Feynman diagram in the full theory containing a RHN at one-loop level. We can directly use the same argument to find the matching as
\begin{equation}
    \Gamma_\text{tree}(Y_{\ell}^\text{EFT}) \simeq \Gamma_\text{tree}(Y_{\ell}^\text{f.t.}) -\frac12 \Gamma_\text{tree}(Y_{\ell}^\text{f.t.}) \Delta L - \frac12 \Gamma_\text{tree}(Y_{\ell}^\text{f.t.}) \Delta \phi + \Gamma_N (Y_{\ell}^\text{f.t.}).
\end{equation}
Using the quantities [see Eqs.~\eqref{eq:GYN}, \eqref{eq:DL}, and \eqref{eq:DP}]
\begin{align}
    \ii \left(\Gamma_\text{tree}(Y_{\ell})\right)^{gf}_{ba} &= -\ii (Y_{\ell})_{gf} \delta_{ba}\PL,\\
    \Delta\phi_{ba} &\simeq \frac{1}{32\pi^2} \Tr(Y_\nu^\dagger Y_\nu)\delta_{ba},\\
    \Delta L^{gf}_{ba} &\simeq \frac{3}{64\pi^2} (Y_\nu^\dagger Y_\nu)_{gf} \delta_{ba}\PL,\\
    \ii \left(\Gamma_N(Y_{\ell})\right)^{gf}_{ba} &\simeq -\frac{\ii}{16\pi^2} (Y_{\ell}Y_\nu^\dagger Y_\nu)_{gf} \delta_{ba}\PL,
\end{align}
we can write the matching as 
\begin{multline}
    (Y_{\ell}^\text{EFT})_{gf} \delta_{ba}\PL \simeq (Y_{\ell}^\text{f.t.})_{gi} \left\{\delta_{if}\delta_{ba}\PL - \frac12 (\delta_{bc}\PL) \left[\frac{3}{64\pi^2} (Y_\nu^\dagger Y_\nu)_{if}\delta_{ca}\PL\right] \right.\\
    -\left. \frac12 \left[\frac{1}{32\pi^2} \Tr(Y_\nu^\dagger Y_\nu)\delta_{bc} \right](\delta_{if}\delta_{ca}\PL) + \frac{1}{16\pi^2}(Y_\nu^\dagger Y_\nu)_{if} \delta_{ba}\PL \right\},
\end{multline}
which simplifies to the matching condition
\begin{equation}
    (Y_{\ell}^\text{EFT})_{gf} \simeq (Y_{\ell}^\text{f.t.})_{gi} \left\{\delta_{if}\left[1 - \frac{1}{64\pi^2}\Tr(Y_\nu^\dagger Y_\nu)\right] + \frac{5}{128\pi^2}(Y_\nu^\dagger Y_\nu)_{if} \right\}.
\end{equation}

\subsection{Higgs Quartic Coupling}

The matching for the Higgs quartic coupling at one-loop level reads
\begin{align}
    \Gamma_\text{tree}^\text{EFT}(\lambda^\text{EFT}) +  \Gamma_\text{loop}^\text{EFT}(\lambda^\text{EFT}) \simeq \; &\Gamma_\text{tree}^\text{f.t.}(\lambda^\text{f.t.}) - \frac12 \Gamma_\text{tree}^\text{f.t.}(\lambda^\text{f.t.}) \Delta \phi - \frac12 \Gamma_\text{tree}^\text{f.t.}(\lambda^\text{f.t.}) \Delta \phi\nonumber\\
    & - \frac12 \Delta \phi \Gamma_\text{tree}^\text{f.t.}(\lambda^\text{f.t.}) - \frac12  \Delta \phi \Gamma_\text{tree}^\text{f.t.}(\lambda^\text{f.t.}) + \Gamma_\text{loop}^\text{f.t.}(\lambda^\text{f.t.}).\label{eq:match_lambda}
\end{align}
In this case, we have found the following quantities [cf.~Eq.~\eqref{eq:DP}]
\begin{align}
    \ii \left(\Gamma_\text{tree}(\lambda)\right)_{abcd} &= -\frac{\ii}{2} \lambda (\delta_{ac}\delta_{bd} + \delta_{ad}\delta_{bc}), \label{eq:HQCgl}\\
    \Delta\phi &\simeq \frac{1}{32\pi^2} \Tr(Y_\nu^\dagger Y_\nu),\\
    \ii \left(\Gamma_\text{loop}^\text{EFT}(\lambda^\text{EFT})\right)_{abcd} &\simeq \frac{\ii}{16\pi^2}p\cdot q_1\log\left(-\frac{M^2}{p\cdot q_1}\right)\Tr(\kappa^\dagger\kappa)(\delta_{ac}\delta_{bd} + \delta_{ad}\delta_{bc}),\\
    \ii \left(\Gamma_\text{loop}^\text{f.t.}(\lambda^\text{f.t.})\right)_{abcd} &\simeq \frac{\ii}{8\pi^2}\left[\Tr(Y_\nu^\dagger Y_\nu^* Y_\nu^T Y_\nu)+\frac{m^2+p\cdot q_1}{p\cdot q_1}\Tr(Y_\nu^\dagger Y_\nu Y_\nu^\dagger Y_\nu)\right](\delta_{ac}\delta_{bd} + \delta_{ad}\delta_{bc}), \label{eq:HQCl}
\end{align}
where the two quantities $\Gamma_\text{loop}^\text{EFT}(\lambda^\text{EFT})$ and $\Gamma_\text{loop}^\text{f.t.}(\lambda^\text{f.t.})$ have been computed in great detail using Mathematica and are based on the nine loop amplitudes listed in Eqs.~\eqref{eq:Glb1}--\eqref{eq:Glk} in App.~\ref{app:lambda_loops}.
Here, we have assumed that $p_1\simeq p_2\equiv p$. Combining the quantities in Eqs.~\eqref{eq:HQCgl}--\eqref{eq:HQCl}, using $\log(-x)= \log(x) + \ii\pi \simeq \log(x)$ for large $x$, and simplifying the group structure, Eq.~\eqref{eq:match_lambda} becomes
\begin{align}\label{eq:lambda_matching}
    \lambda^\text{EFT} &\simeq \lambda^\text{f.t.} \left[1 - \frac{1}{16\pi^2}\Tr(Y_\nu^\dagger Y_\nu)\right] \nonumber\\   
    &\quad +\frac{1}{8\pi^2} \left[p\cdot q_1 \log\left(\frac{M^2}{p\cdot q_1}\right)\Tr(\kappa^\dagger \kappa) - 2\Tr(Y_\nu^\dagger Y_\nu^* Y_\nu^T Y_\nu) - 2 \frac{m^2+p\cdot q_1}{p\cdot q_1}\Tr(Y_\nu^\dagger Y_\nu Y_\nu^\dagger Y_\nu)\right].
\end{align}

Since the coupling in the effective theory is to be matched to the couplings in the full theory, we wish for the right-hand side of Eq.~\eqref{eq:lambda_matching} to only contain couplings of the full theory. As such, we need to substitute $\kappa$ with its expression, $\kappa = 2Y_\nu^T M^{-1}Y_\nu$. This is the tree-level expression, which is the one required in order to retain the correct order in the loop expansion. Then, the trace becomes $\Tr(Y_\nu^\dagger Y_\nu^* Y_\nu^T Y_\nu)/M^2$, resulting in the matching condition being given by
\begin{align}\label{eq:lambda_matching2}
    \lambda^\text{EFT} &\simeq \lambda^\text{f.t.} \left[1 - \frac{1}{16\pi^2}\Tr(Y_\nu^\dagger Y_\nu)\right] \nonumber\\
    &\quad +\frac{1}{4\pi^2} \left\{\left[2\frac{p\cdot q_1}{M^2}\log\left(\frac{M^2}{p\cdot q_1}\right) - 1\right]\Tr(Y_\nu^\dagger Y_\nu^* Y_\nu^T Y_\nu) - \frac{m^2+p\cdot q_1}{p\cdot q_1}\Tr(Y_\nu^\dagger Y_\nu Y_\nu^\dagger Y_\nu)\right\}.
\end{align}
Furthermore, setting $m^2\simeq p\cdot q_1$, since both are considered small quantities, we find that
\begin{align}\label{eq:lambda_matching3}
    \lambda^\text{EFT} &\simeq \lambda^\text{f.t.} \left[1 - \frac{1}{16\pi^2}\Tr(Y_\nu^\dagger Y_\nu)\right] \nonumber\\
    &\quad +\frac{1}{4\pi^2} \left\{\left[2\frac{p\cdot q_1}{M^2}\log\left(\frac{M^2}{p\cdot q_1}\right) - 1\right]\Tr(Y_\nu^\dagger Y_\nu^* Y_\nu^T Y_\nu) - 2\Tr(Y_\nu^\dagger Y_\nu Y_\nu^\dagger Y_\nu)\right\}.
\end{align}
Finally, neglecting the term proportional to $\frac{p\cdot q_1}{M^2}\log\left(\frac{M^2}{p\cdot q_1}\right)$, we obtain that the matching condition can be approximated by
\begin{equation}\label{eq:lambda_matching4}
    \lambda^\text{EFT} \simeq \lambda^\text{f.t.} \left[1 - \frac{1}{16\pi^2}\Tr(Y_\nu^\dagger Y_\nu)\right] - \frac{1}{4\pi^2} \left[\Tr(Y_\nu^\dagger Y_\nu^* Y_\nu^T Y_\nu) + 2\Tr(Y_\nu^\dagger Y_\nu Y_\nu^\dagger Y_\nu)\right].
\end{equation}

\subsection{Effective Neutrino Mass Matrix}

The matching for the effective neutrino mass matrix at one-loop level is given by the following equation
\begin{align}\label{eq:kappa_matching_loop}
    \Gamma_\text{tree}^\text{EFT}(\kappa) +  \Gamma_\text{loop}^\text{EFT}(\kappa) \simeq \; &\Gamma_\text{tree}^\text{f.t.}(\{Y_\nu, M\}) - \frac12 \Gamma_\text{tree}^\text{f.t.}(\{Y_\nu, M\}) \Delta L - \frac12 \Gamma_\text{tree}^\text{f.t.}(\{Y_\nu, M\}) \Delta \phi \nonumber\\
    & - \frac12 \Delta L \Gamma_\text{tree}^\text{f.t.}(\{Y_\nu, M\}) - \frac12  \Delta \phi \Gamma_\text{tree}^\text{f.t.}(\{Y_\nu, M\}) + \Gamma_\text{loop}^\text{f.t.}(\{Y_\nu, M\}).
\end{align}
Using the following quantities for tree-level amplitudes and wave function corrections [see Eqs.~\eqref{eq:Gammaab}, \eqref{eq:Gammac}, \eqref{eq:DL}, and \eqref{eq:DP}], i.e.
\begin{align}
    \ii \left(\Gamma^\text{EFT}_\text{tree}(\kappa)\right)_{abcd} &= \frac{\ii}{2} \kappa (\epsilon_{cd}\epsilon_{ba} + \epsilon_{ca}\epsilon_{bd})\PL,\\
    \ii \left(\Gamma^\text{f.t.}_\text{tree}(\{Y_\nu, M\})\right)_{abcd} &= \ii Y_\nu^T M^{-1} Y_\nu (\epsilon_{cd}\epsilon_{ba} + \epsilon_{ca}\epsilon_{bd})\PL,\\
    \Delta\phi &\simeq \frac{1}{32\pi^2} \Tr(Y_\nu^\dagger Y_\nu),\\
    \Delta L &\simeq \frac{3}{64\pi^2} Y_\nu^\dagger Y_\nu \PL,
\end{align}
we can write the matching condition in Eq.~\eqref{eq:kappa_matching_loop} as
\begin{align}
    \kappa (\epsilon_{cd}\epsilon_{ba} + \epsilon_{ca}\epsilon_{bd}) \PL &\simeq \; Y_\nu^T M^{-1} Y_\nu\left[2 - \frac{1}{16\pi^2} \Tr(Y_\nu^\dagger Y_\nu)\right](\epsilon_{cd}\epsilon_{ba} + \epsilon_{ca}\epsilon_{bd}) \PL \nonumber \\
    &\quad - \frac{3}{64\pi^2} \left(Y_\nu^T M^{-1} Y_\nu Y_\nu^\dagger Y_\nu + Y_\nu^\dagger Y_\nu Y_\nu^T M^{-1} Y_\nu \right)(\epsilon_{cd}\epsilon_{ba} + \epsilon_{ca}\epsilon_{bd}) \PL \nonumber \\
    &\quad + 2\left[\left(\Gamma^\text{f.t.}_\text{loop}(\{Y_\nu, M\})\right)_{abcd} - \left(\Gamma^\text{EFT}_\text{loop}(\kappa)\right)_{abcd}\right]
\label{eq:MCk}
\end{align}
into which the expressions for the two quantities $\Gamma^\text{f.t.}_\text{loop}(\{Y_\nu, M\})$ and $\Gamma^\text{EFT}_\text{loop}(\kappa)$ are to be computed and inserted. Note that we leave the computation of these two expressions for future work. Finally, multiplying Eq.~\eqref{eq:MCk} with the structure $\delta_{ad} \delta_{bc}$ (it is also possible to use $\delta_{ab} \delta_{cd}$ or $\delta_{ac} \delta_{bd}$) and by the projection operator $\PL$ from the right as well as identifying the non-zero contributions for $\PL$, we obtain the matching condition
\begin{align}
    \kappa &\simeq \; Y_\nu^T M^{-1} Y_\nu\left[2 - \frac{1}{16\pi^2} \Tr(Y_\nu^\dagger Y_\nu)\right] - \frac{3}{64\pi^2} \left(Y_\nu^T M^{-1} Y_\nu Y_\nu^\dagger Y_\nu + Y_\nu^\dagger Y_\nu Y_\nu^T M^{-1} Y_\nu \right) \nonumber \\
    &\quad + \frac{1}{2} \left[\left(\Gamma^\text{f.t.}_\text{loop}(\{Y_\nu, M\})\right)_{abcd} - \left(\Gamma^\text{EFT}_\text{loop}(\kappa)\right)_{abcd}\right] \delta_{ad} \delta_{bc}.
\end{align}

\section{Summary and Conclusions}\label{sec:sc}

In summary, using a Feynman diagrammatical approach by computing the loop amplitudes, the matching conditions at one-loop level between the effective and the full theories are given by
\begin{align}\label{eq:summary}
(m_p^\text{EFT})^2 &\simeq (m_p^\text{f.t.})^2 - \frac{1}{8\pi^2} M^2 \Tr(Y_\nu^\dagger Y_\nu), \\
g_A^{\text{EFT}} &\simeq g_A^{\text{f.t.}}, \quad A = 1,2,3 \\
Y_q^\text{EFT} &\simeq Y_q^\text{f.t.}\left[1-\frac{1}{64\pi^2}\Tr(Y_\nu^\dagger Y_\nu)\right], \quad q = u,d, \\
(Y_{\ell}^\text{EFT})_{gf} &\simeq (Y_{\ell}^\text{f.t.})_{gi} \left\{\delta_{if}\left[1 - \frac{1}{64\pi^2}\Tr(Y_\nu^\dagger Y_\nu)\right] + \frac{5}{128\pi^2}(Y_\nu^\dagger Y_\nu)_{if} \right\},\label{eq:Ye_summary} \\
\lambda^\text{EFT} &\simeq \lambda^\text{f.t.} \left[1 - \frac{1}{16\pi^2}\Tr(Y_\nu^\dagger Y_\nu)\right] - \frac{1}{4\pi^2} \left[\Tr(Y_\nu^\dagger Y_\nu^* Y_\nu^T Y_\nu) + 2\Tr(Y_\nu^\dagger Y_\nu Y_\nu^\dagger Y_\nu)\right], \label{eq:lambda_summary}\\
\kappa &\simeq \; Y_\nu^T M^{-1} Y_\nu\left[2 - \frac{1}{16\pi^2} \Tr(Y_\nu^\dagger Y_\nu)\right] - \frac{3}{64\pi^2} \left(Y_\nu^T M^{-1} Y_\nu Y_\nu^\dagger Y_\nu + Y_\nu^\dagger Y_\nu Y_\nu^T M^{-1} Y_\nu \right) \nonumber \\
    &\quad + \frac{1}{2} \left[\left(\Gamma^\text{f.t.}_\text{loop}(\{Y_\nu, M\})\right)_{abcd} - \left(\Gamma^\text{EFT}_\text{loop}(\kappa)\right)_{abcd}\right] \delta_{ad} \delta_{bc}.
    \label{eq:kappa_summary}
\end{align}
We also have the matching wave function corrections for the Higgs and the lepton doublet fields, respectively, namely
\begin{equation}\label{eq:DeltaPhDeltaL}
\Delta \phi \simeq \frac{1}{32\pi^2} \Tr(Y_\nu^\dagger Y_\nu), \quad \Delta L \simeq \frac{3}{64\pi^2} Y_\nu^\dagger Y_\nu \PL .
\end{equation}

In conclusion, comparing our results in Eqs.~\eqref{eq:summary}--\eqref{eq:lambda_summary} and~\eqref{eq:DeltaPhDeltaL} to the results found in Ref.~\cite{Zhang:2021jdf}, we have found that they are consistent to leading order for all parameters except $\lambda$. For $\lambda$ in Eq.~\eqref{eq:lambda_summary}, our coefficients in front of $\Tr(Y_\nu^\dagger Y_\nu Y_\nu^\dagger Y_\nu)$ and $\Tr(Y_\nu^\dagger Y_\nu^* Y_\nu^T Y_\nu)$ are larger than the corresponding coefficient in Ref.~\cite{Zhang:2021jdf}, but with the same sign.

\section*{Acknowledgments}

We would like to thank Di Zhang and Shun Zhou for useful discussions. T.O.~acknowledges support by the Swedish Research Council (Vetenskapsr{\aa}det) through contract No.~2017-03934.

\newpage

\appendix

\section{Loop Amplitudes}

We collect some of the Feynman diagrams and computations of loop amplitudes that are too lengthy to include in the main text.

\subsection{Higgs Quartic Coupling}\label{app:lambda_loops}

In this appendix, we present the 1LPI loop amplitudes for the Higgs quartic coupling, which consist of eight Feynman diagrams in the full theory and one Feynman diagram in the effective theory. 

\begin{figure}[h]
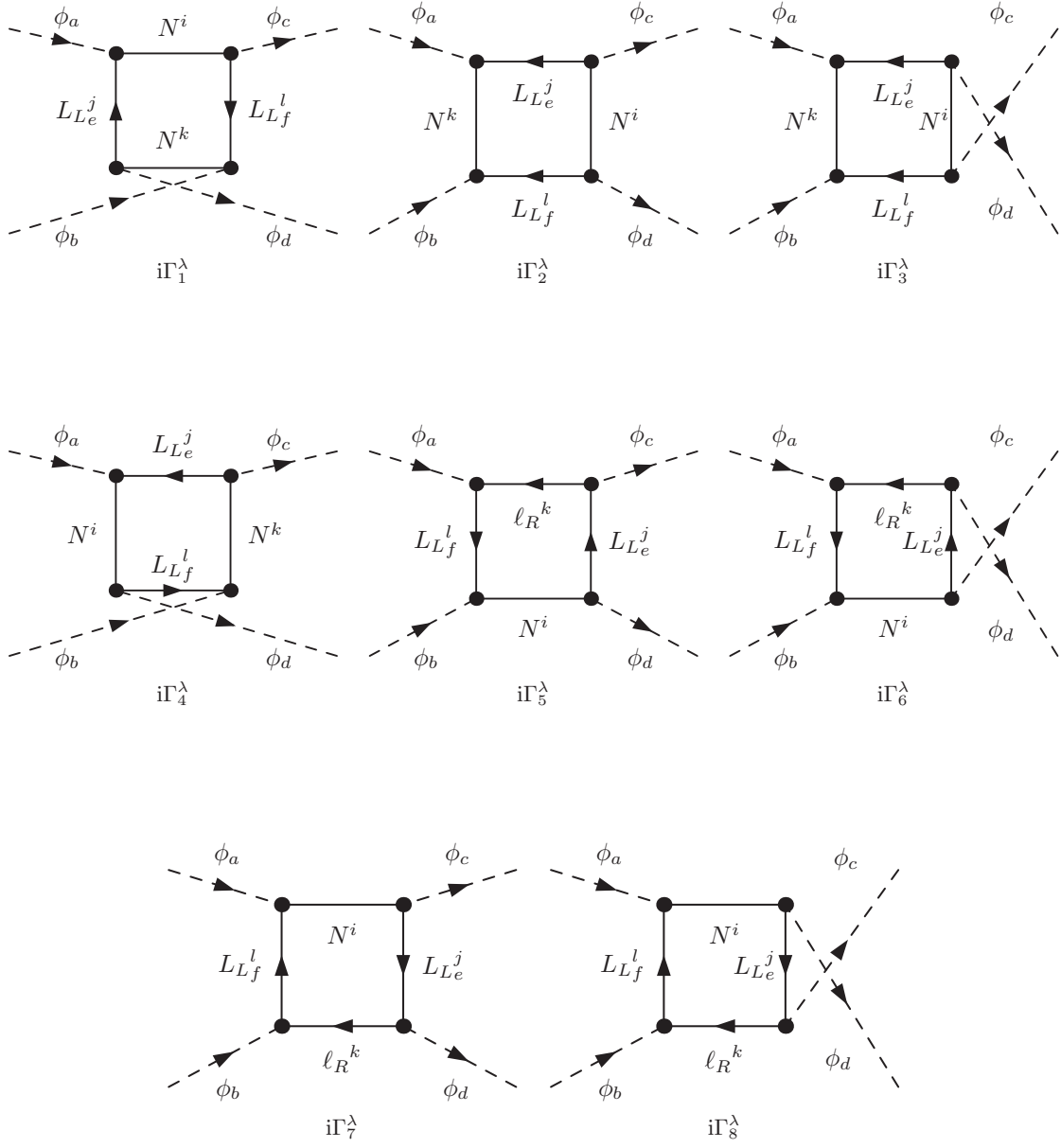

\vspace{-1cm}
\begin{center}
    \begin{feynartspicture}(432,168)(3,1.1)
        
        \FADiagram{$\ii\Gamma^{\lambda}_1$}
        \FAProp(0.,15.)(6.5,13.5)(0.,){ScalarDash}{1}
        \FALabel(3.54232,15.0367)[b]{$\phi_a$}
        \FAProp(0.,2.5)(13.5,6.5)(0.,){ScalarDash}{1}
        \FALabel(3.54232,3.)[t]{$\phi_b$}
        \FAProp(20,15.)(13.5,13.5)(0.,){ScalarDash}{-1}
        \FALabel(16.4577,15.0367)[b]{$\phi_c$}
        \FAProp(20,2.5)(6.5,6.5)(0.,){ScalarDash}{-1}
        \FALabel(16.4577,3.)[t]{$\phi_d$}
        \FAProp(6.5,13.5)(6.5,6.5)(0.,){Straight}{-1}
        \FALabel(5.43,10.)[r]{${L_L}_e^j$}
        \FAProp(6.5,13.5)(13.5,13.5)(0.,){Straight}{0}
        \FALabel(10.,14.57)[b]{$N^i$}
        \FAProp(6.5,6.5)(13.5,6.5)(0.,){Straight}{0}
        \FALabel(10.,9.)[t]{$N^k$}
        \FAProp(13.5,13.5)(13.5,6.5)(0.,){Straight}{1}
        \FALabel(14.57,10.)[l]{${L_L}_f^l$}
        \FAVert(6.5,13.5){0}
        \FAVert(6.5,6.5){0}
        \FAVert(13.5,13.5){0}
        \FAVert(13.5,6.5){0}    

        \FADiagram{$\ii\Gamma^{\lambda}_2$}
        \FAProp(0.,15.)(6.5,13.)(0.,){ScalarDash}{1}
        \FALabel(3.37315,15.28)[b]{$\phi_a$}
        \FAProp(0.,2.5)(6.5,6.)(0.,){ScalarDash}{1}
        \FALabel(3.44767,3.)[t]{$\phi_b$}
        \FAProp(20,15.)(13.5,13.)(0.,){ScalarDash}{-1}
        \FALabel(16.6265,15.2808)[b]{$\phi_c$}
        \FAProp(20,2.5)(13.5,6.)(0.,){ScalarDash}{-1}
        \FALabel(16.5523,3.)[t]{$\phi_d$}
        \FAProp(13.5,13.)(6.5,13.)(0.,){Straight}{1}
        \FALabel(10.,11.93)[t]{${L_L}_e^j$}
        \FAProp(13.5,13.)(13.5,6.)(0.,){Straight}{0}
        \FALabel(14.57,9.5)[l]{$N^i$}
        \FAProp(6.5,6.)(6.5,13.)(0.,){Straight}{0}
        \FALabel(5.43,9.5)[r]{$N^k$}
        \FAProp(6.5,6.)(13.5,6.)(0.,){Straight}{-1}
        \FALabel(10.,4.93)[t]{${L_L}_f^l$}
        \FAVert(13.5,13.){0}
        \FAVert(6.5,6.){0}
        \FAVert(6.5,13.){0}
        \FAVert(13.5,6.){0}

        \FADiagram{$\ii\Gamma^{\lambda}_3$}
        \FAProp(0.,15.)(6.5,13.)(0.,){ScalarDash}{1}
        \FALabel(3.37315,15.28)[b]{$\phi_a$}
        \FAProp(0.,2.5)(6.5,6.)(0.,){ScalarDash}{1}
        \FALabel(3.44767,3.)[t]{$\phi_b$}
        \FAProp(20,15.)(13.5,6.)(0.,){ScalarDash}{-1}
        \FALabel(16.6265,15.2808)[b]{$\phi_c$}
        \FAProp(20,2.5)(13.5,13.)(0.,){ScalarDash}{-1}
        \FALabel(16.5523,4.69512)[t]{$\phi_d$}
        \FAProp(13.5,13.)(6.5,13.)(0.,){Straight}{1}
        \FALabel(10.,11.93)[t]{${L_L}_e^j$}
        \FAProp(13.5,13.)(13.5,6.)(0.,){Straight}{0}
        \FALabel(11.5,9.5)[l]{$N^i$}
        \FAProp(6.5,6.)(6.5,13.)(0.,){Straight}{0}
        \FALabel(5.43,9.5)[r]{$N^k$}
        \FAProp(6.5,6.)(13.5,6.)(0.,){Straight}{-1}
        \FALabel(10.,4.93)[t]{${L_L}_f^l$}
        \FAVert(13.5,13.){0}
        \FAVert(6.5,6.){0}
        \FAVert(6.5,13.){0}
        \FAVert(13.5,6.){0}

        \end{feynartspicture}
        \begin{feynartspicture}(432,168)(3,1.1)

        \FADiagram{$\ii\Gamma^{\lambda}_4$}
        \FAProp(0.,15.)(6.5,13.5)(0.,){ScalarDash}{1}
        \FALabel(3.54232,15.28)[b]{$\phi_a$}
        \FAProp(0.,2.5)(13.5,6.5)(0.,){ScalarDash}{1}
        \FALabel(3.54232,3.)[t]{$\phi_b$}
        \FAProp(20,15.)(13.5,13.5)(0.,){ScalarDash}{-1}
        \FALabel(16.4577,15.0367)[b]{$\phi_c$}
        \FAProp(20,2.5)(6.5,6.5)(0.,){ScalarDash}{-1}
        \FALabel(16.4577,3.)[t]{$\phi_d$}
        \FAProp(6.5,13.5)(6.5,6.5)(0.,){Straight}{0}
        \FALabel(5.43,10.)[r]{$N^i$}
        \FAProp(6.5,13.5)(13.5,13.5)(0.,){Straight}{-1}
        \FALabel(10.,14.57)[b]{${L_L}_e^j$}
        \FAProp(6.5,6.5)(13.5,6.5)(0.,){Straight}{1}
        \FALabel(10.,9.)[t]{${L_L}_f^l$}
        \FAProp(13.5,13.5)(13.5,6.5)(0.,){Straight}{0}
        \FALabel(14.57,10.)[l]{$N^k$}
        \FAVert(6.5,13.5){0}
        \FAVert(6.5,6.5){0}
        \FAVert(13.5,13.5){0}
        \FAVert(13.5,6.5){0}
        
        \FADiagram{$\ii\Gamma^{\lambda}_5$}
        \FAProp(0.,15.)(6.5,13.)(0.,){ScalarDash}{1}
        \FALabel(3.37315,15.28)[b]{$\phi_a$}
        \FAProp(0.,2.5)(6.5,6.)(0.,){ScalarDash}{1}
        \FALabel(3.44767,3.)[t]{$\phi_b$}
        \FAProp(20,15.)(13.5,13.)(0.,){ScalarDash}{-1}
        \FALabel(16.6265,15.2808)[b]{$\phi_c$}
        \FAProp(20,2.5)(13.5,6.)(0.,){ScalarDash}{-1}
        \FALabel(16.5523,3.)[t]{$\phi_d$}
        \FAProp(13.5,13.)(6.5,13.)(0.,){Straight}{1}
        \FALabel(10.,11.93)[t]{${\ell_R}^k$}
        \FAProp(13.5,13.)(13.5,6.)(0.,){Straight}{-1}
        \FALabel(14.57,9.5)[l]{${L_L}_e^j$}
        \FAProp(6.5,6.)(6.5,13.)(0.,){Straight}{-1}
        \FALabel(5.43,9.5)[r]{${L_L}_f^l$}
        \FAProp(6.5,6.)(13.5,6.)(0.,){Straight}{0}
        \FALabel(10.,4.93)[t]{$N^i$}
        \FAVert(13.5,13.){0}
        \FAVert(6.5,6.){0}
        \FAVert(6.5,13.){0}
        \FAVert(13.5,6.){0}
        
        \FADiagram{$\ii\Gamma^{\lambda}_6$}
        \FAProp(0.,15.)(6.5,13.)(0.,){ScalarDash}{1}
        \FALabel(3.37315,15.28)[b]{$\phi_a$}
        \FAProp(0.,2.5)(6.5,6.)(0.,){ScalarDash}{1}
        \FALabel(3.44767,3.)[t]{$\phi_b$}
        \FAProp(20,15.)(13.5,6.)(0.,){ScalarDash}{-1}
        \FALabel(16.6265,15.2808)[b]{$\phi_c$}
        \FAProp(20,2.5)(13.5,13.)(0.,){ScalarDash}{-1}
        \FALabel(16.5523,4.69512)[t]{$\phi_d$}
        \FAProp(13.5,13.)(6.5,13.)(0.,){Straight}{1}
        \FALabel(10.,11.93)[t]{${\ell_R}^k$}
        \FAProp(13.5,13.)(13.5,6.)(0.,){Straight}{-1}
        \FALabel(10.5,9.5)[l]{${L_L}_e^j$}
        \FAProp(6.5,6.)(6.5,13.)(0.,){Straight}{-1}
        \FALabel(5.43,9.5)[r]{${L_L}_f^l$}
        \FAProp(6.5,6.)(13.5,6.)(0.,){Straight}{0}
        \FALabel(10.,4.93)[t]{$N^i$}
        \FAVert(13.5,13.){0}
        \FAVert(6.5,6.){0}
        \FAVert(6.5,13.){0}
        \FAVert(13.5,6.){0}

\end{feynartspicture}
\begin{feynartspicture}(432,168)(2,1.1)

        \FADiagram{$\ii\Gamma^{\lambda}_{7}$}
        \FAProp(0.,15.)(6.5,13.)(0.,){ScalarDash}{1}
        \FALabel(3.37315,15.28)[b]{$\phi_a$}
        \FAProp(0.,2.5)(6.5,6.)(0.,){ScalarDash}{1}
        \FALabel(3.44767,3.)[t]{$\phi_b$}
        \FAProp(20,15.)(13.5,13.)(0.,){ScalarDash}{-1}
        \FALabel(16.6265,15.2808)[b]{$\phi_c$}
        \FAProp(20,2.5)(13.5,6.)(0.,){ScalarDash}{-1}
        \FALabel(16.5523,3.)[t]{$\phi_d$}
        \FAProp(13.5,13.)(6.5,13.)(0.,){Straight}{0}
        \FALabel(10.,11.93)[t]{$N^i$}
        \FAProp(13.5,13.)(13.5,6.)(0.,){Straight}{1}
        \FALabel(14.57,9.5)[l]{${L_L}_e^j$}
        \FAProp(6.5,6.)(6.5,13.)(0.,){Straight}{1}
        \FALabel(5.43,9.5)[r]{${L_L}_f^l$}
        \FAProp(6.5,6.)(13.5,6.)(0.,){Straight}{-1}
        \FALabel(10.,4.93)[t]{${\ell_R}^k$}
        \FAVert(13.5,13.){0}
        \FAVert(6.5,6.){0}
        \FAVert(6.5,13.){0}
        \FAVert(13.5,6.){0}
    
        \FADiagram{$\ii\Gamma^{\lambda}_{8}$}
        \FAProp(0.,15.)(6.5,13.)(0.,){ScalarDash}{1}
        \FALabel(3.37315,15.28)[b]{$\phi_a$}
        \FAProp(0.,2.5)(6.5,6.)(0.,){ScalarDash}{1}
        \FALabel(3.44767,3.)[t]{$\phi_b$}
        \FAProp(20,15.)(13.5,6.)(0.,){ScalarDash}{-1}
        \FALabel(17.,15.)[b]{$\phi_c$}
        \FAProp(20,2.5)(13.5,13.)(0.,){ScalarDash}{-1}
        \FALabel(16.5523,4.69512)[t]{$\phi_d$}
        \FAProp(13.5,13.)(6.5,13.)(0.,){Straight}{0}
        \FALabel(10.,11.93)[t]{$N^i$}
        \FAProp(13.5,13.)(13.5,6.)(0.,){Straight}{1}
        \FALabel(10.5,9.5)[l]{${L_L}_e^j$}
        \FAProp(6.5,6.)(6.5,13.)(0.,){Straight}{1}
        \FALabel(5.43,9.5)[r]{${L_L}_f^l$}
        \FAProp(6.5,6.)(13.5,6.)(0.,){Straight}{-1}
        \FALabel(10.,4.93)[t]{${\ell_R}^k$}
        \FAVert(13.5,13.){0}
        \FAVert(6.5,6.){0}
        \FAVert(6.5,13.){0}
        \FAVert(13.5,6.){0}

    \end{feynartspicture}
    \caption{\label{fig:higgs2}Feynman diagrams for Higgs quartic coupling in full theory.}
    \end{center}
\end{figure}

First, in Fig.~\ref{fig:higgs2}, we display the eight Feynman diagrams in the full theory. Using the corresponding Feynman rules, we compute the loop amplitudes of these eight contributions to the Higgs quartic coupling, which are given in Eqs.~\eqref{eq:Glb1}--\eqref{eq:Glb8}.

\bigskip

\noindent The contribution $\ii \Gamma^{\lambda}_1$ in the full theory to the Higgs quartic coupling:
\begin{align}
    {\bar\mu}^\epsilon \ii (\Gamma^{\lambda}_1)_{abcd} &= \int\frac{\dd^dk}{(2\pi)^d}(-1) \Tr\left[\left(-\ii {\bar\mu}^{\epsilon/2} (Y_\nu)_{ij} (\epsilon^T)_{ae}P_\text{L}\right)\frac{\ii(\slashed{k}-\slashed{q_1})}{(k-q_1)^2} \left(-\ii{\bar\mu}^{\epsilon/2}(Y_\nu^\dagger)_{jk}\epsilon_{ed}P_\text{R}\right)\right.\nonumber \\
    &\quad \times \left.\frac{\ii(\slashed{k}-\slashed{p_1}+\slashed{q_2}+M)}{(k-p_1+q_2)^2-M^2}\left(-\ii{\bar\mu}^{\epsilon/2}(Y_\nu)_{kl}(\epsilon^T)_{bf}P_\text{L}\right)\frac{\ii(\slashed{k}-\slashed{p_1})}{(k-p_1)^2} \right. \nonumber\\
    &\quad \times \left. \left(-\ii{\bar\mu}^{\epsilon/2}(Y_\nu^\dagger)_{li}\epsilon_{fc}P_\text{R}\right) \frac{\ii(\slashed{k}+M)}{k^2-M^2}\right] \nonumber\\
    &= -{\bar\mu}^{2\epsilon}\Tr(Y_\nu^\dagger Y_\nu Y_\nu^\dagger Y_\nu) \delta_{ad}\delta_{bc}\nonumber \\
    &\quad \times \int\frac{\dd^dk}{(2\pi)^d} \frac{\Tr[P_\text{L} (\slashed{k}-\slashed{q_1})P_\text{R} (\slashed{k}-\slashed{p_1}+\slashed{q_2})P_\text{L}(\slashed{k}-\slashed{p_1})P_\text{R}\slashed{k}]}{(k-q_1)^2[(k-p_1+q_2)^2-M^2](k-p_1)^2(k^2-M^2)}.
    \label{eq:Glb1}
\end{align}

\noindent The contribution $\ii \Gamma^{\lambda}_2$ in the full theory to the Higgs quartic coupling:
\begin{align}
    {\bar\mu}^\epsilon \ii (\Gamma^{\lambda}_2)_{abcd} &= \int\frac{\dd^dk}{(2\pi)^d}(-1) \Tr\left[\left(-\ii {\bar\mu}^{\epsilon/2} (Y_\nu^T)_{jk} \epsilon_{ea}P_\text{L}\right)\frac{\ii(\slashed{k}-\slashed{p_2}+\slashed{q_2})+M}{(k-p_2+q_2)^2-M^2} \right. \nonumber\\
    &\quad \times \left. \left(-\ii{\bar\mu}^{\epsilon/2}(Y_\nu)_{kl}(\epsilon^T)_{bf}P_\text{L}\right) \frac{\ii(\slashed{k}-\slashed{p_2})}{(k-p_2)^2}\left(-\ii{\bar\mu}^{\epsilon/2}(Y_\nu^\dagger)_{li}\epsilon_{fd}P_\text{R}\right)\frac{\ii(\slashed{k}+M)}{k^2-M^2} \right. \nonumber\\
    &\quad \times \left. \left(-\ii{\bar\mu}^{\epsilon/2}(Y_\nu^*)_{ij}(\epsilon^T)_{ce}P_\text{R}\right)\frac{\ii(\slashed{k}+\slashed{p_1})}{(k+p_1)^2}\right] \nonumber\\
    &= -{\bar\mu}^{2\epsilon}\Tr(Y_\nu^T Y_\nu Y_\nu^\dagger Y_\nu^*) \delta_{ac}\delta_{bd}\nonumber \\
    &\quad \times \int\frac{\dd^dk}{(2\pi)^d} \frac{\Tr[P_\text{L} M (\slashed{k}-\slashed{p_2})P_\text{R} MP_\text{R}(\slashed{k}+\slashed{p_1})]}{[(k-p_2+q_2)^2-M^2](k-p_2)^2(k^2-M^2)(k+p_1)^2}.
    \label{eq:Glb2}
\end{align}

\noindent The contribution $\ii \Gamma^{\lambda}_3$ in the full theory to the Higgs quartic coupling:
\begin{align}
    {\bar\mu}^\epsilon \ii (\Gamma^{\lambda}_3)_{abcd} &= \int\frac{\dd^dk}{(2\pi)^d}(-1) \Tr\left[\left(-\ii {\bar\mu}^{\epsilon/2} (Y_\nu^T)_{jk} \epsilon_{ea}P_\text{L}\right)\frac{\ii(\slashed{k}-\slashed{p_1}-\slashed{q_2})+M}{(k-p_1+q_2)^2-M^2} \right. \nonumber\\
    &\quad \times \left. \left(-\ii{\bar\mu}^{\epsilon/2}(Y_\nu)_{kl}(\epsilon^T)_{bf} P_\text{L}\right) \frac{\ii(\slashed{k}-\slashed{p_1})}{(k-p_1)^2}\left(-\ii{\bar\mu}^{\epsilon/2}(Y_\nu^\dagger)_{li}\epsilon_{fc}P_\text{R}\right)\frac{\ii(\slashed{k}+M)}{k^2-M^2} \right.\nonumber \\
    &\quad \times \left. \left(-\ii{\bar\mu}^{\epsilon/2}(Y_\nu^*)_{ij}(\epsilon^T)_{de}P_\text{R}\right)\frac{\ii(\slashed{k}+\slashed{p_2})}{(k+p_2)^2}\right] \nonumber\\
    &= -{\bar\mu}^{2\epsilon}\Tr(Y_\nu^T Y_\nu Y_\nu^\dagger Y_\nu^*) \delta_{ad}\delta_{bc}\nonumber \\
    &\quad \times \int\frac{\dd^dk}{(2\pi)^d} \frac{\Tr[P_\text{L} M (\slashed{k}-\slashed{p_1})P_\text{R} MP_\text{R}(\slashed{k}+\slashed{p_2})]}{[(k-p_1+q_2)^2-M^2](k-p_1)^2(k^2-M^2)(k+p_2)^2}.
    \label{eq:Glb3}
\end{align}

\noindent The contribution $\ii \Gamma^{\lambda}_4$ in the full theory to the Higgs quartic coupling:
\begin{align}
    {\bar\mu}^\epsilon \ii (\Gamma^{\lambda}_4)_{abcd} &= \int\frac{\dd^dk}{(2\pi)^d}(-1) \Tr\left[\left(-\ii {\bar\mu}^{\epsilon/2} (Y_\nu)_{ij} (\epsilon^T)_{ae}P_\text{L}\right)\frac{\ii(\slashed{k}-\slashed{q_1})}{(k-q_1)^2} \left(-\ii{\bar\mu}^{\epsilon/2}(Y_\nu^\dagger)_{jk}\epsilon_{ec} P_\text{R}\right)\right.\nonumber \\
    &\quad \times \left.\frac{\ii(\slashed{k}-\slashed{p_2}+\slashed{q_2}+M)}{(k-p_2+q_2)^2-M^2}\left(-\ii{\bar\mu}^{\epsilon/2}(Y_\nu)_{kl}(\epsilon^T)_{bd}P_\text{L}\right)\frac{\ii(\slashed{k}-\slashed{p_2})}{(k-p_2)^2} \right. \nonumber\\ 
    &\quad \times \left. \left(-\ii{\bar\mu}^{\epsilon/2}(Y_\nu^\dagger)_{li}\epsilon_{fd}P_\text{R}\right)\frac{\ii(\slashed{k}+M)}{k^2-M^2}\right] \nonumber\\
    &= -{\bar\mu}^{2\epsilon}\Tr( Y_\nu^\dagger Y_\nu Y_\nu^\dagger Y_\nu) \delta_{ac}\delta_{bd}\nonumber \\
    &\quad \times \int\frac{\dd^dk}{(2\pi)^d} \frac{\Tr[P_\text{L} (\slashed{k}-\slashed{q_1}) P_\text{R} (\slashed{k}-\slashed{p_2}+\slashed{q_2})P_\text{L} (\slashed{k}-\slashed{p_2})P_\text{R}\slashed{k}]}{(k-q_1)^2[(k-p_2+q_2)^2-M^2](k-p_2)^2(k^2-M^2)}.
    \label{eq:Glb4}
\end{align}

\noindent The contribution $\ii \Gamma^{\lambda}_5$ in the full theory to the Higgs quartic coupling:
\begin{align}
    {\bar\mu}^\epsilon \ii (\Gamma^{\lambda}_5)_{abcd} &= \int\frac{\dd^dk}{(2\pi)^d}(-1) \Tr\left[\left(-\ii {\bar\mu}^{\epsilon/2} (Y_{\ell}^\dagger)_{lk} \delta_{af}P_\text{R}\right)\frac{\ii(\slashed{k}-\slashed{p_1}-\slashed{p_2})}{(k-p_1-p_2)^2} \left(-\ii{\bar\mu}^{\epsilon/2}(Y_{\ell})_{kj}\delta_{ce}P_\text{L}\right)\right.\nonumber \\ 
    &\quad \times \left.\frac{\ii(\slashed{k}-\slashed{p_2})}{(k-p_2)^2}\left(-\ii{\bar\mu}^{\epsilon/2}(Y_\nu^\dagger)_{ji}\epsilon_{ed}P_\text{R}\right)\frac{\ii(\slashed{k}+M)}{k^2-M^2}\left(-\ii{\bar\mu}^{\epsilon/2}(Y_\nu)_{il}(\epsilon^T)_{bf}P_\text{L}\right)\frac{\ii(\slashed{k}-\slashed{q_2})}{(k-q_2)^2}\right] \nonumber\\
    &= -{\bar\mu}^{2\epsilon}\Tr(Y_{\ell}^\dagger Y_{\ell} Y_\nu^\dagger Y_\nu) \epsilon_{ab}\epsilon_{cd}\nonumber \\
    &\quad \times \int\frac{\dd^dk}{(2\pi)^d} \frac{\Tr[P_\text{R} (\slashed{k}-\slashed{p_1}-\slashed{p_2}) P_\text{L} (\slashed{k}-\slashed{p_2})P_\text{R} \slashed{k}P_\text{L}(\slashed{k}-\slashed{q_2})]}{(k-p_1-p_2)^2(k-p_2)^2(k^2-M^2)(k-q_2)^2}.
    \label{eq:Glb5}
\end{align}

\noindent The contribution $\ii \Gamma^{\lambda}_6$ in the full theory to the Higgs quartic coupling:
\begin{align}
    {\bar\mu}^\epsilon \ii (\Gamma^{\lambda}_6)_{abcd} &= \int\frac{\dd^dk}{(2\pi)^d}(-1) \Tr\left[\left(-\ii {\bar\mu}^{\epsilon/2} (Y_{\ell}^\dagger)_{lk} \delta_{af}P_\text{R}\right)\frac{\ii(\slashed{k}-\slashed{p_1}-\slashed{p_2})}{(k-p_1-p_2)^2} \left(-\ii{\bar\mu}^{\epsilon/2}(Y_{\ell})_{kj}\delta_{de}P_\text{L}\right)\right.\nonumber \\
    &\quad \times \left.\frac{\ii(\slashed{k}-\slashed{p_1})}{(k-p_1)^2}\left(-\ii{\bar\mu}^{\epsilon/2}(Y_\nu^\dagger)_{ji}\epsilon_{ec}P_\text{R}\right)\frac{\ii(\slashed{k}+M)}{k^2-M^2}\left(-\ii{\bar\mu}^{\epsilon/2}(Y_\nu)_{il}(\epsilon^T)_{bf}P_\text{L}\right)\frac{\ii(\slashed{k}-\slashed{q_2})}{(k-q_2)^2}\right] \nonumber\\
    &= {\bar\mu}^{2\epsilon}\Tr(Y_{\ell}^\dagger Y_{\ell} Y_\nu^\dagger Y_\nu) \epsilon_{ab}\epsilon_{cd}\nonumber \\
    &\quad \times \int\frac{\dd^dk}{(2\pi)^d} \frac{\Tr[P_\text{R} (\slashed{k}-\slashed{p_1}-\slashed{p_2}) P_\text{L} (\slashed{k}-\slashed{p_1})P_\text{R} \slashed{k}P_\text{L}(\slashed{k}-\slashed{q_2})]}{(k-p_1-p_2)^2(k-p_1)^2(k^2-M^2)(k-q_2)^2}.
    \label{eq:Glb6}
\end{align}

\noindent The contribution $\ii \Gamma^{\lambda}_7$ in the full theory to the Higgs quartic coupling:
\begin{align}
    {\bar\mu}^\epsilon \ii (\Gamma^{\lambda}_7)_{abcd} &= \int\frac{\dd^dk}{(2\pi)^d}(-1) \Tr\left[\left(-\ii {\bar\mu}^{\epsilon/2} (Y_\nu)_{il} (\epsilon^T)_{af}P_\text{L}\right)\frac{\ii(\slashed{k}-\slashed{q_1})}{(k-q_1)^2} \left(-\ii{\bar\mu}^{\epsilon/2}(Y_{\ell}^\dagger)_{lk}\delta_{fb}P_\text{R}\right)\right.\nonumber \\
    &\quad \times \left.\frac{\ii(\slashed{k}-\slashed{p_1}-\slashed{p_2})}{(k-p_1-p_2)^2}\left(-\ii{\bar\mu}^{\epsilon/2}(Y_{\ell})_{kj}\delta_{ed}P_\text{L}\right)\frac{\ii(\slashed{k}-\slashed{p_1})}{(k-p_1)^2}\left(-\ii{\bar\mu}^{\epsilon/2}(Y_\nu^\dagger)_{ji}\epsilon_{ec}P_\text{R}\right)\frac{\ii(\slashed{k}+M)}{k^2-M^2}\right] \nonumber\\
    &= -{\bar\mu}^{2\epsilon}\Tr(Y_{\ell}^\dagger Y_{\ell} Y_\nu^\dagger Y_\nu) \epsilon_{ab}\epsilon_{cd}\nonumber \\
    &\quad \times \int\frac{\dd^dk}{(2\pi)^d} \frac{\Tr[P_\text{L} (\slashed{k}-\slashed{q_1}) P_\text{R} (\slashed{k}-\slashed{p_1}-\slashed{p_2})P_\text{L} (\slashed{k}-\slashed{p_1})P_\text{R}\slashed{k}]}{(k-q_1)^2(k-p_1-p_2)^2(k-p_1)^2(k^2-M^2)}.
    \label{eq:Glb7}
\end{align}

\noindent The contribution $\ii \Gamma^{\lambda}_8$ in the full theory to the Higgs quartic coupling:
\begin{align}
    {\bar\mu}^\epsilon \ii (\Gamma^{\lambda}_8)_{abcd} &= \int\frac{\dd^dk}{(2\pi)^d}(-1) \Tr\left[\left(-\ii {\bar\mu}^{\epsilon/2} (Y_\nu)_{il} (\epsilon^T)_{af}P_\text{L}\right)\frac{\ii(\slashed{k}-\slashed{q_1})}{(k-q_1)^2} \left(-\ii{\bar\mu}^{\epsilon/2}(Y_{\ell}^\dagger)_{lk}\delta_{fb}P_\text{R}\right)\right.\nonumber \\
    &\quad \times \left.\frac{\ii(\slashed{k}-\slashed{p_1}-\slashed{p_2})}{(k-p_1-p_2)^2}\left(-\ii{\bar\mu}^{\epsilon/2}(Y_{\ell})_{kj}\delta_{ec}P_\text{L}\right)\frac{\ii(\slashed{k}-\slashed{p_2})}{(k-p_2)^2}\left(-\ii{\bar\mu}^{\epsilon/2}(Y_\nu^\dagger)_{ji}\epsilon_{ed}P_\text{R}\right)\frac{\ii(\slashed{k}+M)}{k^2-M^2}\right] \nonumber\\
    &= {\bar\mu}^{2\epsilon}\Tr(Y_{\ell}^\dagger Y_{\ell} Y_\nu^\dagger Y_\nu) \epsilon_{ab}\epsilon_{cd}\nonumber \\
    &\quad \times \int\frac{\dd^dk}{(2\pi)^d} \frac{\Tr[P_\text{L} (\slashed{k}-\slashed{q_1}) P_\text{R} (\slashed{k}-\slashed{p_1}-\slashed{p_2})P_\text{L} (\slashed{k}-\slashed{p_2})P_\text{R}\slashed{k}]}{(k-q_1)^2(k-p_1-p_2)^2(k-p_2)^2(k^2-M^2)}.
    \label{eq:Glb8}
\end{align}

\begin{figure}[h]
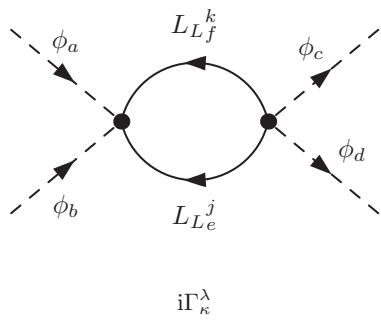

\vspace{-1cm}
\begin{center}
    \begin{feynartspicture}(432,168)(1,1.1)

        \FADiagram{$\ii\Gamma^{\lambda}_{\kappa}$}
        \FAProp(0.,15.)(6.,10.)(0.,){ScalarDash}{1}
        \FALabel(3.,15.)[t]{$\phi_a$}
        \FAProp(0.,5.)(6.,10.)(0.,){ScalarDash}{1}
        \FALabel(3.,6.)[t]{$\phi_b$}
        \FAProp(20,15.)(14.,10.)(0.,){ScalarDash}{-1}
        \FALabel(17.0811,13.3193)[br]{$\phi_c$}
        \FAProp(20,5.)(14.,10.)(0.,){ScalarDash}{-1}
        \FALabel(17.8203,7.77045)[bl]{$\phi_d$}
        \FAProp(6.,10.)(14.,10.)(0.8,){Straight}{-1}
        \FALabel(10.,5.73)[t]{${L_L}_e^j$}
        \FAProp(6.,10.)(14.,10.)(-0.8,){Straight}{-1}
        \FALabel(10.,14.27)[b]{${L_L}_f^k$}
        \FAVert(6.,10.){0}
        \FAVert(14.,10.){0}
        
    \end{feynartspicture}
    \caption{\label{fig:higgs_eff}Feynman diagram for Higgs quartic coupling in effective theory.}
    \end{center}
\end{figure}

Second, in Fig.~\ref{fig:higgs_eff}, we display the sole Feynman diagram in the effective theory. Using the corresponding Feynman rules, we compute the loop amplitude of this contribution to the Higgs quartic coupling, which is given in Eq.~\eqref{eq:Glk}.

\bigskip

\noindent The contribution $\ii \Gamma^{\lambda}_\kappa$ in the effective theory to the Higgs quartic coupling:
\begin{align}
    {\bar\mu}^\epsilon \ii (\Gamma^{\lambda}_\kappa)_{abcd} &= \frac12 \int\frac{\dd^dk}{(2\pi)^d}(-1) \Tr\left\{\left[\ii {\bar\mu}^{\epsilon} \kappa_{jk} \frac12(\epsilon_{fa}\epsilon_{eb} + \epsilon_{fb}\epsilon_{ea})P_\text{L}\right]\frac{\ii\slashed{k}}{k^2}\right.\nonumber \\
    &\quad\times \left. \left[\ii{\bar\mu}^{\epsilon}(\kappa^\dagger)_{kj}\frac12(\epsilon_{fc}\epsilon_{ed} + \epsilon_{ec}\epsilon_{fd})P_\text{R}\right]\frac{\ii(\slashed{k}+\slashed{q_1}+\slashed{q_2})}{(k+q_1+q_2)^2}\right\} \nonumber\\
    &= -{\bar\mu}^{2\epsilon}\frac14 \Tr(\kappa^\dagger \kappa) (\delta_{ac}\delta_{bd}+\delta_{ad}\delta_{bc}) \int\frac{\dd^dk}{(2\pi)^d} \frac{\Tr[P_\text{L} \slashed{k} P_\text{R} (\slashed{k}+\slashed{q_1}+\slashed{q_2})]}{k^2(k+q_1+q_2)^2},
    \label{eq:Glk}
\end{align}
where the first factor of $\frac12$ is a symmetry factor.

\FloatBarrier

\subsection{Neutrino Mass Matrix}

In this appendix, we present the loop amplitudes for the neutrino mass matrix, which consist of 29 Feynman diagrams in the full theory (with six counterterm diagrams) and eleven Feynman diagrams in the effective theory (with one counterterm diagram).

\FloatBarrier

\subsubsection{Full Theory}\label{app:kappa_loops_full}

\begin{figure}[h]
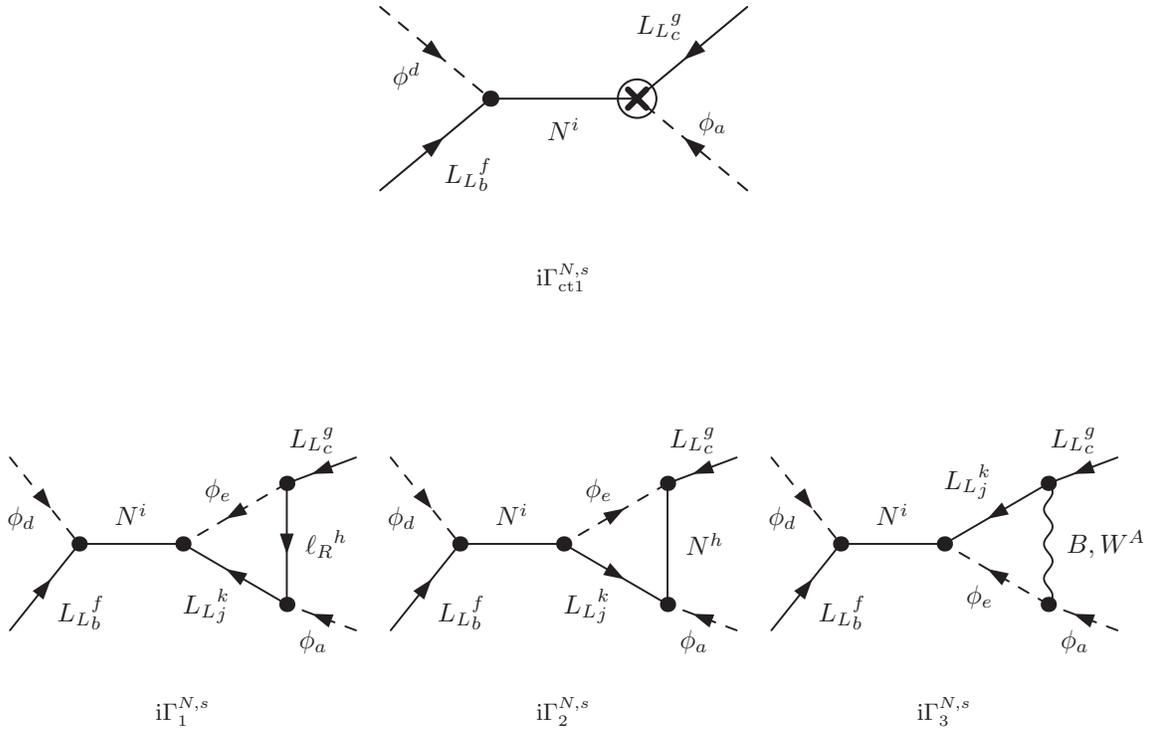

\vspace{-1cm}
\begin{center}
    \begin{feynartspicture}(432,168)(1,1.1)
        
        \FADiagram{$\ii\Gamma^{N,s}_\text{ct1}$}
        \FAProp(0.,15.)(6.,10.)(0.,){ScalarDash}{1}
        \FALabel(2.2,11.9813)[tr]{$\phi^d$}
        \FAProp(0.,5.)(6.,10.)(0.,){Straight}{1}
        \FALabel(3.51229,6.78926)[tl]{${L_L}_b^f$}
        \FAProp(20,15.)(14.,10.)(0.,){Straight}{1}
        \FALabel(16.4877,13.2107)[br]{${L_L}_c^g$}
        \FAProp(20,5.)(14.,10.)(0.,){ScalarDash}{1}
        \FALabel(17.3522,8.01869)[bl]{$\phi_a$}
        \FAProp(6.,10.)(14.,10.)(0.,){Straight}{0}
        \FALabel(10.,8.93)[t]{$N^i$}
        \FAVert(6.,10.){0}
        \FAVert(14.,10.){2}
        
        \end{feynartspicture} 
        \begin{feynartspicture}(432,168)(3,1.1)
        
        \FADiagram{$\ii\Gamma^{N,s}_1$}
        \FAProp(0.,15.)(4.,10.)(0.,){ScalarDash}{1}
        \FALabel(1.46487,12.1679)[tr]{$\phi_d$}
        \FAProp(0.,5.)(4.,10.)(0.,){Straight}{1}
        \FALabel(2.73035,7.01172)[tl]{${L_L}_b^f$}
        \FAProp(20,15.)(16.,13.5)(0.,){Straight}{1}
        \FALabel(17.4558,15.2213)[b]{${L_L}_c^g$}
        \FAProp(20,5.)(16.,6.5)(0.,){ScalarDash}{1}
        \FALabel(17.5435,5.01277)[t]{$\phi_a$}
        \FAProp(4.,10.)(10.,10.)(0.,){Straight}{0}
        \FALabel(7.,11.07)[b]{$N^i$}
        \FAProp(16.,13.5)(16.,6.5)(0.,){Straight}{1}
        \FALabel(17.07,10.)[l]{${\ell_R}^h$}
        \FAProp(16.,13.5)(10.,10.)(0.,){ScalarDash}{1}
        \FALabel(12.825,12.3929)[br]{$\phi_e$}
        \FAProp(16.,6.5)(10.,10.)(0.,){Straight}{1}
        \FALabel(12.699,7.39114)[tr]{${L_L}_j^k$}
        \FAVert(4.,10.){0}
        \FAVert(16.,13.5){0}
        \FAVert(16.,6.5){0}
        \FAVert(10.,10.){0}
        
        \FADiagram{$\ii\Gamma^{N,s}_2$}
        \FAProp(0.,15.)(4.,10.)(0.,){ScalarDash}{1}
        \FALabel(1.46487,12.1679)[tr]{$\phi_d$}
        \FAProp(0.,5.)(4.,10.)(0.,){Straight}{1}
        \FALabel(2.73035,7.01172)[tl]{${L_L}_b^f$}
        \FAProp(20,15.)(16.,13.5)(0.,){Straight}{1}
        \FALabel(17.4558,15.2213)[b]{${L_L}_c^g$}
        \FAProp(20,5.)(16.,6.5)(0.,){ScalarDash}{1}
        \FALabel(17.5435,5.01277)[t]{$\phi_a$}
        \FAProp(4.,10.)(10.,10.)(0.,){Straight}{0}
        \FALabel(7.,11.07)[b]{$N^i$}
        \FAProp(16.,13.5)(16.,6.5)(0.,){Straight}{0}
        \FALabel(17.07,10.)[l]{$N^h$}
        \FAProp(16.,13.5)(10.,10.)(0.,){ScalarDash}{-1}
        \FALabel(12.825,12.3929)[br]{$\phi_e$}
        \FAProp(16.,6.5)(10.,10.)(0.,){Straight}{-1}
        \FALabel(12.699,7.39114)[tr]{${L_L}_j^k$}
        \FAVert(4.,10.){0}
        \FAVert(16.,13.5){0}
        \FAVert(16.,6.5){0}
        \FAVert(10.,10.){0}
        
        \FADiagram{$\ii\Gamma^{N,s}_3$}
        \FAProp(0.,15.)(4.,10.)(0.,){ScalarDash}{1}
        \FALabel(1.46487,12.1679)[tr]{$\phi_d$}
        \FAProp(0.,5.)(4.,10.)(0.,){Straight}{1}
        \FALabel(2.73035,7.01172)[tl]{${L_L}_b^f$}
        \FAProp(20,15.)(16.,13.5)(0.,){Straight}{1}
        \FALabel(17.4558,15.2213)[b]{${L_L}_c^g$}
        \FAProp(20,5.)(16.,6.5)(0.,){ScalarDash}{1}
        \FALabel(17.5435,5.01277)[t]{$\phi_a$}
        \FAProp(4.,10.)(10.,10.)(0.,){Straight}{0}
        \FALabel(7.,11.07)[b]{$N^i$}
        \FAProp(16.,13.5)(16.,6.5)(0.,){Sine}{0}
        \FALabel(17.07,10.)[l]{$B,W^A$}
        \FAProp(16.,13.5)(10.,10.)(0.,){Straight}{1}
        \FALabel(12.699,12.6089)[br]{${L_L}_j^k$}
        \FAProp(16.,6.5)(10.,10.)(0.,){ScalarDash}{1}
        \FALabel(12.825,7.60709)[tr]{$\phi_e$}
        \FAVert(4.,10.){0}
        \FAVert(16.,13.5){0}
        \FAVert(16.,6.5){0}
        \FAVert(10.,10.){0}
        
    \end{feynartspicture} 
    \caption{\label{fig:RHN_s_right_ct}$s$-channel Feynman diagrams with loop corrections on the right-hand vertex.}
    \end{center}
    \end{figure}

In Figs.~\ref{fig:RHN_s_right_ct}--\ref{fig:RHN_s_mid_ct}, we display the eight $s$-channel Feynman diagrams (including also three counterterm diagrams) with loop corrections on the right-hand vertex, the left-hand vertex, and the RHN propagator, respectively. In general, note that the computations of the loop amplitudes in the full theory are too lengthy, so we have chosen not to present the analytical results of these computations in this work. However, all computations have been performed and checked, and can be presented on request.
    
\begin{figure}[h]
\vspace{-1cm}
\begin{center}
    \begin{feynartspicture}(432,168)(1,1.1)

    \FADiagram{$\ii\Gamma^{N,s}_\text{ct2}$}
    \FAProp(0.,15.)(6.,10.)(0.,){ScalarDash}{1}
    \FALabel(2.64776,11.9813)[tr]{$\phi^d$}
    \FAProp(0.,5.)(6.,10.)(0.,){Straight}{1}
    \FALabel(3.51229,6.78926)[tl]{${L_L}_b^f$}
    \FAProp(20,15.)(14.,10.)(0.,){Straight}{1}
    \FALabel(16.4877,13.2107)[br]{${L_L}_c^g$}
    \FAProp(20,5.)(14.,10.)(0.,){ScalarDash}{1}
    \FALabel(17.3522,8.01869)[bl]{$\phi_a$}
    \FAProp(6.,10.)(14.,10.)(0.,){Straight}{0}
    \FALabel(10.,8.93)[t]{$N^i$}
    \FAVert(6.,10.){2}
    \FAVert(14.,10.){0}
    
    \end{feynartspicture}
    \begin{feynartspicture}(432,168)(3,1.1)
    
    \FADiagram{$\ii\Gamma^{N,s}_4$}
    \FAProp(0.,15.)(4.,13.5)(0.,){ScalarDash}{1}
    \FALabel(2.45646,14.9872)[b]{$\phi_d$}
    \FAProp(0.,5.)(4.,6.5)(0.,){Straight}{1}
    \FALabel(2.54424,4.77869)[t]{${L_L}_b^f$}
    \FAProp(20,15.)(16.,10.)(0.,){Straight}{1}
    \FALabel(17.2697,12.9883)[br]{${L_L}_c^g$}
    \FAProp(20,5.)(16.,10.)(0.,){ScalarDash}{1}
    \FALabel(18.5351,7.8321)[bl]{$\phi_a$}
    \FAProp(16.,10.)(10.,10.)(0.,){Straight}{0}
    \FALabel(13.,8.93)[t]{$N^i$}
    \FAProp(4.,13.5)(4.,6.5)(0.,){Straight}{-1}
    \FALabel(2.93,10.)[r]{${\ell_R}^h$}
    \FAProp(4.,13.5)(10.,10.)(0.,){Straight}{1}
    \FALabel(7.301,12.6089)[bl]{${L_L}_j^k$}
    \FAProp(4.,6.5)(10.,10.)(0.,){ScalarDash}{1}
    \FALabel(7.17503,7.60709)[tl]{$\phi_e$}
    \FAVert(4.,13.5){0}
    \FAVert(4.,6.5){0}
    \FAVert(16.,10.){0}
    \FAVert(10.,10.){0}
    
    \FADiagram{$\ii\Gamma^{N,s}_5$}
    \FAProp(0.,15.)(4.,13.5)(0.,){ScalarDash}{1}
    \FALabel(2.45646,14.9872)[b]{$\phi_d$}
    \FAProp(0.,5.)(4.,6.5)(0.,){Straight}{1}
    \FALabel(2.54424,4.77869)[t]{${L_L}_b^f$}
    \FAProp(20,15.)(16.,10.)(0.,){Straight}{1}
    \FALabel(17.2697,12.9883)[br]{${L_L}_c^g$}
    \FAProp(20,5.)(16.,10.)(0.,){ScalarDash}{1}
    \FALabel(18.5351,7.8321)[bl]{$\phi_a$}
    \FAProp(16.,10.)(10.,10.)(0.,){Straight}{0}
    \FALabel(13.,8.93)[t]{$N^i$}
    \FAProp(4.,13.5)(4.,6.5)(0.,){Straight}{0}
    \FALabel(2.93,10.)[r]{$N^h$}
    \FAProp(4.,13.5)(10.,10.)(0.,){Straight}{-1}
    \FALabel(7.301,12.6089)[bl]{${L_L}_j^k$}
    \FAProp(4.,6.5)(10.,10.)(0.,){ScalarDash}{-1}
    \FALabel(7.17503,7.60709)[tl]{$\phi_e$}
    \FAVert(4.,13.5){0}
    \FAVert(4.,6.5){0}
    \FAVert(16.,10.){0}
    \FAVert(10.,10.){0}
    
    \FADiagram{$\ii\Gamma^{N,s}_6$}
    \FAProp(0.,15.)(4.,13.5)(0.,){ScalarDash}{1}
    \FALabel(2.45646,14.9872)[b]{$\phi_d$}
    \FAProp(0.,5.)(4.,6.5)(0.,){Straight}{1}
    \FALabel(2.54424,4.77869)[t]{${L_L}_b^f$}
    \FAProp(20,15.)(16.,10.)(0.,){Straight}{1}
    \FALabel(17.2697,12.9883)[br]{${L_L}_c^g$}
    \FAProp(20,5.)(16.,10.)(0.,){ScalarDash}{1}
    \FALabel(18.5351,7.8321)[bl]{$\phi_a$}
    \FAProp(16.,10.)(10.,10.)(0.,){Straight}{0}
    \FALabel(13.,8.93)[t]{$N^i$}
    \FAProp(4.,13.5)(4.,6.5)(0.,){Sine}{0}
    \FALabel(2.93,10.)[r]{$B,W^A$}
    \FAProp(4.,13.5)(10.,10.)(0.,){ScalarDash}{1}
    \FALabel(7.17503,12.3929)[bl]{$\phi_e$}
    \FAProp(4.,6.5)(10.,10.)(0.,){Straight}{1}
    \FALabel(7.301,7.39114)[tl]{${L_L}_j^k$}
    \FAVert(4.,13.5){0}
    \FAVert(4.,6.5){0}
    \FAVert(16.,10.){0}
    \FAVert(10.,10.){0}
    
    \end{feynartspicture}
    \caption{\label{fig:RHN_s_left_ct}$s$-channel Feynman diagrams with loop corrections on the left-hand vertex.}
    \end{center}
    \end{figure}
    
\begin{figure}[h]
\vspace{-1cm}
\begin{center}
    \begin{feynartspicture}(432,168)(1,1.1)
    
    \FADiagram{$\ii\Gamma^{N,s}_\text{ct3}$}
    \FAProp(0.,15.)(6.,10.)(0.,){ScalarDash}{1}
    \FALabel(2.64776,11.9813)[tr]{$\phi_d$}
    \FAProp(0.,5.)(6.,10.)(0.,){Straight}{1}
    \FALabel(3.51229,6.78926)[tl]{${L_L}_b^f$}
    \FAProp(20,15.)(14.,10.)(0.,){Straight}{1}
    \FALabel(16.4877,13.2107)[br]{${L_L}_c^g$}
    \FAProp(20,5.)(14.,10.)(0.,){ScalarDash}{1}
    \FALabel(17.3522,8.01869)[bl]{$\phi_a$}
    \FAProp(6.,10.)(14.,10.)(0.,){Straight}{0}
    \FALabel(10.,8.93)[t]{$N^i$}
    \FAVert(6.,10.){0}
    \FAVert(14.,10.){0}
    \FAVert(10.,10.){2}
    
    \end{feynartspicture}
    \begin{feynartspicture}(432,168)(2,1.1)
    
    \FADiagram{$\ii\Gamma^{N,s}_7$}
    \FAProp(0.,15.)(3.,10.)(0.,){ScalarDash}{1}
    \FALabel(0.865259,12.3112)[tr]{$\phi_d$}
    \FAProp(0.,5.)(3.,10.)(0.,){Straight}{1}
    \FALabel(2.34911,7.18253)[tl]{${L_L}_b^f$}
    \FAProp(20,15.)(17.,10.)(0.,){Straight}{1}
    \FALabel(17.6509,12.8175)[br]{${L_L}_c^g$}
    \FAProp(20,5.)(17.,10.)(0.,){ScalarDash}{1}
    \FALabel(19.1347,7.68884)[bl]{$\phi_a$}
    \FAProp(3.,10.)(7.,10.)(0.,){Straight}{0}
    \FALabel(5.,11.07)[b]{$N^i$}
    \FAProp(17.,10.)(13.,10.)(0.,){Straight}{0}
    \FALabel(15.,8.93)[t]{$N^j$}
    \FAProp(7.,10.)(13.,10.)(0.8,){Straight}{-1}
    \FALabel(10.,6.53)[t]{${L_L}_h^k$}
    \FAProp(7.,10.)(13.,10.)(-0.8,){ScalarDash}{-1}
    \FALabel(10.,13.22)[b]{$\phi_e$}
    \FAVert(3.,10.){0}
    \FAVert(17.,10.){0}
    \FAVert(7.,10.){0}
    \FAVert(13.,10.){0}
        
    \FADiagram{$\ii\Gamma^{N,s}_8$}
    \FAProp(0.,15.)(3.,10.)(0.,){ScalarDash}{1}
    \FALabel(0.865259,12.3112)[tr]{$\phi_d$}
    \FAProp(0.,5.)(3.,10.)(0.,){Straight}{1}
    \FALabel(2.34911,7.18253)[tl]{${L_L}_b^f$}
    \FAProp(20,15.)(17.,10.)(0.,){Straight}{1}
    \FALabel(17.6509,12.8175)[br]{${L_L}_c^g$}
    \FAProp(20,5.)(17.,10.)(0.,){ScalarDash}{1}
    \FALabel(19.1347,7.68884)[bl]{$\phi_a$}
    \FAProp(3.,10.)(7.,10.)(0.,){Straight}{0}
    \FALabel(5.,11.07)[b]{$N^i$}
    \FAProp(17.,10.)(13.,10.)(0.,){Straight}{0}
    \FALabel(15.,8.93)[t]{$N^j$}
    \FAProp(7.,10.)(13.,10.)(0.8,){Straight}{1}
    \FALabel(10.,6.53)[t]{${L_L}_h^k$}
    \FAProp(7.,10.)(13.,10.)(-0.8,){ScalarDash}{1}
    \FALabel(10.,13.22)[b]{$\phi_e$}
    \FAVert(3.,10.){0}
    \FAVert(17.,10.){0}
    \FAVert(7.,10.){0}
    \FAVert(13.,10.){0}
    
    \end{feynartspicture}
    \caption{\label{fig:RHN_s_mid_ct}$s$-channel Feynman diagrams with loop corrections on the RHN propagator.}
    \end{center}
    \end{figure}

Then, in Figs.~\ref{fig:RHN_t_top_ct}--\ref{fig:RHN_t_mid_ct}, we display the eight $t$-channel Feynman diagrams (including also three counterterm diagrams) with loop corrections on the top vertex, the bottom vertex, and the RHN propagator, respectively.     
\begin{figure}[h]
\vspace{-1cm}
\begin{center}
    \begin{feynartspicture}(432,168)(1,1.1)

    \FADiagram{$\ii\Gamma^{N,t}_\text{ct1}$}
    \FAProp(0.,15.)(10.,14.)(0.,){ScalarDash}{1}
    \FALabel(4.87065,13.6865)[t]{$\phi_d$}
    \FAProp(0.,5.)(10.,6.)(0.,){Straight}{1}
    \FALabel(5.15423,4.43769)[t]{${L_L}_b^f$}
    \FAProp(20,15.)(10.,14.)(0.,){Straight}{1}
    \FALabel(14.8458,15.5623)[b]{${L_L}_c^g$}
    \FAProp(20,5.)(10.,6.)(0.,){ScalarDash}{1}
    \FALabel(15.1294,6.31355)[b]{$\phi_a$}
    \FAProp(10.,14.)(10.,6.)(0.,){Straight}{0}
    \FALabel(8.93,10.)[r]{$N^i$}
    \FAVert(10.,14.){2}
    \FAVert(10.,6.){0}
    
    \end{feynartspicture}
    \begin{feynartspicture}(432,168)(3,1.1)
    
    \FADiagram{$\ii\Gamma^{N,t}_1$}
    \FAProp(0.,15.)(6.5,14.5)(0.,){ScalarDash}{1}
    \FALabel(3.34971,15.5662)[b]{$\phi_d$}
    \FAProp(0.,5.)(10.,5.5)(0.,){Straight}{1}
    \FALabel(5.0774,4.18193)[t]{${L_L}_b^f$}
    \FAProp(20,15.)(13.5,14.5)(0.,){Straight}{1}
    \FALabel(16.6311,15.8154)[b]{${L_L}_c^g$}
    \FAProp(20,5.)(10.,5.5)(0.,){ScalarDash}{1}
    \FALabel(14.9351,4.43162)[t]{$\phi_a$}
    \FAProp(10.,5.5)(10.,8.5)(0.,){Straight}{0}
    \FALabel(8.93,7.)[r]{$N^i$}
    \FAProp(6.5,14.5)(13.5,14.5)(0.,){Straight}{-1}
    \FALabel(10.,15.57)[b]{${\ell_R}^h$}
    \FAProp(6.5,14.5)(10.,8.5)(0.,){Straight}{1}
    \FALabel(7.39114,11.199)[tr]{${L_L}_j^k$}
    \FAProp(13.5,14.5)(10.,8.5)(0.,){ScalarDash}{1}
    \FALabel(12.3929,11.325)[tl]{$\phi_e$}
    \FAVert(6.5,14.5){0}
    \FAVert(10.,5.5){0}
    \FAVert(13.5,14.5){0}
    \FAVert(10.,8.5){0}
    
    \FADiagram{$\ii\Gamma^{N,t}_2$}
    \FAProp(0.,15.)(6.5,14.5)(0.,){ScalarDash}{1}
    \FALabel(3.34971,15.5662)[b]{$\phi_d$}
    \FAProp(0.,5.)(10.,5.5)(0.,){Straight}{1}
    \FALabel(5.0774,4.18193)[t]{${L_L}_b^f$}
    \FAProp(20,15.)(13.5,14.5)(0.,){Straight}{1}
    \FALabel(16.6311,15.8154)[b]{${L_L}_c^g$}
    \FAProp(20,5.)(10.,5.5)(0.,){ScalarDash}{1}
    \FALabel(14.9351,4.43162)[t]{$\phi_a$}
    \FAProp(10.,5.5)(10.,8.5)(0.,){Straight}{0}
    \FALabel(8.93,7.)[r]{$N^i$}
    \FAProp(6.5,14.5)(13.5,14.5)(0.,){Straight}{0}
    \FALabel(10.,15.57)[b]{$N^h$}
    \FAProp(6.5,14.5)(10.,8.5)(0.,){Straight}{-1}
    \FALabel(7.39114,11.199)[tr]{${L_L}_j^k$}
    \FAProp(13.5,14.5)(10.,8.5)(0.,){ScalarDash}{-1}
    \FALabel(12.3929,11.325)[tl]{$\phi_e$}
    \FAVert(6.5,14.5){0}
    \FAVert(10.,5.5){0}
    \FAVert(13.5,14.5){0}
    \FAVert(10.,8.5){0}
    
    \FADiagram{$\ii\Gamma^{N,t}_3$}
    \FAProp(0.,15.)(6.5,14.5)(0.,){ScalarDash}{1}
    \FALabel(3.34971,15.5662)[b]{$\phi_d$}
    \FAProp(0.,5.)(10.,5.5)(0.,){Straight}{1}
    \FALabel(5.0774,4.18193)[t]{${L_L}_b^f$}
    \FAProp(20,15.)(13.5,14.5)(0.,){Straight}{1}
    \FALabel(16.6311,15.8154)[b]{${L_L}_c^g$}
    \FAProp(20,5.)(10.,5.5)(0.,){ScalarDash}{1}
    \FALabel(14.9351,4.43162)[t]{$\phi_a$}
    \FAProp(10.,5.5)(10.,8.5)(0.,){Straight}{0}
    \FALabel(8.93,7.)[r]{$N^i$}
    \FAProp(6.5,14.5)(13.5,14.5)(0.,){Sine}{0}
    \FALabel(10.,15.57)[b]{$B,W^A$}
    \FAProp(6.5,14.5)(10.,8.5)(0.,){ScalarDash}{1}
    \FALabel(7.60709,11.325)[tr]{$\phi_e$}
    \FAProp(13.5,14.5)(10.,8.5)(0.,){Straight}{1}
    \FALabel(12.6089,11.199)[tl]{${L_L}_j^k$}
    \FAVert(6.5,14.5){0}
    \FAVert(10.,5.5){0}
    \FAVert(13.5,14.5){0}
    \FAVert(10.,8.5){0} 
    
    \end{feynartspicture}
    \caption{\label{fig:RHN_t_top_ct}$t$-channel Feynman diagrams with loop corrections on the top vertex.}
    \end{center}
    \end{figure}
    
\begin{figure}[h]
\vspace{-1cm}
\begin{center}
    \begin{feynartspicture}(432,168)(1,1.1)

    \FADiagram{$\ii\Gamma^{N,t}_\text{ct2}$}
    \FAProp(0.,15.)(10.,14.)(0.,){ScalarDash}{1}
    \FALabel(4.87065,13.6865)[t]{$\phi_d$}
    \FAProp(0.,5.)(10.,6.)(0.,){Straight}{1}
    \FALabel(5.15423,4.43769)[t]{${L_L}_b^f$}
    \FAProp(20,15.)(10.,14.)(0.,){Straight}{1}
    \FALabel(14.8458,15.5623)[b]{${L_L}_c^g$}
    \FAProp(20,5.)(10.,6.)(0.,){ScalarDash}{1}
    \FALabel(15.1294,6.31355)[b]{$\phi_a$}
    \FAProp(10.,14.)(10.,6.)(0.,){Straight}{0}
    \FALabel(8.93,10.)[r]{$N^i$}
    \FAVert(10.,14.){0}
    \FAVert(10.,6.){2}
    
    \end{feynartspicture}
    \begin{feynartspicture}(432,168)(3,1.1)
    
    \FADiagram{$\ii\Gamma^{N,t}_4$}
    \FAProp(0.,15.)(10.,14.5)(0.,){ScalarDash}{1}
    \FALabel(5.06492,15.5684)[b]{$\phi_d$}
    \FAProp(0.,5.)(6.5,5.5)(0.,){Straight}{1}
    \FALabel(3.36888,4.18457)[t]{${L_L}_b^f$}
    \FAProp(20,15.)(10.,14.5)(0.,){Straight}{1}
    \FALabel(14.9226,15.8181)[b]{${L_L}_c^g$}
    \FAProp(20,5.)(13.5,5.5)(0.,){ScalarDash}{1}
    \FALabel(16.6503,4.43383)[t]{$\phi_a$}
    \FAProp(10.,14.5)(10.,11.)(0.,){Straight}{0}
    \FALabel(11.07,12.75)[l]{$N^i$}
    \FAProp(6.5,5.5)(13.5,5.5)(0.,){Straight}{1}
    \FALabel(10.,4.43)[t]{${\ell_R}^h$}
    \FAProp(6.5,5.5)(10.,11.)(0.,){ScalarDash}{1}
    \FALabel(7.63324,8.46794)[br]{$\phi_e$}
    \FAProp(13.5,5.5)(10.,11.)(0.,){Straight}{1}
    \FALabel(12.5777,8.60216)[bl]{${L_L}_j^k$}
    \FAVert(10.,14.5){0}
    \FAVert(6.5,5.5){0}
    \FAVert(13.5,5.5){0}
    \FAVert(10.,11.){0}
    
    \FADiagram{$\ii\Gamma^{N,t}_5$}
    \FAProp(0.,15.)(10.,14.5)(0.,){ScalarDash}{1}
    \FALabel(5.06492,15.5684)[b]{$\phi_d$}
    \FAProp(0.,5.)(6.5,5.5)(0.,){Straight}{1}
    \FALabel(3.36888,4.18457)[t]{${L_L}_b^f$}
    \FAProp(20,15.)(10.,14.5)(0.,){Straight}{1}
    \FALabel(14.9226,15.8181)[b]{${L_L}_c^g$}
    \FAProp(20,5.)(13.5,5.5)(0.,){ScalarDash}{1}
    \FALabel(16.6503,4.43383)[t]{$\phi_a$}
    \FAProp(10.,14.5)(10.,11.)(0.,){Straight}{0}
    \FALabel(11.07,12.75)[l]{$N^i$}
    \FAProp(6.5,5.5)(13.5,5.5)(0.,){Straight}{0}
    \FALabel(10.,4.43)[t]{$N^h$}
    \FAProp(6.5,5.5)(10.,11.)(0.,){ScalarDash}{-1}
    \FALabel(7.63324,8.46794)[br]{$\phi_e$}
    \FAProp(13.5,5.5)(10.,11.)(0.,){Straight}{-1}
    \FALabel(12.5777,8.60216)[bl]{${L_L}_j^k$}
    \FAVert(10.,14.5){0}
    \FAVert(6.5,5.5){0}
    \FAVert(13.5,5.5){0}
    \FAVert(10.,11.){0}
    
    \FADiagram{$\ii\Gamma^{N,t}_6$}
    \FAProp(0.,15.)(10.,14.5)(0.,){ScalarDash}{1}
    \FALabel(5.06492,15.5684)[b]{$\phi_d$}
    \FAProp(0.,5.)(6.5,5.5)(0.,){Straight}{1}
    \FALabel(3.36888,4.18457)[t]{${L_L}_b^f$}
    \FAProp(20,15.)(10.,14.5)(0.,){Straight}{1}
    \FALabel(14.9226,15.8181)[b]{${L_L}_c^g$}
    \FAProp(20,5.)(13.5,5.5)(0.,){ScalarDash}{1}
    \FALabel(16.6503,4.43383)[t]{$\phi_a$}
    \FAProp(10.,14.5)(10.,11.)(0.,){Straight}{0}
    \FALabel(11.07,12.75)[l]{$N^i$}
    \FAProp(6.5,5.5)(13.5,5.5)(0.,){Sine}{0}
    \FALabel(10.,4.43)[t]{$B,W^A$}
    \FAProp(6.5,5.5)(10.,11.)(0.,){Straight}{1}
    \FALabel(7.42232,8.60216)[br]{${L_L}_j^k$}
    \FAProp(13.5,5.5)(10.,11.)(0.,){ScalarDash}{1}
    \FALabel(12.3668,8.46794)[bl]{$\phi_e$}
    \FAVert(10.,14.5){0}
    \FAVert(6.5,5.5){0}
    \FAVert(13.5,5.5){0}
    \FAVert(10.,11.){0}
    
    \end{feynartspicture}
    \caption{\label{fig:RHN_t_bottom_ct}$t$-channel Feynman diagrams with loop corrections on the bottom vertex.}
    \end{center}
    \end{figure}
    
\begin{figure}[h]
\vspace{-1cm}
\begin{center}
    \begin{feynartspicture}(432,168)(1,1.1)

    \FADiagram{$\ii\Gamma^{N,t}_\text{ct3}$}
    \FAProp(0.,15.)(10.,14.)(0.,){ScalarDash}{1}
    \FALabel(4.87065,13.6865)[t]{$\phi_d$}
    \FAProp(0.,5.)(10.,6.)(0.,){Straight}{1}
    \FALabel(5.15423,4.43769)[t]{${L_L}_b^f$}
    \FAProp(20,15.)(10.,14.)(0.,){Straight}{1}
    \FALabel(14.8458,15.5623)[b]{${L_L}_c^g$}
    \FAProp(20,5.)(10.,6.)(0.,){ScalarDash}{1}
    \FALabel(15.1294,6.31355)[b]{$\phi_a$}
    \FAProp(10.,14.)(10.,6.)(0.,){Straight}{0}
    \FALabel(8.93,10.)[r]{$N^i$}
    \FAVert(10.,14.){0}
    \FAVert(10.,6.){0}
    \FAVert(10.,10.){2}
    
    \end{feynartspicture}
    \begin{feynartspicture}(432,168)(2,1.1)
    
    \FADiagram{$\ii\Gamma^{N,t}_7$}
    \FAProp(0.,15.)(10.,14.5)(0.,){ScalarDash}{1}
    \FALabel(5.06492,15.5684)[b]{$\phi_d$}
    \FAProp(0.,5.)(10.,5.5)(0.,){Straight}{1}
    \FALabel(5.0774,4.18193)[t]{${L_L}_b^f$}
    \FAProp(20,15.)(10.,14.5)(0.,){Straight}{1}
    \FALabel(14.9226,15.8181)[b]{${L_L}_c^g$}
    \FAProp(20,5.)(10.,5.5)(0.,){ScalarDash}{1}
    \FALabel(14.9351,4.43162)[t]{$\phi_a$}
    \FAProp(10.,14.5)(10.,12.)(0.,){Straight}{0}
    \FALabel(11.07,13.25)[l]{$N^i$}
    \FAProp(10.,5.5)(10.,8.)(0.,){Straight}{0}
    \FALabel(8.93,6.75)[r]{$N^j$}
    \FAProp(10.,12.)(10.,8.)(1.,){Straight}{-1}
    \FALabel(6.93,10.)[r]{${L_L}_h^k$}
    \FAProp(10.,12.)(10.,8.)(-1.,){ScalarDash}{-1}
    \FALabel(12.82,10.)[l]{$\phi_e$}
    \FAVert(10.,14.5){0}
    \FAVert(10.,5.5){0}
    \FAVert(10.,12.){0}
    \FAVert(10.,8.){0}
    
    \FADiagram{$\ii\Gamma^{N,t}_8$}
    \FAProp(0.,15.)(10.,14.5)(0.,){ScalarDash}{1}
    \FALabel(5.06492,15.5684)[b]{$\phi_d$}
    \FAProp(0.,5.)(10.,5.5)(0.,){Straight}{1}
    \FALabel(5.0774,4.18193)[t]{${L_L}_b^f$}
    \FAProp(20,15.)(10.,14.5)(0.,){Straight}{1}
    \FALabel(14.9226,15.8181)[b]{${L_L}_c^g$}
    \FAProp(20,5.)(10.,5.5)(0.,){ScalarDash}{1}
    \FALabel(14.9351,4.43162)[t]{$\phi_a$}
    \FAProp(10.,14.5)(10.,12.)(0.,){Straight}{0}
    \FALabel(11.07,13.25)[l]{$N^i$}
    \FAProp(10.,5.5)(10.,8.)(0.,){Straight}{0}
    \FALabel(8.93,6.75)[r]{$N^j$}
    \FAProp(10.,12.)(10.,8.)(1.,){Straight}{1}
    \FALabel(6.93,10.)[r]{${L_L}_h^k$}
    \FAProp(10.,12.)(10.,8.)(-1.,){ScalarDash}{1}
    \FALabel(12.82,10.)[l]{$\phi_e$}
    \FAVert(10.,14.5){0}
    \FAVert(10.,5.5){0}
    \FAVert(10.,12.){0}
    \FAVert(10.,8.){0}
        
    \end{feynartspicture}
    \caption{\label{fig:RHN_t_mid_ct}$t$-channel Feynman diagrams with loop corrections on the RHN propagator.}
    \end{center}
    \end{figure}

\begin{figure}[h]
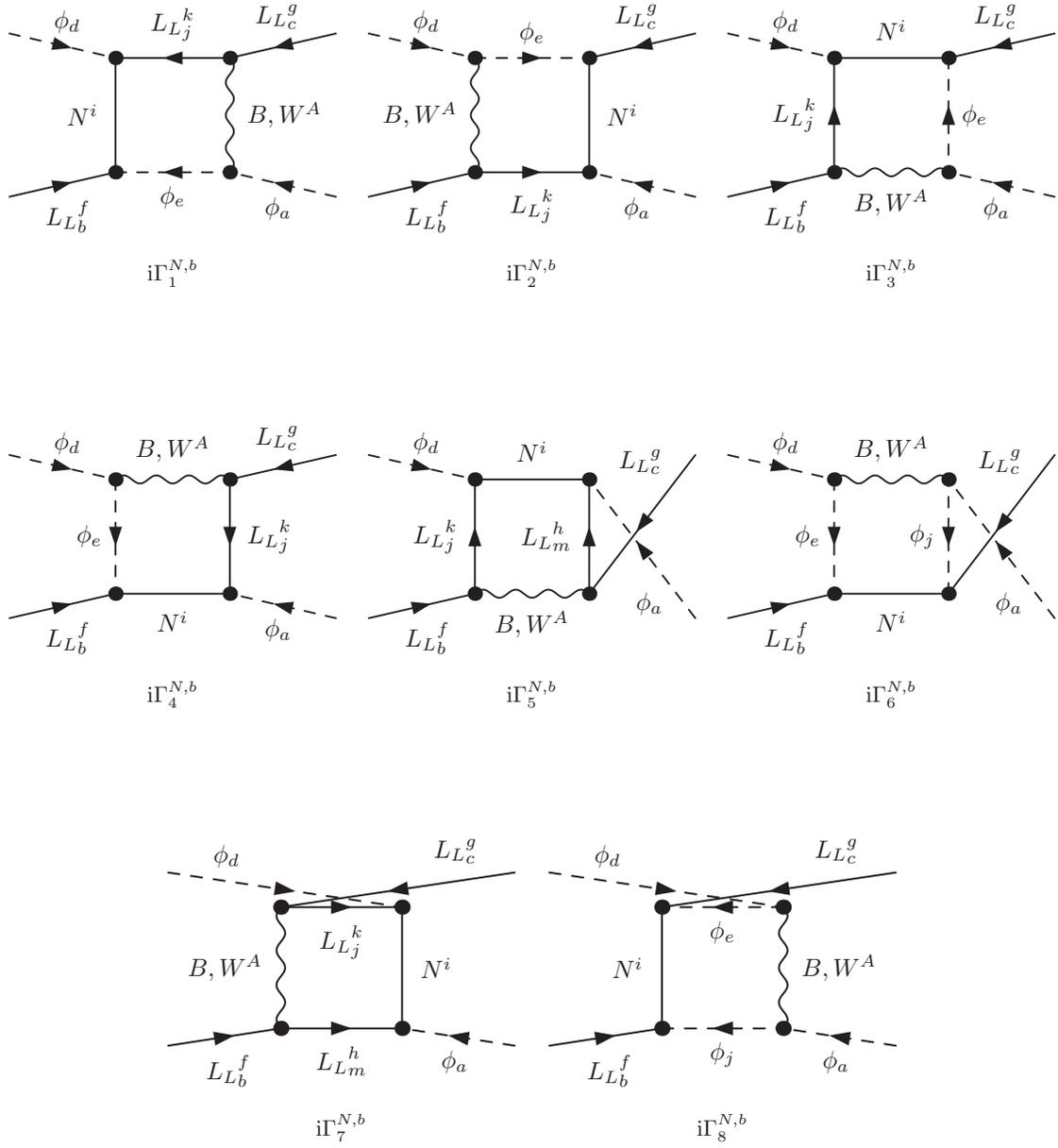

\vspace{-1cm}
\begin{center}
    \begin{feynartspicture}(432,168)(3,1.1)
    
    \FADiagram{$\ii\Gamma^{N,b}_1$}
    \FAProp(0.,15.)(6.5,13.5)(0.,){ScalarDash}{1}
    \FALabel(3.54232,15.0367)[b]{$\phi_d$}
    \FAProp(0.,5.)(6.5,6.5)(0.,){Straight}{1}
    \FALabel(3.59853,4.71969)[t]{${L_L}_b^f$}
    \FAProp(20,15.)(13.5,13.5)(0.,){Straight}{1}
    \FALabel(16.4015,15.2803)[b]{${L_L}_c^g$}
    \FAProp(20,5.)(13.5,6.5)(0.,){ScalarDash}{1}
    \FALabel(16.4577,4.96329)[t]{$\phi_a$}
    \FAProp(6.5,13.5)(6.5,6.5)(0.,){Straight}{0}
    \FALabel(5.43,10.)[r]{$N^i$}
    \FAProp(6.5,13.5)(13.5,13.5)(0.,){Straight}{-1}
    \FALabel(10.,14.57)[b]{${L_L}_j^k$}
    \FAProp(6.5,6.5)(13.5,6.5)(0.,){ScalarDash}{-1}
    \FALabel(10.,5.68)[t]{$\phi_e$}
    \FAProp(13.5,13.5)(13.5,6.5)(0.,){Sine}{0}
    \FALabel(14.57,10.)[l]{$B,W^A$}
    \FAVert(6.5,13.5){0}
    \FAVert(6.5,6.5){0}
    \FAVert(13.5,13.5){0}
    \FAVert(13.5,6.5){0}
    
    \FADiagram{$\ii\Gamma^{N,b}_2$}
    \FAProp(0.,15.)(6.5,13.5)(0.,){ScalarDash}{1}
    \FALabel(3.54232,15.0367)[b]{$\phi_d$}
    \FAProp(0.,5.)(6.5,6.5)(0.,){Straight}{1}
    \FALabel(3.59853,4.71969)[t]{${L_L}_b^f$}
    \FAProp(20,15.)(13.5,13.5)(0.,){Straight}{1}
    \FALabel(16.4015,15.2803)[b]{${L_L}_c^g$}
    \FAProp(20,5.)(13.5,6.5)(0.,){ScalarDash}{1}
    \FALabel(16.4577,4.96329)[t]{$\phi_a$}
    \FAProp(6.5,13.5)(6.5,6.5)(0.,){Sine}{0}
    \FALabel(5.43,10.)[r]{$B,W^A$}
    \FAProp(6.5,13.5)(13.5,13.5)(0.,){ScalarDash}{1}
    \FALabel(10.,14.32)[b]{$\phi_e$}
    \FAProp(6.5,6.5)(13.5,6.5)(0.,){Straight}{1}
    \FALabel(10.,5.43)[t]{${L_L}_j^k$}
    \FAProp(13.5,13.5)(13.5,6.5)(0.,){Straight}{0}
    \FALabel(14.57,10.)[l]{$N^i$}
    \FAVert(6.5,13.5){0}
    \FAVert(6.5,6.5){0}
    \FAVert(13.5,13.5){0}
    \FAVert(13.5,6.5){0}
    
    \FADiagram{$\ii\Gamma^{N,b}_3$}
    \FAProp(0.,15.)(6.5,13.5)(0.,){ScalarDash}{1}
    \FALabel(3.54232,15.0367)[b]{$\phi_d$}
    \FAProp(0.,5.)(6.5,6.5)(0.,){Straight}{1}
    \FALabel(3.59853,4.71969)[t]{${L_L}_b^f$}
    \FAProp(20,15.)(13.5,13.5)(0.,){Straight}{1}
    \FALabel(16.4015,15.2803)[b]{${L_L}_c^g$}
    \FAProp(20,5.)(13.5,6.5)(0.,){ScalarDash}{1}
    \FALabel(16.4577,4.96329)[t]{$\phi_a$}
    \FAProp(6.5,13.5)(6.5,6.5)(0.,){Straight}{-1}
    \FALabel(5.43,10.)[r]{${L_L}_j^k$}
    \FAProp(6.5,13.5)(13.5,13.5)(0.,){Straight}{0}
    \FALabel(10.,14.57)[b]{$N^i$}
    \FAProp(6.5,6.5)(13.5,6.5)(0.,){Sine}{0}
    \FALabel(10.,5.43)[t]{$B,W^A$}
    \FAProp(13.5,13.5)(13.5,6.5)(0.,){ScalarDash}{-1}
    \FALabel(14.32,10.)[l]{$\phi_e$}
    \FAVert(6.5,13.5){0}
    \FAVert(6.5,6.5){0}
    \FAVert(13.5,13.5){0}
    \FAVert(13.5,6.5){0}
    
    \end{feynartspicture}
    \begin{feynartspicture}(432,168)(3,1.1)
    
    \FADiagram{$\ii\Gamma^{N,b}_4$}
    \FAProp(0.,15.)(6.5,13.5)(0.,){ScalarDash}{1}
    \FALabel(3.54232,15.0367)[b]{$\phi_d$}
    \FAProp(0.,5.)(6.5,6.5)(0.,){Straight}{1}
    \FALabel(3.59853,4.71969)[t]{${L_L}_b^f$}
    \FAProp(20,15.)(13.5,13.5)(0.,){Straight}{1}
    \FALabel(16.4015,15.2803)[b]{${L_L}_c^g$}
    \FAProp(20,5.)(13.5,6.5)(0.,){ScalarDash}{1}
    \FALabel(16.4577,4.96329)[t]{$\phi_a$}
    \FAProp(6.5,13.5)(6.5,6.5)(0.,){ScalarDash}{1}
    \FALabel(5.68,10.)[r]{$\phi_e$}
    \FAProp(6.5,13.5)(13.5,13.5)(0.,){Sine}{0}
    \FALabel(10.,14.57)[b]{$B,W^A$}
    \FAProp(6.5,6.5)(13.5,6.5)(0.,){Straight}{0}
    \FALabel(10.,5.43)[t]{$N^i$}
    \FAProp(13.5,13.5)(13.5,6.5)(0.,){Straight}{1}
    \FALabel(14.57,10.)[l]{${L_L}_j^k$}
    \FAVert(6.5,13.5){0}
    \FAVert(6.5,6.5){0}
    \FAVert(13.5,13.5){0}
    \FAVert(13.5,6.5){0}
    
    \FADiagram{$\ii\Gamma^{N,b}_5$}
    \FAProp(0.,15.)(6.5,13.5)(0.,){ScalarDash}{1}
    \FALabel(3.54232,15.0367)[b]{$\phi_d$}
    \FAProp(0.,5.)(6.5,6.5)(0.,){Straight}{1}
    \FALabel(3.59853,4.71969)[t]{${L_L}_b^f$}
    \FAProp(20,15.)(13.5,6.5)(0.,){Straight}{1}
    \FALabel(17.9806,13.8219)[br]{${L_L}_c^g$}
    \FAProp(20,5.)(13.5,13.5)(0.,){ScalarDash}{1}
    \FALabel(18.0298,6.57995)[tr]{$\phi_a$}
    \FAProp(6.5,13.5)(6.5,6.5)(0.,){Straight}{-1}
    \FALabel(5.43,10.)[r]{${L_L}_j^k$}
    \FAProp(6.5,13.5)(13.5,13.5)(0.,){Straight}{0}
    \FALabel(10.,14.57)[b]{$N^i$}
    \FAProp(6.5,6.5)(13.5,6.5)(0.,){Sine}{0}
    \FALabel(10.,5.43)[t]{$B,W^A$}
    \FAProp(13.5,6.5)(13.5,13.5)(0.,){Straight}{1}
    \FALabel(12.43,10.)[r]{${L_L}_m^h$}
    \FAVert(6.5,13.5){0}
    \FAVert(6.5,6.5){0}
    \FAVert(13.5,6.5){0}
    \FAVert(13.5,13.5){0}
    
    \FADiagram{$\ii\Gamma^{N,b}_6$}
    \FAProp(0.,15.)(6.5,13.5)(0.,){ScalarDash}{1}
    \FALabel(3.54232,15.0367)[b]{$\phi_d$}
    \FAProp(0.,5.)(6.5,6.5)(0.,){Straight}{1}
    \FALabel(3.59853,4.71969)[t]{${L_L}_b^f$}
    \FAProp(20,15.)(13.5,6.5)(0.,){Straight}{1}
    \FALabel(17.9806,13.8219)[br]{${L_L}_c^g$}
    \FAProp(20,5.)(13.5,13.5)(0.,){ScalarDash}{1}
    \FALabel(18.0298,6.57995)[tr]{$\phi_a$}
    \FAProp(6.5,13.5)(6.5,6.5)(0.,){ScalarDash}{1}
    \FALabel(5.68,10.)[r]{$\phi_e$}
    \FAProp(6.5,13.5)(13.5,13.5)(0.,){Sine}{0}
    \FALabel(10.,14.57)[b]{$B,W^A$}
    \FAProp(6.5,6.5)(13.5,6.5)(0.,){Straight}{0}
    \FALabel(10.,5.43)[t]{$N^i$}
    \FAProp(13.5,6.5)(13.5,13.5)(0.,){ScalarDash}{-1}
    \FALabel(12.68,10.)[r]{$\phi_j$}
    \FAVert(6.5,13.5){0}
    \FAVert(6.5,6.5){0}
    \FAVert(13.5,6.5){0}
    \FAVert(13.5,13.5){0}
    
    \end{feynartspicture}
    \begin{feynartspicture}(432,168)(2,1.1)
    
    \FADiagram{$\ii\Gamma^{N,b}_7$}
    \FAProp(0.,15.)(13.5,13.)(0.,){ScalarDash}{1}
    \FALabel(3.37315,15.28)[b]{$\phi_d$}
    \FAProp(0.,5.)(6.5,6.)(0.,){Straight}{1}
    \FALabel(3.48569,4.44802)[t]{${L_L}_b^f$}
    \FAProp(20,15.)(6.5,13.)(0.,){Straight}{1}
    \FALabel(16.5898,15.5281)[b]{${L_L}_c^g$}
    \FAProp(20,5.)(13.5,6.)(0.,){ScalarDash}{1}
    \FALabel(16.5523,4.69512)[t]{$\phi_a$}
    \FAProp(13.5,13.)(6.5,13.)(0.,){Straight}{-1}
    \FALabel(10.,11.93)[t]{${L_L}_j^k$}
    \FAProp(13.5,13.)(13.5,6.)(0.,){Straight}{0}
    \FALabel(14.57,9.5)[l]{$N^i$}
    \FAProp(6.5,6.)(6.5,13.)(0.,){Sine}{0}
    \FALabel(5.43,9.5)[r]{$B,W^A$}
    \FAProp(6.5,6.)(13.5,6.)(0.,){Straight}{1}
    \FALabel(10.,4.93)[t]{${L_L}_m^h$}
    \FAVert(13.5,13.){0}
    \FAVert(6.5,6.){0}
    \FAVert(6.5,13.){0}
    \FAVert(13.5,6.){0}
    
    \FADiagram{$\ii\Gamma^{N,b}_8$}
    \FAProp(0.,15.)(13.5,13.)(0.,){ScalarDash}{1}
    \FALabel(3.37315,15.28)[b]{$\phi_d$}
    \FAProp(0.,5.)(6.5,6.)(0.,){Straight}{1}
    \FALabel(3.48569,4.44802)[t]{${L_L}_b^f$}
    \FAProp(20,15.)(6.5,13.)(0.,){Straight}{1}
    \FALabel(16.5898,15.5281)[b]{${L_L}_c^g$}
    \FAProp(20,5.)(13.5,6.)(0.,){ScalarDash}{1}
    \FALabel(16.5523,4.69512)[t]{$\phi_a$}
    \FAProp(13.5,13.)(6.5,13.)(0.,){ScalarDash}{1}
    \FALabel(10.,12.18)[t]{$\phi_e$}
    \FAProp(13.5,13.)(13.5,6.)(0.,){Sine}{0}
    \FALabel(14.57,9.5)[l]{$B,W^A$}
    \FAProp(6.5,6.)(6.5,13.)(0.,){Straight}{0}
    \FALabel(5.43,9.5)[r]{$N^i$}
    \FAProp(6.5,6.)(13.5,6.)(0.,){ScalarDash}{-1}
    \FALabel(10.,5.18)[t]{$\phi_j$}
    \FAVert(13.5,13.){0}
    \FAVert(6.5,6.){0}
    \FAVert(6.5,13.){0}
    \FAVert(13.5,6.){0}
    
    \end{feynartspicture}
    \caption{\label{fig:RHN_finite_gauge}Finite Feynman diagrams involving gauge bosons.}
    \end{center}
    \end{figure}

\begin{figure}[h]
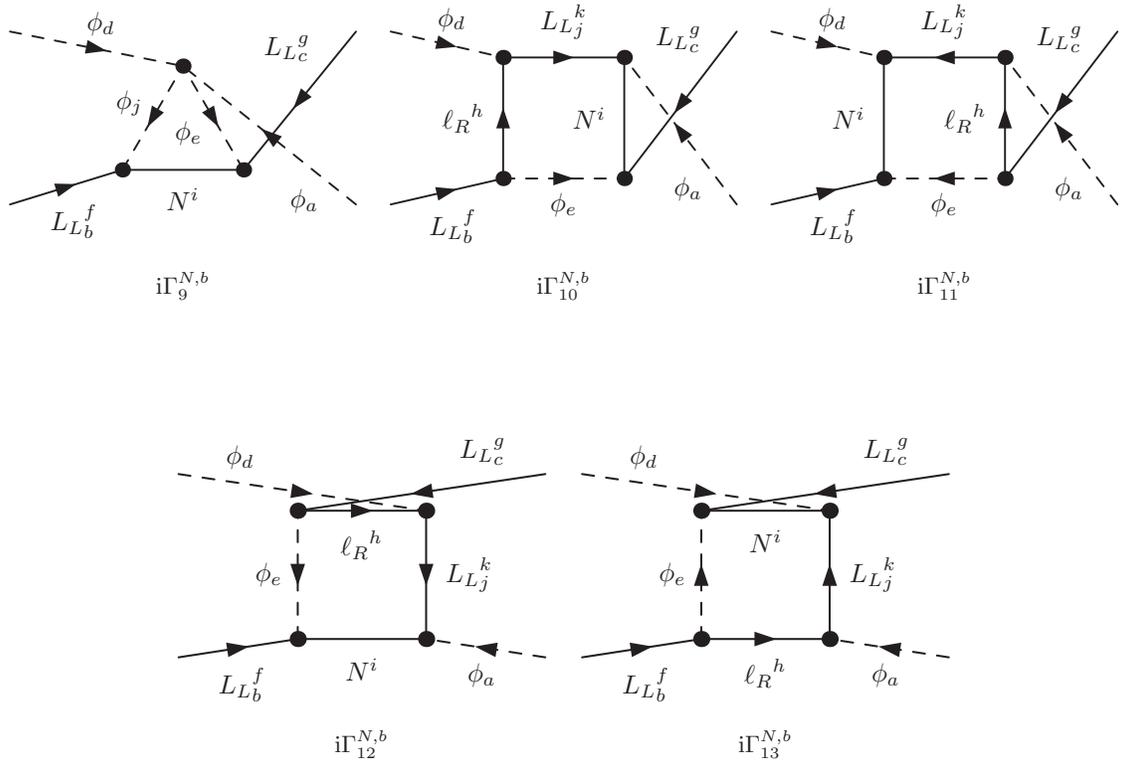

\vspace{-1cm}
\begin{center}
    \begin{feynartspicture}(432,168)(3,1.1)
    
    \FADiagram{$\ii\Gamma^{N,b}_9$}
    \FAProp(0.,15.)(10.,13.)(0.,){ScalarDash}{1}
    \FALabel(5.25495,14.7948)[b]{$\phi_d$}
    \FAProp(0.,5.)(6.5,7.)(0.,){Straight}{1}
    \FALabel(3.70583,4.99854)[t]{${L_L}_b^f$}
    \FAProp(20,15.)(13.5,7.)(0.,){Straight}{1}
    \FALabel(17.3355,13.2775)[br]{${L_L}_c^g$}
    \FAProp(20,5.)(10.,13.)(0.,){ScalarDash}{1}
    \FALabel(17.8282,5.87899)[tr]{$\phi_a$}
    \FAProp(6.5,7.)(13.5,7.)(0.,){Straight}{0}
    \FALabel(10.,5.93)[t]{$N^i$}
    \FAProp(6.5,7.)(10.,13.)(0.,){ScalarDash}{-1}
    \FALabel(7.69355,10.1246)[br]{$\phi_j$}
    \FAProp(13.5,7.)(10.,13.)(0.,){ScalarDash}{-1}
    \FALabel(11.1071,9.82497)[tr]{$\phi_e$}
    \FAVert(6.5,7.){0}
    \FAVert(13.5,7.){0}
    \FAVert(10.,13.){0}
    
    \FADiagram{$\ii\Gamma^{N,b}_{10}$}
    \FAProp(0.,15.)(6.5,13.5)(0.,){ScalarDash}{1}
    \FALabel(3.54232,15.0367)[b]{$\phi_d$}
    \FAProp(0.,5.)(6.5,6.5)(0.,){Straight}{1}
    \FALabel(3.59853,4.71969)[t]{${L_L}_b^f$}
    \FAProp(20,15.)(13.5,6.5)(0.,){Straight}{1}
    \FALabel(17.9806,13.8219)[br]{${L_L}_c^g$}
    \FAProp(20,5.)(13.5,13.5)(0.,){ScalarDash}{1}
    \FALabel(18.0298,6.57995)[tr]{$\phi_a$}
    \FAProp(6.5,13.5)(6.5,6.5)(0.,){Straight}{-1}
    \FALabel(5.43,10.)[r]{${\ell_R}^h$}
    \FAProp(6.5,13.5)(13.5,13.5)(0.,){Straight}{1}
    \FALabel(10.,14.57)[b]{${L_L}_j^k$}
    \FAProp(6.5,6.5)(13.5,6.5)(0.,){ScalarDash}{1}
    \FALabel(10.,5.68)[t]{$\phi_e$}
    \FAProp(13.5,6.5)(13.5,13.5)(0.,){Straight}{0}
    \FALabel(12.43,10.)[r]{$N^i$}
    \FAVert(6.5,13.5){0}
    \FAVert(6.5,6.5){0}
    \FAVert(13.5,6.5){0}
    \FAVert(13.5,13.5){0}
    
    \FADiagram{$\ii\Gamma^{N,b}_{11}$}
    \FAProp(0.,15.)(6.5,13.5)(0.,){ScalarDash}{1}
    \FALabel(3.54232,15.0367)[b]{$\phi_d$}
    \FAProp(0.,5.)(6.5,6.5)(0.,){Straight}{1}
    \FALabel(3.59853,4.71969)[t]{${L_L}_b^f$}
    \FAProp(20,15.)(13.5,6.5)(0.,){Straight}{1}
    \FALabel(17.9806,13.8219)[br]{${L_L}_c^g$}
    \FAProp(20,5.)(13.5,13.5)(0.,){ScalarDash}{1}
    \FALabel(18.0298,6.57995)[tr]{$\phi_a$}
    \FAProp(6.5,13.5)(6.5,6.5)(0.,){Straight}{0}
    \FALabel(5.43,10.)[r]{$N^i$}
    \FAProp(6.5,13.5)(13.5,13.5)(0.,){Straight}{-1}
    \FALabel(10.,14.57)[b]{${L_L}_j^k$}
    \FAProp(6.5,6.5)(13.5,6.5)(0.,){ScalarDash}{-1}
    \FALabel(10.,5.68)[t]{$\phi_e$}
    \FAProp(13.5,6.5)(13.5,13.5)(0.,){Straight}{1}
    \FALabel(12.43,10.)[r]{${\ell_R}^h$}
    \FAVert(6.5,13.5){0}
    \FAVert(6.5,6.5){0}
    \FAVert(13.5,6.5){0}
    \FAVert(13.5,13.5){0}
    
    \end{feynartspicture}
    \begin{feynartspicture}(432,168)(2,1.1)
    
    \FADiagram{$\ii\Gamma^{N,b}_{12}$}
    \FAProp(0.,15.)(13.5,13.)(0.,){ScalarDash}{1}
    \FALabel(3.37315,15.28)[b]{$\phi_d$}
    \FAProp(0.,5.)(6.5,6.)(0.,){Straight}{1}
    \FALabel(3.48569,4.44802)[t]{${L_L}_b^f$}
    \FAProp(20,15.)(6.5,13.)(0.,){Straight}{1}
    \FALabel(16.5898,15.5281)[b]{${L_L}_c^g$}
    \FAProp(20,5.)(13.5,6.)(0.,){ScalarDash}{1}
    \FALabel(16.5523,4.69512)[t]{$\phi_a$}
    \FAProp(13.5,13.)(6.5,13.)(0.,){Straight}{-1}
    \FALabel(10.,11.93)[t]{${\ell_R}^h$}
    \FAProp(13.5,13.)(13.5,6.)(0.,){Straight}{1}
    \FALabel(14.57,9.5)[l]{${L_L}_j^k$}
    \FAProp(6.5,6.)(6.5,13.)(0.,){ScalarDash}{-1}
    \FALabel(5.68,9.5)[r]{$\phi_e$}
    \FAProp(6.5,6.)(13.5,6.)(0.,){Straight}{0}
    \FALabel(10.,4.93)[t]{$N^i$}
    \FAVert(13.5,13.){0}
    \FAVert(6.5,6.){0}
    \FAVert(6.5,13.){0}
    \FAVert(13.5,6.){0}
    
    \FADiagram{$\ii\Gamma^{N,b}_{13}$}
    \FAProp(0.,15.)(13.5,13.)(0.,){ScalarDash}{1}
    \FALabel(3.37315,15.28)[b]{$\phi_d$}
    \FAProp(0.,5.)(6.5,6.)(0.,){Straight}{1}
    \FALabel(3.48569,4.44802)[t]{${L_L}_b^f$}
    \FAProp(20,15.)(6.5,13.)(0.,){Straight}{1}
    \FALabel(16.5898,15.5281)[b]{${L_L}_c^g$}
    \FAProp(20,5.)(13.5,6.)(0.,){ScalarDash}{1}
    \FALabel(16.5523,4.69512)[t]{$\phi_a$}
    \FAProp(13.5,13.)(6.5,13.)(0.,){Straight}{0}
    \FALabel(10.,11.93)[t]{$N^i$}
    \FAProp(13.5,13.)(13.5,6.)(0.,){Straight}{-1}
    \FALabel(14.57,9.5)[l]{${L_L}_j^k$}
    \FAProp(6.5,6.)(6.5,13.)(0.,){ScalarDash}{1}
    \FALabel(5.68,9.5)[r]{$\phi_e$}
    \FAProp(6.5,6.)(13.5,6.)(0.,){Straight}{1}
    \FALabel(10.,4.93)[t]{${\ell_R}^h$}
    \FAVert(13.5,13.){0}
    \FAVert(6.5,6.){0}
    \FAVert(6.5,13.){0}
    \FAVert(13.5,6.){0}
    
    \end{feynartspicture}
    \caption{\label{fig:RHN_finite_other}Finite Feynman diagrams without gauge bosons.}
    \end{center}
    \end{figure}

Next, in Figs.~\ref{fig:RHN_finite_gauge}--\ref{fig:RHN_finite_other}, we display the 13 finite Feynman diagrams including box diagrams with and without gauge bosons, respectively. 
    
\FloatBarrier

\subsubsection{Effective Theory}\label{app:kappa_loops_EFT}

In Fig.~\ref{fig:EFT_divergent}, we display the eleven divergent Feynman diagrams (including also one counterterm diagram) in the effective theory. Using the corresponding Feynman rules, we compute the loop amplitudes of these eleven contributions to the neutrino mass matrix, which are given in Eqs.~\eqref{eq:Gk1}--\eqref{eq:Gk11}.
\begin{figure}[h]
\vspace{-2cm}
\begin{center}
    \begin{feynartspicture}(432,168)(3,1.1)
        
        \FADiagram{$\ii\Gamma^\kappa_\text{ct}$}
        \FAProp(0.,15.)(10.,10.)(0.,){ScalarDash}{1}
        \FALabel(4.89862,11.8172)[tr]{$\phi_d$}
        \FAProp(0.,5.)(10.,10.)(0.,){Straight}{1}
        \FALabel(5.21318,6.59364)[tl]{${L_L}_b^f$}
        \FAProp(20,15.)(10.,10.)(0.,){Straight}{1}
        \FALabel(14.7868,13.4064)[br]{${L_L}_c^g$}
        \FAProp(20,5.)(10.,10.)(0.,){ScalarDash}{1}
        \FALabel(15.1014,8.18276)[bl]{$\phi_a$}
        \FAVert(10.,10.){2}
        
        \FADiagram{$\ii\Gamma^\kappa_1$}
        \FAProp(0.,15.)(6.,13.5)(0.,){ScalarDash}{1}
        \FALabel(3.3153,15.0312)[b]{$\phi_d$}
        \FAProp(0.,5.)(6.,6.5)(0.,){Straight}{1}
        \FALabel(3.37593,4.72628)[t]{${L_L}_b^f$}
        \FAProp(20,15.)(12.,10.)(0.,){Straight}{1}
        \FALabel(15.6585,13.3344)[br]{${L_L}_c^g$}
        \FAProp(20,5.)(12.,10.)(0.,){ScalarDash}{1}
        \FALabel(16.209,8.1224)[bl]{$\phi_a$}
        \FAProp(6.,13.5)(6.,6.5)(0.,){Straight}{-1}
        \FALabel(4.93,10.)[r]{${\ell_R}^h$}
        \FAProp(6.,13.5)(12.,10.)(0.,){Straight}{1}
        \FALabel(9.301,12.6089)[bl]{${L_L}_j^k$}
        \FAProp(6.,6.5)(12.,10.)(0.,){ScalarDash}{1}
        \FALabel(9.17503,7.60709)[tl]{$\phi_e$}
        \FAVert(6.,13.5){0}
        \FAVert(6.,6.5){0}
        \FAVert(12.,10.){0}
        
        \FADiagram{$\ii\Gamma^\kappa_2$}
        \FAProp(0.,15.)(8.,10.)(0.,){ScalarDash}{1}
        \FALabel(3.791,11.8776)[tr]{$\phi_d$}
        \FAProp(0.,5.)(8.,10.)(0.,){Straight}{1}
        \FALabel(4.3415,6.6656)[tl]{${L_L}_b^f$}
        \FAProp(20,15.)(14.,13.5)(0.,){Straight}{1}
        \FALabel(16.6241,15.2737)[b]{${L_L}_c^g$}
        \FAProp(20,5.)(14.,6.5)(0.,){ScalarDash}{1}
        \FALabel(16.6847,4.96881)[t]{$\phi_a$}
        \FAProp(14.,13.5)(14.,6.5)(0.,){Straight}{1}
        \FALabel(15.07,10.)[l]{${\ell_R}^h$}
        \FAProp(14.,13.5)(8.,10.)(0.,){ScalarDash}{1}
        \FALabel(10.825,12.3929)[br]{$\phi_e$}
        \FAProp(14.,6.5)(8.,10.)(0.,){Straight}{1}
        \FALabel(10.699,7.39114)[tr]{${L_L}_j^k$}
        \FAVert(14.,13.5){0}
        \FAVert(14.,6.5){0}
        \FAVert(8.,10.){0}
    
    \end{feynartspicture}
    \begin{feynartspicture}(432,168)(3,1.1)
    
        \FADiagram{$\ii\Gamma^\kappa_3$}
        \FAProp(5.,17.5)(10.,13.)(0.,){ScalarDash}{1}
        \FALabel(3.88965,15.7363)[bl]{$\phi_d$}
        \FAProp(0.,7.)(6.5,7.)(0.,){Straight}{1}
        \FALabel(3.25,5.93)[t]{${L_L}_b^f$}
        \FAProp(14.,17.5)(10.,13.)(0.,){Straight}{1}
        \FALabel(12.3215,16.1998)[br]{${L_L}_c^g$}
        \FAProp(20,7.)(13.5,7.)(0.,){ScalarDash}{1}
        \FALabel(16.4998,5.47958)[t]{$\phi_a$}
        \FAProp(6.5,7.)(13.5,7.)(0.,){Straight}{1}
        \FALabel(10.,5.93)[t]{${\ell_R}^h$}
        \FAProp(6.5,7.)(10.,13.)(0.,){ScalarDash}{1}
        \FALabel(7.60709,10.175)[br]{$\phi_e$}
        \FAProp(13.5,7.)(10.,13.)(0.,){Straight}{1}
        \FALabel(12.6089,10.301)[bl]{${L_L}_j^k$}
        \FAVert(6.5,7.){0}
        \FAVert(13.5,7.){0}
        \FAVert(10.,13.){0}    
    
        \FADiagram{$\ii\Gamma^\kappa_4$}
        \FAProp(0.,15.)(6.5,13.)(0.,){ScalarDash}{1}
        \FALabel(3.63231,14.7625)[b]{$\phi_d$}
        \FAProp(0.,5.)(10.,7.)(0.,){Straight}{1}
        \FALabel(5.30398,4.9601)[t]{${L_L}_b^f$}
        \FAProp(20,15.)(13.5,13.)(0.,){Straight}{1}
        \FALabel(16.2942,15.0015)[b]{${L_L}_c^g$}
        \FAProp(20,5.)(10.,7.)(0.,){ScalarDash}{1}
        \FALabel(14.745,5.20525)[t]{$\phi_a$}
        \FAProp(6.5,13.)(13.5,13.)(0.,){Straight}{-1}
        \FALabel(10.,14.07)[b]{${\ell_R}^h$}
        \FAProp(6.5,13.)(10.,7.)(0.,){Straight}{1}
        \FALabel(7.39114,9.699)[tr]{${L_L}_j^k$}
        \FAProp(13.5,13.)(10.,7.)(0.,){ScalarDash}{1}
        \FALabel(12.3929,9.82497)[tl]{$\phi_e$}
        \FAVert(6.5,13.){0}
        \FAVert(13.5,13.){0}
        \FAVert(10.,7.){0}

        \FADiagram{$\ii\Gamma^\kappa_5$}
        \FAProp(0.,15.)(10.,13.)(0.,){ScalarDash}{1}
        \FALabel(5.25495,14.7948)[b]{$\phi_d$}
        \FAProp(0.,5.)(6.5,7.)(0.,){Straight}{1}
        \FALabel(3.70583,4.99854)[t]{${L_L}_b^f$}
        \FAProp(20,15.)(13.5,7.)(0.,){Straight}{1}
        \FALabel(17.3355,13.2775)[br]{${L_L}_c^g$}
        \FAProp(20,5.)(10.,13.)(0.,){ScalarDash}{1}
        \FALabel(17.8282,5.87899)[tr]{$\phi_a$}
        \FAProp(6.5,7.)(13.5,7.)(0.,){Sine}{0}
        \FALabel(10.,5.93)[t]{$B,W^A$}
        \FAProp(6.5,7.)(10.,13.)(0.,){Straight}{1}
        \FALabel(7.47761,10.2506)[br]{${L_L}_j^k$}
        \FAProp(13.5,7.)(10.,13.)(0.,){Straight}{1}
        \FALabel(11.8911,9.699)[tr]{${L_L}_m^h$}
        \FAVert(6.5,7.){0}
        \FAVert(13.5,7.){0}
        \FAVert(10.,13.){0}
    
    \end{feynartspicture}
    \begin{feynartspicture}(432,168)(3,1.1)
   
        \FADiagram{$\ii\Gamma^\kappa_{6}$}
        \FAProp(0.,15.)(6.5,13.)(0.,){ScalarDash}{1}
        \FALabel(3.63231,14.7625)[b]{$\phi_d$}
        \FAProp(0.,5.)(10.,7.)(0.,){Straight}{1}
        \FALabel(5.30398,4.9601)[t]{${L_L}_b^f$}
        \FAProp(20,15.)(10.,7.)(0.,){Straight}{1}
        \FALabel(17.6235,14.3154)[br]{${L_L}_c^g$}
        \FAProp(20,5.)(13.5,13.)(0.,){ScalarDash}{1}
        \FALabel(17.1171,7.66351)[tr]{$\phi_a$}
        \FAProp(6.5,13.)(13.5,13.)(0.,){Sine}{0}
        \FALabel(10.,14.07)[b]{$B,W^A$}
        \FAProp(6.5,13.)(10.,7.)(0.,){ScalarDash}{1}
        \FALabel(7.60709,9.82497)[tr]{$\phi_e$}
        \FAProp(13.5,13.)(10.,7.)(0.,){ScalarDash}{1}
        \FALabel(11.1071,10.175)[br]{$\phi_j$}
        \FAVert(6.5,13.){0}
        \FAVert(13.5,13.){0}
        \FAVert(10.,7.){0}    
    
        \FADiagram{$\ii\Gamma^\kappa_7$}
        \FAProp(0.,15.)(6.,13.5)(0.,){ScalarDash}{1}
        \FALabel(3.3153,15.0312)[b]{$\phi_d$}
        \FAProp(0.,5.)(6.,6.5)(0.,){Straight}{1}
        \FALabel(3.37593,4.72628)[t]{${L_L}_b^f$}
        \FAProp(20,15.)(12.,10.)(0.,){Straight}{1}
        \FALabel(15.6585,13.3344)[br]{${L_L}_c^g$}
        \FAProp(20,5.)(12.,10.)(0.,){ScalarDash}{1}
        \FALabel(16.209,8.1224)[bl]{$\phi_a$}
        \FAProp(6.,13.5)(6.,6.5)(0.,){Sine}{0}
        \FALabel(4.93,10.)[r]{$B,W^A$}
        \FAProp(6.,13.5)(12.,10.)(0.,){ScalarDash}{1}
        \FALabel(9.17503,12.3929)[bl]{$\phi_e$}
        \FAProp(6.,6.5)(12.,10.)(0.,){Straight}{1}
        \FALabel(9.301,7.39114)[tl]{${L_L}_j^k$}
        \FAVert(6.,13.5){0}
        \FAVert(6.,6.5){0}
        \FAVert(12.,10.){0}
    
        \FADiagram{$\ii\Gamma^\kappa_8$}
        \FAProp(0.,15.)(8.,10.)(0.,){ScalarDash}{1}
        \FALabel(3.791,11.8776)[tr]{$\phi_d$}
        \FAProp(0.,5.)(8.,10.)(0.,){Straight}{1}
        \FALabel(4.3415,6.6656)[tl]{${L_L}_b^f$}
        \FAProp(20,15.)(14.,13.5)(0.,){Straight}{1}
        \FALabel(16.6241,15.2737)[b]{${L_L}_c^g$}
        \FAProp(20,5.)(14.,6.5)(0.,){ScalarDash}{1}
        \FALabel(16.6847,4.96881)[t]{$\phi_a$}
        \FAProp(14.,13.5)(14.,6.5)(0.,){Sine}{0}
        \FALabel(15.07,10.)[l]{$B,W^A$}
        \FAProp(14.,13.5)(8.,10.)(0.,){Straight}{1}
        \FALabel(10.699,12.6089)[br]{${L_L}_j^k$}
        \FAProp(14.,6.5)(8.,10.)(0.,){ScalarDash}{1}
        \FALabel(10.825,7.60709)[tr]{$\phi_e$}
        \FAVert(14.,13.5){0}
        \FAVert(14.,6.5){0}
        \FAVert(8.,10.){0}
    
    \end{feynartspicture}
    \begin{feynartspicture}(432,168)(3,1.1)
    
        \FADiagram{$\ii\Gamma^\kappa_{9}$}
        \FAProp(5.,17.5)(10.,13.)(0.,){ScalarDash}{1}
        \FALabel(6.88965,16.2363)[bl]{$\phi_d$}
        \FAProp(0.,7.)(6.5,7.)(0.,){Straight}{1}
        \FALabel(3.25,5.93)[t]{${L_L}_b^f$}
        \FAProp(14.,17.5)(10.,13.)(0.,){Straight}{1}
        \FALabel(12.3215,16.2998)[br]{${L_L}_c^g$}
        \FAProp(20,7.)(13.5,7.)(0.,){ScalarDash}{1}
        \FALabel(16.4998,5.47958)[t]{$\phi_a$}
        \FAProp(6.5,7.)(13.5,7.)(0.,){Sine}{0}
        \FALabel(10.,5.3)[t]{$B,W^A$}
        \FAProp(6.5,7.)(10.,13.)(0.,){Straight}{1}
        \FALabel(7.39114,10.301)[br]{${L_L}_j^k$}
        \FAProp(13.5,7.)(10.,13.)(0.,){ScalarDash}{1}
        \FALabel(12.3929,10.175)[bl]{$\phi_e$}
        \FAVert(6.5,7.){0}
        \FAVert(13.5,7.){0}
        \FAVert(10.,13.){0}    
    
        \FADiagram{$\ii\Gamma^\kappa_{10}$}
        \FAProp(0.,15.)(6.5,13.)(0.,){ScalarDash}{1}
        \FALabel(3.63231,14.7625)[b]{$\phi_d$}
        \FAProp(0.,5.)(10.,7.)(0.,){Straight}{1}
        \FALabel(5.30398,4.9601)[t]{${L_L}_b^f$}
        \FAProp(20,15.)(13.5,13.)(0.,){Straight}{1}
        \FALabel(16.2942,15.0015)[b]{${L_L}_c^g$}
        \FAProp(20,5.)(10.,7.)(0.,){ScalarDash}{1}
        \FALabel(14.745,5.20525)[t]{$\phi_a$}
        \FAProp(6.5,13.)(13.5,13.)(0.,){Sine}{0}
        \FALabel(10.,14.07)[b]{$B,W^A$}
        \FAProp(6.5,13.)(10.,7.)(0.,){ScalarDash}{1}
        \FALabel(7.60709,9.82497)[tr]{$\phi_e$}
        \FAProp(13.5,13.)(10.,7.)(0.,){Straight}{1}
        \FALabel(12.6089,9.699)[tl]{${L_L}_j^k$}
        \FAVert(6.5,13.){0}
        \FAVert(13.5,13.){0}
        \FAVert(10.,7.){0}
    
        \FADiagram{$\ii\Gamma^\kappa_{11}$}
        \FAProp(0.,15.)(9.5,14.5)(0.,){ScalarDash}{1}
        \FALabel(4.81833,15.5682)[b]{$\phi_d$}
        \FAProp(0.,5.)(9.5,5.5)(0.,){Straight}{1}
        \FALabel(4.83147,4.18214)[t]{${L_L}_b^f$}
        \FAProp(20,15.)(9.5,5.5)(0.,){Straight}{1}
        \FALabel(16.9134,13.5046)[br]{${L_L}_c^g$}
        \FAProp(20,5.)(9.5,14.5)(0.,){ScalarDash}{1}
        \FALabel(17.8203,7.77045)[bl]{$\phi_a$}
        \FAProp(9.5,14.5)(9.5,5.5)(0.8,){ScalarDash}{1}
        \FALabel(5.08,10.)[r]{$\phi_e$}
        \FAProp(9.5,14.5)(9.5,5.5)(-0.444,){ScalarDash}{1}
        \FALabel(10.678,10.)[r]{$\phi_j$}
        \FAVert(9.5,14.5){0}
        \FAVert(9.5,5.5){0}
    
    \end{feynartspicture}
    \caption{\label{fig:EFT_divergent}Divergent Feynman diagrams in effective theory.}
    \end{center}
    \end{figure}    

\bigskip

\noindent The contribution $\ii \Gamma^{\kappa}_1$ in the effective theory to the loop amplitude of the neutrino mass matrix:    
    \begin{align}
        {\bar\mu}^\epsilon \ii (\Gamma^\kappa_1)^{gf}_{abcd} &= \int \frac{\dd^dk}{(2\pi)^d} \left[\frac{\ii}{2}{\bar\mu}^\epsilon \kappa_{gk}(\epsilon_{ca}\epsilon_{je}+\epsilon_{ce}\epsilon_{ja})P_\text{L}\right]\frac{\ii{\slashed k}}{k^2}\left(-\ii{\bar\mu}^{\epsilon/2}\left(Y_{\ell}^\dagger\right)_{kh}\delta_{dj}P_\text{R}\right)\nonumber\\
        &\quad \times\frac{\ii({\slashed k}-{\slashed q_2})}{(k-q_2)^2}\left(-\ii{\bar\mu}^{\epsilon/2}\left(Y_{\ell}\right)_{hf}\delta_{be}P_\text{L}\right)\frac{\ii}{(k-p_1-p_2)^2-m^2} \nonumber\\
        &= -{\bar\mu}^{2\epsilon}\frac12 \left(\kappa Y_{\ell}^\dagger Y_{\ell}\right)_{gf} (\epsilon_{ac}\epsilon_{bd}+\epsilon_{ad}\epsilon_{bc})P_\text{L} \int\frac{\dd^dk}{(2\pi)^d} \frac{{\slashed k}}{k^2}\frac{{\slashed k}-{\slashed q_2}}{(k-q_2)^2}\frac{1}{(k-p_1-p_2)^2-m^2}.
        \label{eq:Gk1}
    \end{align}
    
\noindent The contribution $\ii \Gamma^{\kappa}_2$ in the effective theory to the loop amplitude of the neutrino mass matrix:   
    \begin{align}
        {\bar\mu}^\epsilon \ii (\Gamma^\kappa_2)^{gf}_{abcd} &= \int \frac{\dd^dk}{(2\pi)^d} \left(-\ii{\bar\mu}^{\epsilon/2}\left(Y_{\ell}^T\right)_{gh}\delta_{ce}P_\text{L}\right)\frac{-\ii{\slashed k}}{k^2}\left(-\ii{\bar\mu}^{\epsilon/2}\left(Y_{\ell}^*\right)_{hk}\delta_{aj}P_\text{R}\right)\nonumber\\
        &\quad\times\frac{-\ii({\slashed k}-{\slashed p_2})}{(k-p_2)^2} \left[\frac{\ii}{2}{\bar\mu}^\epsilon \kappa_{kf}(\epsilon_{je}\epsilon_{bd}+\epsilon_{jd}\epsilon_{be})P_\text{L}\right]\frac{\ii}{(k+p_1)^2-m^2} \nonumber\\
        &= -{\bar\mu}^{2\epsilon}\frac12 \left(Y_{\ell}^T Y_{\ell}^* \kappa\right)_{gf} (\epsilon_{ac}\epsilon_{bd}+\epsilon_{ad}\epsilon_{bc})P_\text{L} \int\frac{\dd^dk}{(2\pi)^d} \frac{{\slashed k}}{k^2}\frac{{\slashed k}-{\slashed p_2}}{(k-p_2)^2}\frac{1}{(k+p_1)^2-m^2}.
        \label{eq:Gk2}
    \end{align}

\noindent The contribution $\ii \Gamma^{\kappa}_3$ in the effective theory to the loop amplitude of the neutrino mass matrix:    
    \begin{align}
        {\bar\mu}^\epsilon \ii (\Gamma^\kappa_3)^{gf}_{abcd} &= \int \frac{\dd^dk}{(2\pi)^d} \left[\frac{\ii}{2}{\bar\mu}^\epsilon \kappa_{gk}(\epsilon_{cd}\epsilon_{je}+\epsilon_{ce}\epsilon_{jd})P_\text{L}\right] \frac{\ii({\slashed k}-{\slashed p_2})}{(k-p_2)^2} \left(-\ii{\bar\mu}^{\epsilon/2}\left(Y_{\ell}^\dagger\right)_{kh}\delta_{aj}P_\text{R}\right) \nonumber\\
        &\quad \times \frac{\ii{\slashed k}}{k^2} \left(-\ii{\bar\mu}^{\epsilon/2}\left(Y_{\ell}\right)_{hf}\delta_{be}P_\text{L}\right) \frac{\ii}{(k-q_1)^2-m^2} \nonumber\\
        &= -{\bar\mu}^{2\epsilon}\frac12 \left(\kappa Y_{\ell}^\dagger Y_{\ell}\right)_{gf} (\epsilon_{ab}\epsilon_{cd}-\epsilon_{ad}\epsilon_{bc})P_\text{L} \int\frac{\dd^dk}{(2\pi)^d} \frac{{\slashed k}-{\slashed p_2}}{(k-p_2)^2}\frac{{\slashed k}}{k^2}\frac{1}{(k-q_1)^2-m^2}.
        \label{eq:Gk3}
    \end{align}
    
\noindent The contribution $\ii \Gamma^{\kappa}_4$ in the effective theory to the loop amplitude of the neutrino mass matrix:    
    \begin{align}
        {\bar\mu}^\epsilon \ii (\Gamma^\kappa_4)^{gf}_{abcd} &= \int \frac{\dd^dk}{(2\pi)^d} \left(-\ii{\bar\mu}^{\epsilon/2}\left(Y_{\ell}^T\right)_{gh}\delta_{ce}P_\text{L}\right)\frac{-\ii{\slashed k}}{k^2} \left(-\ii{\bar\mu}^{\epsilon/2}\left(Y_{\ell}^*\right)_{hk}\delta_{dj}P_\text{R}\right)\nonumber\\
        &\quad\times\frac{-\ii({\slashed k}+{\slashed q_2})}{(k+q_2)^2} \left[\frac{\ii}{2}{\bar\mu}^\epsilon \kappa_{kf}(\epsilon_{je}\epsilon_{ba}+\epsilon_{ja}\epsilon_{be})P_\text{L}\right]\frac{\ii}{(k+p_1)^2-m^2} \nonumber\\
        &= -{\bar\mu}^{2\epsilon}\frac12 \left(Y_{\ell}^T Y_{\ell}^* \kappa \right)_{gf} (\epsilon_{ab}\epsilon_{cd}-\epsilon_{ad}\epsilon_{bc})P_\text{L} \int\frac{\dd^dk}{(2\pi)^d} \frac{{\slashed k}}{k^2}\frac{{\slashed k}+{\slashed q_2}}{(k+q_2)^2}\frac{1}{(k+p_1)^2-m^2}.
        \label{eq:Gk4}
    \end{align}

\noindent The contribution $\ii \Gamma^{\kappa}_5$ in the effective theory to the loop amplitude of the neutrino mass matrix:    
    \begin{align}
        {\bar\mu}^\epsilon \ii (\Gamma^\kappa_5)^{gf}_{abcd} &= \int \frac{\dd^dk}{(2\pi)^d} \left(\ii {\bar\mu}^{\epsilon/2} g_A (T^A)^T_{cm}\gamma_\mu P_\text{R}\right)\frac{-\ii{\slashed k}}{k^2} \left[ \frac{\ii}{2} {\bar\mu}^\epsilon \kappa_{gf}(\epsilon_{am}\epsilon_{dj}+\epsilon_{aj}\epsilon_{dm})P_\text{L} \right] \nonumber\\
        &\quad \times\frac{-\ii({\slashed k} - {\slashed p_2} + {\slashed q_2})}{(k-p_2+q_2)^2} \left(-\ii{\bar\mu}^{\epsilon/2}g_A T^A_{jb}\gamma_\nu P_\text{L}\right) \ii\frac{-g^{\mu\nu}+(1-\xi_A)\frac{l^\mu l^\nu}{l^2}}{l^2} \nonumber\\
        &= -{\bar\mu}^{2\epsilon} \frac18 \kappa_{gf} g_A^2 (\epsilon_{ab}\epsilon_{cd}+\epsilon_{ac}\epsilon_{bd})P_\text{L}\int \frac{\dd^dk}{(2\pi)^d} \gamma_\mu \frac{{\slashed k}}{k^2} \frac{({\slashed k} - {\slashed p_2} + {\slashed q_2})}{(k-p_2+q_2)^2} \gamma_\nu \frac{-g^{\mu\nu}+(1-\xi_A)\frac{l^\mu l^\nu}{l^2}}{l^2},
        \label{eq:Gk5}
    \end{align}
    where $l=k+p_1$.

\bigskip

\noindent The contribution $\ii \Gamma^{\kappa}_{6}$ in the effective theory to the loop amplitude of the neutrino mass matrix:    
    \begin{align}
        {\bar\mu}^\epsilon \ii (\Gamma^\kappa_{6})^{gf}_{abcd} &= \int\frac{\dd^dk}{(2\pi)^d} \left[ {\bar\mu}^\epsilon \ii \kappa_{gf} \frac12 (\epsilon_{ce}\epsilon_{bj}+\epsilon_{cj}\epsilon_{be})P_\text{L} \right] \left[-\ii{\bar\mu}^{\epsilon/2} g_A (-p_2-k-q_1+p_1)_\mu T^A_{ja}\right] \nonumber \\
        &\quad \times \left[-\ii{\bar\mu}^{\epsilon/2} g_A (q_2+k)_\nu T^A_{ed}\right]\frac{\ii}{(k+q_1-p_1)^2-m^2}\frac{\ii}{k^2-m^2} \ii \frac{-g^{\mu\nu}+(1-\xi_A)\frac{l^\mu l^\nu}{l^2}}{l^2} \nonumber\\
        &= {\bar\mu}^{2\epsilon} \frac{1}{8} \kappa_{gf} g_A^2(\epsilon_{ab}\epsilon_{cd}+\epsilon_{ac}\epsilon_{bd})P_\text{L}\nonumber \\
        &\quad\times \int \frac{\dd^dk}{(2\pi)^d}\frac{(p_1-p_2-q_1-k)_\mu}{(k+q_1-p_1)^2-m^2}\frac{(q_2+k)_\nu}{k^2-m^2}\frac{-g^{\mu\nu}+(1-\xi_A)\frac{l^\mu l^\nu}{l^2}}{l^2},
        \label{eq:Gk6}
    \end{align}
    where $l=k-q_2$.

\bigskip

\noindent The contribution $\ii \Gamma^{\kappa}_7$ in the effective theory to the loop amplitude of the neutrino mass matrix:    
    \begin{align}
        {\bar\mu}^\epsilon \ii (\Gamma^\kappa_7)^{gf}_{abcd} &= \int\frac{\dd^dk}{(2\pi)^d} \left[ {\bar\mu}^\epsilon \ii \kappa_{gf}\frac12 (\epsilon_{ce}\epsilon_{ja}+\epsilon_{ca}\epsilon_{je})P_\text{L} \right] \frac{\ii\slashed{k}}{k^2}
        \left(-\ii{\bar\mu}^{\epsilon/2} g_A T^A_{bj} \gamma_\mu P_\text{L}\right) \nonumber\\
        &\quad \times \ii \frac{-g^{\mu\nu}+(1-\xi_A)\frac{l^\mu l^\nu}{l^2}}{l^2} \left[-\ii {\bar\mu}^{\epsilon/2} g_A (q_2-k+p_1+p_2)_\nu T^A_{de}\right] \frac{\ii}{(k-p_1-p_2)^2-m^2} \nonumber\\
        &= -{\bar\mu}^{2\epsilon}\frac18 \kappa_{gf} P_\text{L}\begin{cases}
            g_1^2(\epsilon_{ab}\epsilon_{cd}+\epsilon_{ac}\epsilon_{bd}), & \text{U}(1)\\
            g_2^2(\epsilon_{ab}\epsilon_{cd}+3\epsilon_{ac}\epsilon_{bd}+2\epsilon_{ad}\epsilon_{bc}), & \SU(2)
        \end{cases}\nonumber\\
        &\quad \times\int\frac{\dd^dk}{(2\pi)^d} \frac{\slashed{k}}{k^2}\gamma_\mu \frac{-g^{\mu\nu}+(1-\xi_A)\frac{l^\mu l^\nu}{l^2}}{l^2}\frac{(q_2-k+p_1+p_2)_\nu}{(k-p_1-p_2)^2-m^2},
        \label{eq:Gk7}
    \end{align}
    where $l=k-q_1$.

\bigskip

\noindent The contribution $\ii \Gamma^{\kappa}_8$ in the effective theory to the loop amplitude of the neutrino mass matrix:    
    \begin{align}
        {\bar\mu}^\epsilon \ii (\Gamma^\kappa_{8})^{gf}_{abcd} &= \int\frac{\dd^dk}{(2\pi)^d} \left(\ii {\bar\mu}^{\epsilon/2} g_A (T^A)^T_{cj} \gamma_\mu P_\text{R}\right) \frac{\ii\slashed{k}}{k^2} \left[{\bar\mu}^\epsilon \ii \kappa_{gf}\frac12 (\epsilon_{dj}\epsilon_{eb}+\epsilon_{db}\epsilon_{ej})P_\text{L} \right] \nonumber\\
        &\quad \times \left[-\ii {\bar\mu}^{\epsilon/2} g_A (-p_2+k-p_1-p_2)_\nu T^A_{ea}\right]
        \ii \frac{-g^{\mu\nu}+(1-\xi_A)\frac{l^\mu l^\nu}{l^2}}{l^2}
        \frac{\ii}{(k-p_1-p_2)^2-m^2} \nonumber\\
        &= {\bar\mu}^{2\epsilon}\frac18 \kappa_{gf} P_\text{L}\begin{cases}
            g_1^2 (\epsilon_{ab}\epsilon_{cd}+\epsilon_{ac}\epsilon_{bd}), & \text{U}(1)\\
            g_2^2 (\epsilon_{ab}\epsilon_{cd}+3\epsilon_{ac}\epsilon_{bd}+2\epsilon_{ad}\epsilon_{bc}), & \SU(2)
        \end{cases}\nonumber\\
        &\quad \times \int\frac{\dd^dk}{(2\pi)^d} \gamma_\mu \frac{\slashed{k}}{k^2}
        \frac{-g^{\mu\nu}+(1-\xi_A)\frac{l^\mu l^\nu}{l^2}}{l^2} \frac{(k-p_1-2p_2)_\nu}{(k-p_1-p_2)^2-m^2},
        \label{eq:Gk8}
    \end{align}
    where $l=k-p_1$.

\bigskip

\noindent The contribution $\ii \Gamma^{\kappa}_{9}$ in the effective theory to the loop amplitude of the neutrino mass matrix:    
    \begin{align}
        {\bar\mu}^\epsilon \ii (\Gamma^\kappa_{9})^{gf}_{abcd} &= \int\frac{\dd^dk}{(2\pi)^d} \left[ {\bar\mu}^\epsilon \ii \kappa_{gf}\frac12 (\epsilon_{cd}\epsilon_{je}+\epsilon_{ce}\epsilon_{jd})P_\text{L} \right] \frac{\ii\slashed{k}}{k^2}
        \left(-\ii {\bar\mu}^{\epsilon/2} g_A T^A_{jb} \gamma_\mu P_\text{L}\right) \nonumber\\
        &\quad \times \ii \frac{-g^{\mu\nu}+(1-\xi_A)\frac{l^\mu l^\nu}{l^2}}{l^2} \left[-\ii {\bar\mu}^{\epsilon/2} g_A (-p_2-k-p_2+q_1)_\nu T^A_{ea}\right] \frac{\ii}{(k+p_2-q_1)^2-m^2} \nonumber\\
        &= -{\bar\mu}^{2\epsilon}\frac18 \kappa_{gf} P_\text{L}\begin{cases}
            g_1^2(\epsilon_{ab}\epsilon_{cd}+\epsilon_{ac}\epsilon_{bd}), & \text{U}(1)\\
            g_2^2(3\epsilon_{ab}\epsilon_{cd}+\epsilon_{ac}\epsilon_{bd}-2\epsilon_{ad}\epsilon_{bc}), & \SU(2)
        \end{cases}\nonumber\\
        &\quad \times \int\frac{\dd^dk}{(2\pi)^d} \frac{\slashed{k}}{k^2}\gamma_\mu \frac{-g^{\mu\nu}+(1-\xi_A)\frac{l^\mu l^\nu}{l^2}}{l^2} \frac{(-k-2p_2+q_1)_\nu}{(k+p_2-q_1)^2-m^2},
        \label{eq:Gk9}
    \end{align}
    where $l=k-q_1$.

\bigskip

\noindent The contribution $\ii \Gamma^{\kappa}_{10}$ in the effective theory to the loop amplitude of the neutrino mass matrix:    
    \begin{align}
        {\bar\mu}^\epsilon \ii (\Gamma^\kappa_{10})^{gf}_{abcd} &=  \int\frac{\dd^dk}{(2\pi)^d} \left(\ii {\bar\mu}^{\epsilon/2} g_A (T^A)^T_{cj} \gamma_\mu P_\text{R}\right) \frac{\ii\slashed{k}}{k^2} \left[ {\bar\mu}^\epsilon \ii \kappa_{gf}\frac12 (\epsilon_{je}\epsilon_{ba}+\epsilon_{be}\epsilon_{ja})P_\text{L} \right] \nonumber\\
        &\quad \times \frac{\ii}{(k-p_1+q_2)^2-m^2}\left[-\ii {\bar\mu}^{\epsilon/2} g_A (q_2+k-p_1+q_2)_\nu T^A_{ed}\right]
        \ii \frac{-g^{\mu\nu}+(1-\xi_A)\frac{l^\mu l^\nu}{l^2}}{l^2} \nonumber\\
        &= {\bar\mu}^{2\epsilon}\frac18 \kappa_{gf} P_\text{L}\begin{cases}
            g_1^2 (\epsilon_{ab}\epsilon_{cd}+\epsilon_{ac}\epsilon_{bd}), & \text{U}(1)\\
            g_2^2 (3\epsilon_{ab}\epsilon_{cd}+\epsilon_{ac}\epsilon_{bd}-2\epsilon_{ad}\epsilon_{bc}), & \SU(2)
        \end{cases}\nonumber\\
        &\quad \times\int \frac{\dd^dk}{(2\pi)^d}\gamma_\mu \frac{\slashed k}{k^2} \frac{(k-p_1+2q_2)_\nu}{(k-p_1+q_2)^2-m^2}\frac{-g^{\mu\nu}+(1-\xi_A)\frac{l^\mu l^\nu}{l^2}}{l^2},
        \label{eq:Gk10}
    \end{align}
    where $l=k-p_1$.

\bigskip

\noindent The contribution $\ii \Gamma^{\kappa}_{11}$ in the effective theory to the loop amplitude of the neutrino mass matrix:   
    \begin{align}
        {\bar\mu}^\epsilon \ii (\Gamma^\kappa_{11})^{gf}_{abcd} &= \frac12 \int\frac{\dd^dk}{(2\pi)^d}\left[ {\bar\mu}^\epsilon \ii \kappa_{gf}\frac12 (\epsilon_{ce}\epsilon_{bj}+\epsilon_{cj}\epsilon_{be})P_\text{L} \right] \left[ - {\bar\mu}^\epsilon \ii \frac{\lambda}{2} (\delta_{aj}\delta_{de} + \delta_{ae}\delta_{dj}) \right] \nonumber\\
        &\quad \times  \frac{\ii}{k^2-m^2}\frac{\ii}{(k+q_1-p_1)^2-m^2} \nonumber\\
        &= {\bar\mu}^{2\epsilon} \frac{\lambda}{4} \kappa_{gf}(\epsilon_{ab}\epsilon_{cd}+\epsilon_{ac}\epsilon_{bd})P_\text{L}\int\frac{\dd^dk}{(2\pi)^d}\frac{1}{k^2-m^2}\frac{1}{(k+q_1-p_1)^2-m^2},
        \label{eq:Gk11}
    \end{align}
    where the first factor of $\frac12$ is a symmetry factor.

\FloatBarrier

\section{Counterterm}\label{sec:counterterm}

Using the computed loop amplitudes in App.~\ref{app:kappa_loops_EFT}, we can derive the counterterm for $\kappa$. The counterterm has been computed in the literature and therefore provides a verification of our results. 

The counterterm is defined to cancel the divergent parts of the amplitudes. The Feynman diagram that results from the Feynman rule corresponding to the counterterm gives
\begin{equation}
    {\bar\mu}^\epsilon\ii(\Gamma^\kappa_\text{ct})^{gf}_{abcd} = {\bar\mu}^\epsilon (\delta \kappa)_{gf} \frac{\ii}{2} (\epsilon_{ab}\epsilon_{cd}+\epsilon_{ac}\epsilon_{bd})P_\text{L},
\end{equation}
which is defined to cancel the divergent parts of the Feynman diagrams. Thus, we have
\begin{equation}
    \Gamma^\kappa_\text{ct} + \sum_{i=1}^{11}\left(\Gamma^\kappa_i\right)|_\text{div} = 0.
\end{equation}
Extracting the divergent parts of the integrals using the Passarino--Veltman functions defined in the Mathematica package \texttt{PackageX}~\cite{Patel:2015tea}. Working in $d=4-\epsilon$ dimensions, we use the $\overline{\text{MS}}$ scheme in which only the $\epsilon$-pole is taken as the divergence. Thus, this gives the counterterm
\begin{equation}
    \delta\kappa = -\frac{1}{16\pi^2} \left[2 Y_{\ell}^T Y_{\ell}^* \kappa + 2 \kappa Y_{\ell}^\dagger Y_{\ell} - \kappa \lambda + g_1^2\kappa  \left(\xi_1 - \frac32\right) + 3 g_2^2 \kappa  \left(\xi_2 - \frac12 \right)\right],
\end{equation}
which is the one that is known from the literature, see~e.g.~Ref.~\cite{Antusch:2002rr}.

\end{document}